\documentclass[usenatbib,useAMS]{mnras}  
\usepackage{float}
\usepackage{amssymb}

\usepackage{graphicx}
\usepackage{epstopdf}

\title[WASP-121\,b]{WASP-121\,b: a hot Jupiter in a polar orbit and close to tidal disruption}

\author[L. Delrez et al.]{L.~Delrez,$^1$\thanks{E-mail: \url{ldelrez@ulg.ac.be}} A.~Santerne,$^{2}$ J.-M.~Almenara,$^{3,4}$ D. R.~Anderson,$^5$ A.~Collier-Cameron,$^6$   
          \newauthor
	  R. F.~D\'iaz,$^7$ M.~Gillon,$^1$ C.~Hellier,$^5$ E.~Jehin,$^1$ M.~Lendl,$^{1,7}$ P. F. L.~Maxted,$^5$
	  \newauthor
	  M.~Neveu-VanMalle,$^{7,8}$ F.~Pepe,$^7$ D.~Pollacco,$^9$ D.~Queloz,$^{8,7}$ D.~S\'{e}gransan,$^7$
	  \newauthor
	  B.~Smalley,$^5$  A. M. S.~Smith,$^{10,5}$ A. H. M. J.~Triaud,$^{11,12,7}$\thanks{Fellow of the Swiss National Science Foundation.} S.~Udry,$^7$
	  \newauthor
	  V.~Van Grootel,$^1$  and R. G.~West$^9$\vspace{0.5cm}\\
            $^1$ Institut d'Astrophysique et G\'eophysique, Universit\'{e} de Li\`{e}ge, all\'{e}e du 6 Ao\^{u}t 17, B-4000 Li\`{e}ge, Belgium\\
            $^2$ Instituto de Astrof\'isica e Ci\^{e}ncias do Espa\c co, Universidade do Porto, CAUP, Rua das Estrelas, PT4150-762 Porto, Portugal\\ 
            $^3$ Univ. Grenoble Alpes, IPAG, F-38000 Grenoble, France\\
            $^4$ CNRS, IPAG, F-38000 Grenoble, France\\  
	    $^5$ Astrophysics Group, Keele University, Staffordshire, ST5 5BG, UK\\
	    $^6$ SUPA, School of Physics and Astronomy, University of St. Andrews, North Haugh, Fife, KY16 9SS, UK\\	    
	    $^7$ Observatoire de Gen\`eve, Universit\'{e} de Gen\`{e}ve, 51 Chemin des Maillettes, 1290 Sauverny, Switzerland\\   
	    $^8$ Cavendish Laboratory, Department of Physics, University of Cambridge, JJ Thomson Avenue, Cambridge, CB3 0HE, UK\\
	    $^9$ Department of Physics, University of Warwick, Coventry CV4 7AL, UK\\    
	    $^{10}$ N. Copernicus Astronomical Centre, Polish Academy of Sciences, Bartycka 18, 00-716 Warsaw, Poland\\
	    $^{11}$ Centre for Planetary Sciences, University of Toronto at Scarborough, 1265 Military Trail, Toronto, ON, M1C 1A4, Canada\\ 
	    $^{12}$ Department of Astronomy \& Astrophysics, University of Toronto, Toronto, ON M5S 3H4, Canada\\       
	    }

\begin{document}

\date{Received date / accepted date}

\pagerange{\pageref{firstpage}--\pageref{lastpage}} \pubyear{2014}

  \maketitle
  
  \label{firstpage}

  \begin{abstract}
We present the discovery by the WASP-South survey, in close collaboration with the Euler and TRAPPIST telescopes, of WASP-121\,b, a new remarkable short-period transiting hot Jupiter, whose planetary nature has been statistically validated by the \texttt{PASTIS} software. The planet has a mass of \hbox{$1.183_{-0.062}^{+0.064}$ $M_{\mathrm{Jup}}$}, a radius of \hbox{1.865 $\pm$ 0.044 $R_{\mathrm{Jup}}$}, and transits every $1.2749255_{-0.0000025}^{+0.0000020}$ days an active F6-type main-sequence star ($V$=10.4, $1.353_{-0.079}^{+0.080}$ $M_{\odot}$, 1.458 $\pm$ 0.030 $R_{\odot}$, \hbox{$T_{\mathrm{eff}}$ = 6460 $\pm$ 140 K}). A notable property of WASP-121\,b is that its orbital semi-major axis is only $\sim$1.15 times larger than its Roche limit, which suggests that the planet might be close to tidal disruption. Furthermore, its large size and extreme irradiation \hbox{($\sim$$7.1\:10^{9}$ erg $\mathrm{s}^{-1} \mathrm{cm}^{-2}$)} make it an excellent target for atmospheric studies via secondary eclipse observations. Using the TRAPPIST telescope, we indeed detect its emission in the \hbox{$z'$-band} at better than $\sim$4$\sigma$, the measured occultation depth being \hbox{603 $\pm$ 130 ppm}. Finally, from a measurement of the Rossiter-McLaughlin effect with the CORALIE spectrograph, we infer a sky-projected spin-orbit angle of \hbox{$257.8_{-5.5}^{+5.3}$ deg}. This result indicates a significant misalignment between the spin axis of the host star and the orbital plane of the planet, the planet being in a nearly polar orbit. Such a high misalignment suggests a migration of the planet involving strong dynamical events with a third body.\vspace{0.5cm}
  \end{abstract}


%________________________________________________________________

\vspace{2cm}

%%%%%%%%%%%%%%%%%%%%%%%%%%%%%%%%%%%%%%
\section{Introduction}

\indent
Most of the transiting exoplanets found by ground-based transit surveys (e.g. WASP, \citealt{pollacco}; HATNet, \citealt{bakos}) are Jovian-type planets with orbital periods of just a few days, these planets being the easiest to detect for such surveys. The orbital period distribution of these so-called ``hot Jupiters'' is not smooth and presents a pile-up around periods of \hbox{$\sim$3-4 days} (see e.g. \citealt{cumming}). While the long-period drop-off can be explained by a lower transit probability for these systems combined to a selection effect, the reduced number of planets in orbital periods less than 2 days is definitely real, being seen in both ground- (e.g. WASP, \citealt{hellierpile}) and space-based (e.g. \textit{Kepler}, \citealt{howard}) transit surveys, as well as in radial velocity surveys (see e.g. \citealt{marcy}).\\
\indent 
\cite{fordrasio} suggested that the lower edge of the pile-up is defined not by an orbital period, but rather by a tidal limit, and found that the inner cutoff is actually close to twice the Roche limit ($a_{\mathrm{R}}$)\footnote{I.e. the critical orbital separation inside which a planet would lose mass via Roche lobe overflow.}. This can be naturally explained if planets were initially scattered into highly eccentric orbits with short pericenter distances from much further out, due e.g. to planet-planet interactions (e.g. \citealt{rasio96}, \citealt{marzari}, \citealt{moorhead}, \citealt{chatterjee}) and/or Kozai cycles (e.g. \citealt{kozai}, \citealt{lidov}, \citealt{wu}, \citealt{fabtrem}), and later circularized via tidal dissipation. On the contrary, they argued that this result is inconsistent with a disk-driven migration scenario (e.g. \citealt{goldreichtrem}, \citealt{linpapa}, \citealt{tanaka}, \citealt{lubow}), as the inner edge of the orbital period distribution should then be right at the Roche limit. The observed distribution of orbital obliquities, with many planets found on misaligned or retrograde orbits (e.g. \citealt{triaudrm}), also supports dynamical migration processes involving a third perturbing body, rather than disk migration.\\ 
\indent
The finding of several hot Jupiters with orbital separations $a$ lower than 2 $a_{\mathrm{R}}$, such as \hbox{WASP-12\,b} ($a$/$a_{\mathrm{R}}$$\sim$1.09, \citealt{w12}), \hbox{WASP-19\,b} ($a$/$a_{\mathrm{R}}$$\sim$1.08, \citealt{w19}), \hbox{WASP-103\,b} ($a$/$a_{\mathrm{R}}$$\sim$1.16, \citealt{w103}), \hbox{OGLE-TR-56\,b} ($a$/$a_{\mathrm{R}}$$\sim$1.23, \citealt{ogle56a}, \citealt{ogle56}), and \hbox{WTS-2\,b} ($a$/$a_{\mathrm{R}}$$\sim$1.27, \citealt{birkby}), challenged the scattering scenario as these planets would have been destroyed or completely ejected from their systems if they had been directly scattered to such short pericenter distances (see e.g. \citealt{guillochon}). However, \cite{matsumura} showed that these extreme orbits can still result from the scattering model, assuming first a scattering into an eccentric orbit beyond \hbox{2 $a_{\mathrm{R}}$}, followed by a slow inward migration and circularization through tidal dissipation inside the planet mainly until reaching $\sim$2 $a_{\mathrm{R}}$, and from then tidal decay through tidal dissipation inside the star only. The speed of the final tidal decay depends on the tidal dissipation efficiency of the star, which is parameterized by $Q'_{\star}$, the stellar tidal dissipation factor. Despite being an essential parameter in the theory of stellar tides, $Q'_{\star}$ is still poorly constrained, with estimates based on theoretical and observational studies ranging from to $10^5$ to $10^9$ (see e.g. \citealt{jackson2}, \citealt{ogilvie}, \citealt{penev}).\\
\indent
Planets in the $a$/$a_{\mathrm{R}}$$<$2 regime are thus key objects to further advance our understanding of how tidal forces influence the orbital evolution of close-in giant planets. Furthermore, these planets being highly irradiated due to their proximity to their host stars, they are also generally favorable targets for atmospheric studies via secondary eclipse observations (see e.g. \citealt{seagerdeming}, \citealt{anderson19}, \citealt{gilloncorot}). They thus provide us with a unique opportunity to study the relationship between the observed atmospheric properties of hot Jupiters and their tidal evolution stage. In this paper, we report the discovery of a new hot Jupiter of this rare kind by the WASP survey, \hbox{WASP-121\,b}, which orbits a 10.4 $V$-magnitude F-type star at just $\sim$1.15 times its Roche limit.\\
\indent
Section \ref{obs} presents the WASP discovery photometry, as well as the follow-up photometric and spectroscopic observations that we used to confirm and characterize the system. In \hbox{Section \ref{analysis}}, we describe the spectroscopic determination of the stellar properties and the derivation of the system parameters through a combined analysis of our photometric and spectroscopic data. The statistical validation of the planet is then described in Section \ref{pastis}. Finally, we discuss our results in Section \ref{discussion}.

\vspace{-0.4cm}
\section{Observations}
\label{obs}

%------------------------------------------------------------------------
\subsection{WASP transit detection photometry}
\label{waspphot}

\begin{figure}
\centering                   
\includegraphics[bb=15 310 563 550, width=0.48\textwidth]{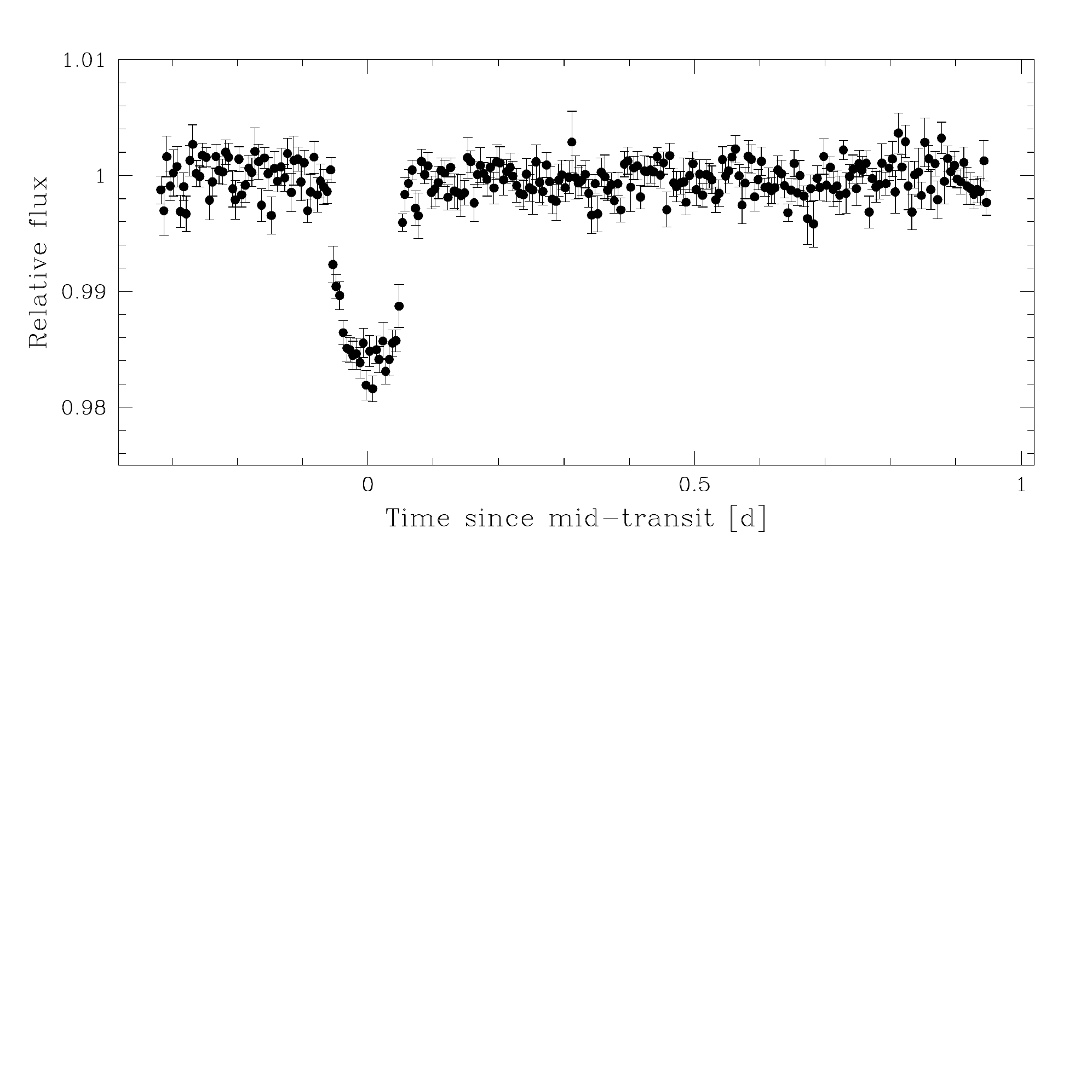}
\vspace{-0.1cm}
\caption{WASP photometry for WASP-121 folded on the best-fit transit ephemeris from the transit search algorithm presented in Collier Cameron et al. (2006), and binned per 0.005d intervals.} 
\vspace{-0.2cm}
\label{wasp_lc}
\end{figure}

The WASP transit survey is operated from two sites with one for each hemisphere: the Observatorio del Roque de los Muchachos in the Canary Islands in the North and the Sutherland Station of the South African Astronomical Observatory (SAAO) in the South. Each facility consists of eight Canon 200mm f/1.8 focal lenses coupled to e2v 2048$\times$2048 pixels CCDs, which yields a field of view of 450 $\mathrm{deg}^{2}$ for each site with a corresponding pixel scale of 13.7''/pixel. Further details of the instruments, survey, and data reduction procedures can be found in \cite{pollacco}, while details of the candidate selection process can be found in \cite{collier2}.\\ 
\indent
The host star WASP-121 (1SWASPJ071024.05-390550.5 = 2MASS07102406-3905506, $V$=10.4, $K$=9.4) was observed by the WASP-South station (\citealt{hellierws}) from 2011 Oct 28 to 2012 Mar 29, leading to the collection of 9642 photometric measurements. These data were processed and searched for transit signals, as described in \cite{collier}, leading to the detection of periodic dimmings of about 1.6\% with a period of \hbox{1.27 days}. \hbox{Fig. \ref{wasp_lc}} presents the WASP photometry folded on the best-fit transit ephemeris.\\
\indent
The sine-wave fitting method described in \cite{maxted} was used to search for periodic modulation in the WASP photometry of WASP-121 that would be caused by the combination of stellar activity and rotation, but no periodic signal was found above the mmag amplitude. This analysis was performed over the frequency interval \hbox{0-1.5 cycles/day} at 8192 evenly spaced frequencies.

\begin{table}
\centering
\begin{tabular}{ccccc}
  \hline
  \hline
  $\mathrm{HJD}$ & RV & $\mathrm{\sigma}_{\mathrm{RV}}$ & BS & FWHM \\
  - 2 450 000 & (km $\mathrm{s}^{-1}$) & (km $\mathrm{s}^{-1}$) & (km $\mathrm{s}^{-1}$) & (km $\mathrm{s}^{-1}$) \\
  \hline
6546.907310 & 38.16352 & 0.01712 & 0.16642 & 19.62220\\	
6567.885240 & 38.55488 & 0.01996 & -0.27337 & 19.61806\\	
6577.854124 & 38.54414 & 0.01899 & -0.17261 & 19.73738\\	
... & ... & ... & ... & ...\\
\hline
\hline
\end{tabular}
\vspace{-0.1cm}
\caption{CORALIE radial-velocity (RV) measurements for WASP-121. The uncertainties ($\mathrm{\sigma}_{\mathrm{RV}}$) are the formal errors (i.e. with no added jitter). The uncertainties on the CCF bisector span (BS) and FWHM values are \hbox{2.5 $\mathrm{\sigma}_{\mathrm{RV}}$}. This table is available in its entirety via the CDS.}
\label{rvs121}
\end{table}

\begin{figure}
\centering   
\includegraphics[bb=112 55 455 550, width=0.32\textwidth, height=12.0cm]{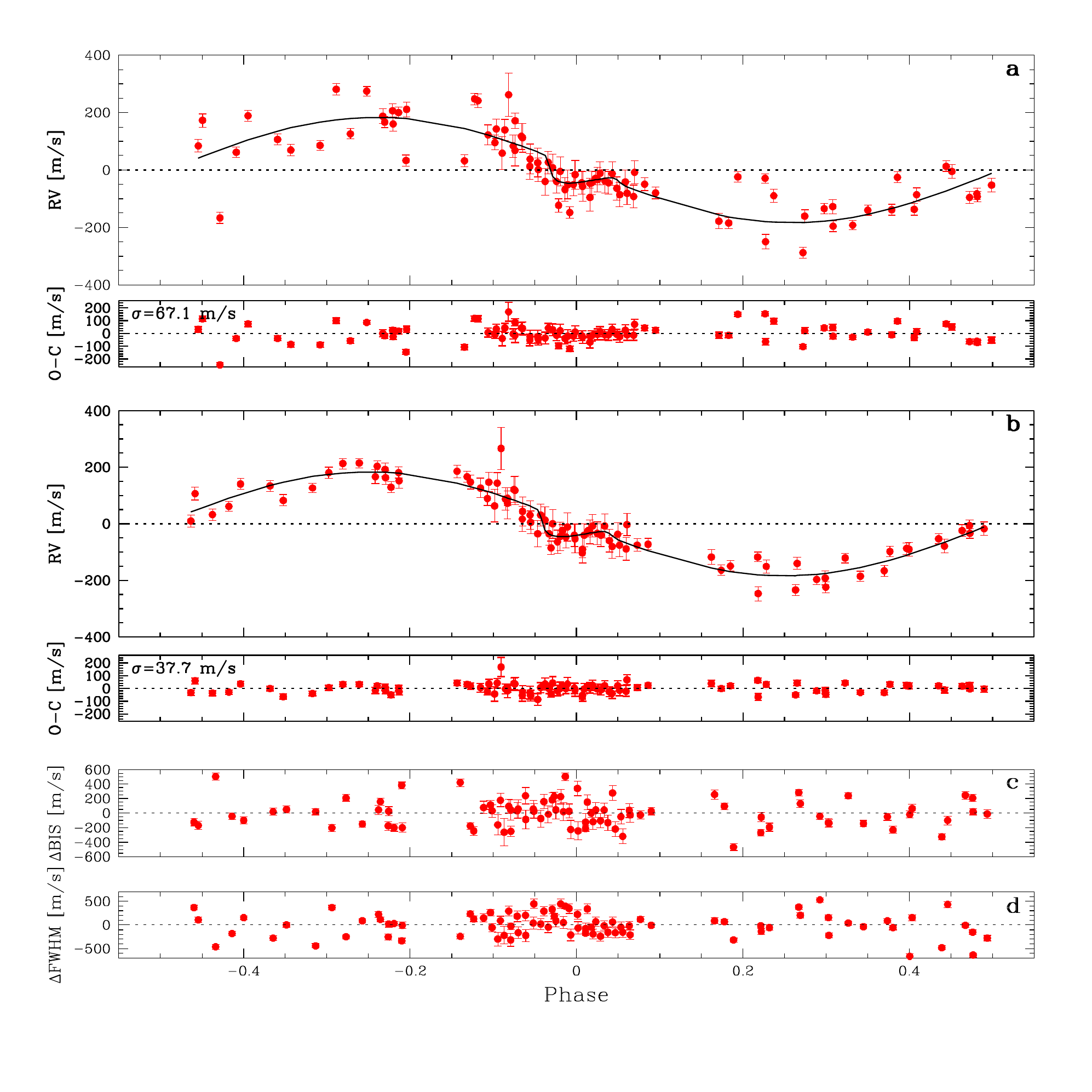} 
\caption{\textit{a)} CORALIE radial velocities (RVs) for WASP-121 phase-folded on the best-fit transit ephemeris, along with the best-fit circular model and residuals (jitter is not included in the error bars). \textit{b)} Same as the top panel but here first-order polynomial functions of the CCF bisector span and FWHM were subtracted from the RVs (see Section \ref{mcmc}). The scatter in the RV residuals is significantly reduced. \textit{c)} Change in the CCF bisector span as a function of orbital phase. \textit{d)} Change in the CCF FWHM as a function of orbital phase.}       
\label{rv}
\end{figure}

\begin{figure}
\centering                    
\includegraphics[bb=15 125 540 525, width=0.45\textwidth, height=5.6cm]{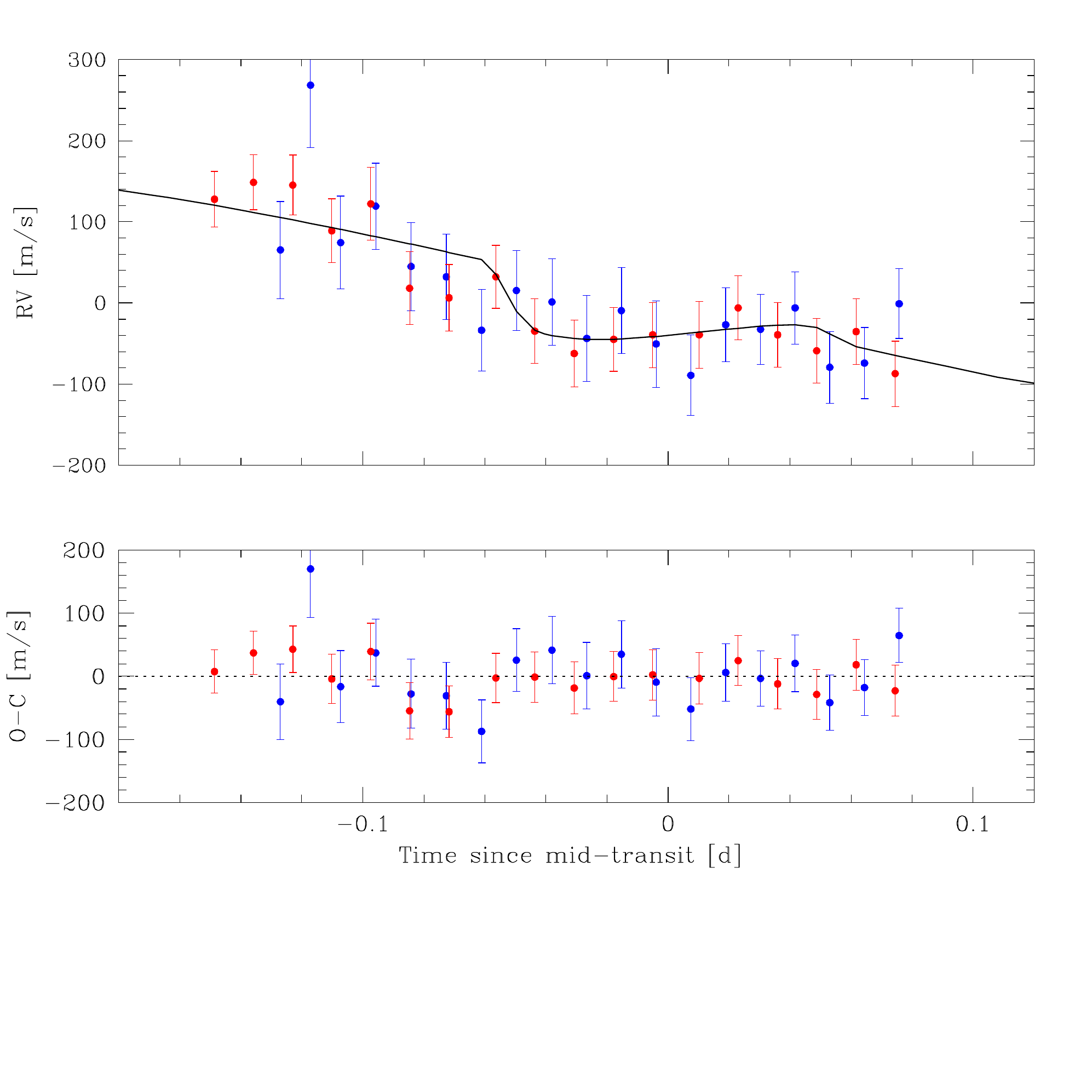}
\caption{\textit{Top:} Zoom on the Rossiter-McLaughlin effect observed with CORALIE. The RVs obtained during the transit of 2014 Dec 24 are plotted in blue, while the RVs obtained during the transit of 2015 Jan 12 are plotted in red. The superimposed, solid black line is our best-fit model. \textit{Bottom:} Corresponding residuals.} 
\label{rossiterfig}
\end{figure}

\vspace{-0.3cm}
%------------------------------------------------------------------------
\subsection{Spectroscopy and radial velocities}
\label{rvdata}

The CORALIE spectrograph, mounted on the 1.2m Euler-Swiss telescope at the ESO La Silla Observatory (Chile), was used to gather eighty-nine spectroscopic measurements of \hbox{WASP-121} between 2013 Sep 11 and 2015 Jan 12 (we note that the optical fibre feeding the instrument was replaced in Nov 2014). Among these spectra, nineteen were obtained during the transit of 2014 Dec 24 and eighteen during the transit of 2015 Jan 12, with the aim of measuring the Rossiter-McLaughlin (RM) effect (\citealt{rossiter}, \citealt{mclaughlin}). WASP-121 was indeed considered an interesting target for such measurements, as its high projected rotation velocity $v_{\star}$ sin $i_{\star}$ of 13.5 $\pm$ 0.7 km $\mathrm{s}^{-1}$ (see Section \ref{barry}), combined with the observed transit depth, was expected to yield a RM effect with a semi-amplitude $\sim$135 $\mathrm{m}\,\mathrm{s}^{-1}$.\\
\indent
Radial velocities (RVs) were computed from the spectra by weighted cross-correlation (\citealt{pepe2}), using a numerical G2-spectral template that provides optimal precisions for late-F to early-K dwarfs (Table \ref{rvs121}). A preliminary orbital analysis of the RV time-series revealed a 1.27 days periodic variation (see the top panel of Fig. \ref{periodograms}), in phase with the WASP photometry, and with a semi-amplitude $\sim$\hbox{180 $\mathrm{m}\,\mathrm{s}^{-1}$} compatible with a planetary-mass companion (see Fig. \ref{rv}a or b). The RM effect was found to have a surprisingly low total amplitude $\sim$70 $\mathrm{m}\,\mathrm{s}^{-1}$, suggesting that the planetary orbit was likely polar (see Fig. \ref{rossiterfig}). It also appeared that the star exhibits an especially high scatter in its RV residuals: the standard deviation of the best-fit residuals is \hbox{67.1 $\mathrm{m}\,\mathrm{s}^{-1}$} for a circular model and \hbox{66.0 $\mathrm{m}\,\mathrm{s}^{-1}$} for an eccentric model, while the average RV error is \hbox{30.7 $\mathrm{m}\,\mathrm{s}^{-1}$}. We consider further the origin of this high jitter in Section \ref{jitter}.

\begin{table*}
\centering
\caption{Summary of the follow-up eclipse photometry obtained for WASP-121. For each lightcurve, this table shows the date of acquisition (UT), the used instrument, the eclipse nature, the filter and exposure time, the number of data points, the selected baseline function, the standard deviation of the best-fit residuals (unbinned and binned per intervals of 2 min), and the deduced values for $\beta_{w}$, $\beta_{r}$ and $CF=\beta_{w} \times \beta_{r}$ (see Section \ref{mcmc} for details). For the baseline function, p($\epsilon^{N}$) denotes, respectively, a $N$-order polynomial function of time ($\epsilon=t$), airmass ($\epsilon=a$), PSF full-width at half maximum ($\epsilon=f$), background ($\epsilon=b$), and $x$ and $y$ positions ($\epsilon=xy$). For the TRAPPIST data, the symbol $o$ denotes an offset fixed at the time of the meridian flip.}
\begin{tabular}{cccccccccccc}
  \hline
  \hline 
 Date (UT) & Instrument & Eclipse nature & Filter & $T_{\mathrm{exp}}$ & $N_{p}$ & Baseline function & $\sigma$ & $\sigma_{120\mathrm{s}}$ & $\beta_{w}$ & $\beta_{r}$ & $CF$  \\
  & & & & (s) & & & (\%) & (\%) & & & \\
  \hline
  2013 Dec 09 & TRAPPIST & Transit & Sloan-$z'$ & 13 & 763 & p($t^{1}$) + $o$ & 0.26 & 0.12 & 1.12 & 1.52 & 1.70 \\
  2013 Dec 25 & TRAPPIST & Occultation & Sloan-$z'$ & 13 & 902 & p($a^{1}$+$xy^{1}$) + $o$ & 0.37 & 0.17 & 1.44 & 1.07 & 1.54 \\
  2013 Dec 30 & TRAPPIST & Occultation & Sloan-$z'$ & 13 & 653 & p($t^{1}$+$xy^{1}$) + $o$ & 0.29 & 0.14 & 1.44 & 1.44 & 2.06 \\
  2014 Jan 01 & TRAPPIST & Transit & Sloan-$z'$ & 13 & 765 & p($t^{1}$+$xy^{1}$) + $o$ & 0.25 & 0.13 & 1.13 & 2.35 & 2.66 \\
  2014 Jan 13 & TRAPPIST & Occultation & Sloan-$z'$ & 13 & 867 & p($a^{1}$+$xy^{1}$) + $o$ & 0.27 & 0.12 & 1.07 & 1.56 & 1.66 \\
  2014 Jan 20 & EulerCam & Transit & Gunn-$r'$ & 50 & 235 & p($t^{1}$+$f^{2}$) & 0.10 & 0.07 & 1.38 & 1.28 & 1.76 \\
  2014 Jan 24 & EulerCam & Transit & Gunn-$r'$ & 50 & 195 & p($t^{1}$+$f^{2}$+$xy^{1}$) & 0.14 & 0.09 & 2.45 & 1.10 & 2.70 \\
  2014 Jan 31 & TRAPPIST & Occultation & Sloan-$z'$ & 12 & 947 & p($a^{1}$) + $o$ & 0.25 & 0.10 & 1.12 & 1.00 & 1.12 \\
  2014 Feb 05 & TRAPPIST & Occultation & Sloan-$z'$ & 12 & 1033 & p($t^{1}$+$xy^{1}$) + $o$ & 0.39 & 0.17 & 1.31 & 1.49 & 1.95 \\
  2014 Mar 22 & TRAPPIST & Occultation & Sloan-$z'$ & 12 & 1007 & p($t^{1}$+$xy^{1}$) + $o$ & 0.38 & 0.16 & 1.17 & 1.18 & 1.38 \\
  2014 Apr 07 & TRAPPIST & Transit & Sloan-$z'$ & 13 & 700 & p($a^{1}$+$xy^{1}$) & 0.48 & 0.19 & 1.61 & 1.45 & 2.32 \\
  2014 Apr 14 & TRAPPIST & Occultation & Sloan-$z'$ & 11 & 851 & p($a^{1}$) & 0.36 & 0.16 & 1.02 & 1.58 & 1.61 \\
  2014 Nov 08 & TRAPPIST & Transit & Johnson-$B$ & 7 & 966 & p($t^{2}$+$b^{2}$+$xy^{1}$) & 0.55 & 0.23 & 1.46 & 1.09 & 1.59 \\
  2014 Dec 01 & EulerCam & Transit & Geneva-$B$ & 90 & 162 & p($a^{1}$+$f^{1}$+$xy^{1}$) & 0.11 & 0.11 & 1.51 & 1.05 & 1.59 \\
  2014 Dec 24 & TRAPPIST & Transit & Sloan-$z'$ & 8 & 961 & p($a^{1}$) & 0.29 & 0.12 & 0.91 & 2.07 & 1.87 \\
  2014 Dec 29 & EulerCam & Transit & Geneva-$B$ & 60 & 223 & p($a^{1}$+$f^{1}$) & 0.08 & 0.06 & 1.15 & 1.19 & 1.36 \\
  \hline
  \hline
\end{tabular}
\label{obstable}
\end{table*}

\indent
The cross-correlation function (CCF) bisector span (\citealt{queloz}) and FWHM values are plotted in Fig. \ref{rv}c and \ref{rv}d, respectively. Both present large variations, their standard deviations being \hbox{190.2 $\mathrm{m}\,\mathrm{s}^{-1}$} and \hbox{245.9 $\mathrm{m}\,\mathrm{s}^{-1}$}, respectively, while their average error (calculated as 2.5 times the average RV error, see \citealt{santerne2015}) is \hbox{76.7 $\mathrm{m}\,\mathrm{s}^{-1}$}. These variations do not phase with the transit ephemeris, as one might expect if the observed RV signal was originating from a false-positive scenario, such as a blended eclipsing binary (see e.g. \citealt{santos}). However, as the scatter in the bisector span values is comparable to the semi-amplitude of the RV signal, we were not able to discard blend scenarios based on the traditional bisector span technique (\citealt{queloz}). Instead, we performed a detailed blend analysis and used the \texttt{PASTIS} Bayesian software (\citealt{diaz}, \citealt{santerne14}) to statistically validate the planet, as described in Section \ref{pastis}.

\vspace{-0.2cm}
%------------------------------------------------------------------------
\subsection{Follow-up eclipse photometry}

\begin{figure}
\centering   
\includegraphics[bb=55 40 265 550, width=0.38\textwidth, height=17cm]{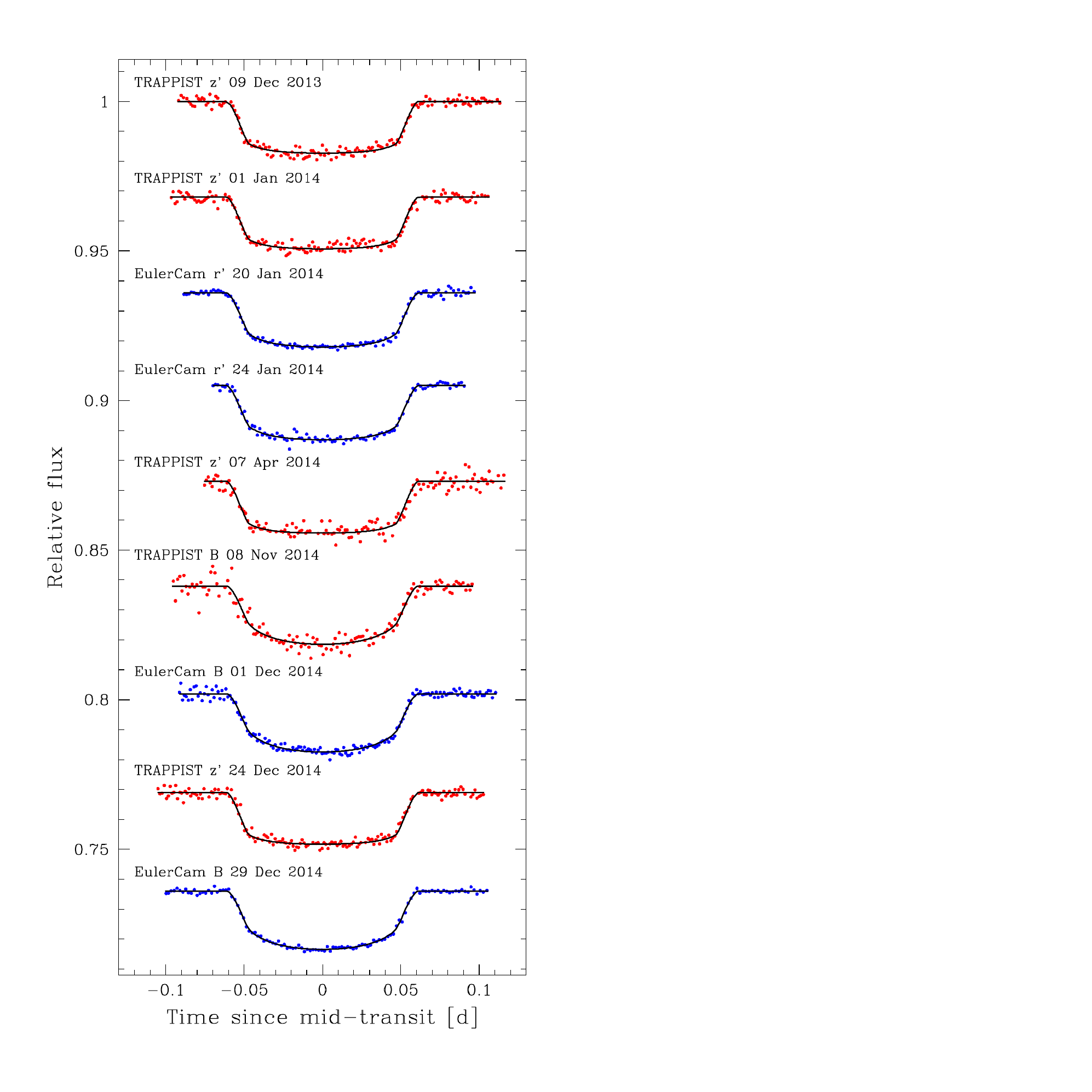}    
\vspace{0.6cm} 
\caption{Follow-up transit photometry for \hbox{WASP-121 b}. The observations are binned per 2 min and period-folded on the best-fit transit ephemeris (see Section \ref{mcmc}). Each lightcurve has been divided by the respective best-fit photometric baseline model. For each filter, the superimposed, solid black line is our best-fit transit model. The lightcurves are shifted along the \textit{y}-axis for clarity.}    
\vspace{-0.3cm}  
\label{lcstransits}
\end{figure}

\begin{figure}
\centering   
\includegraphics[bb=60 50 255 550, width=0.35\textwidth, height=15cm]{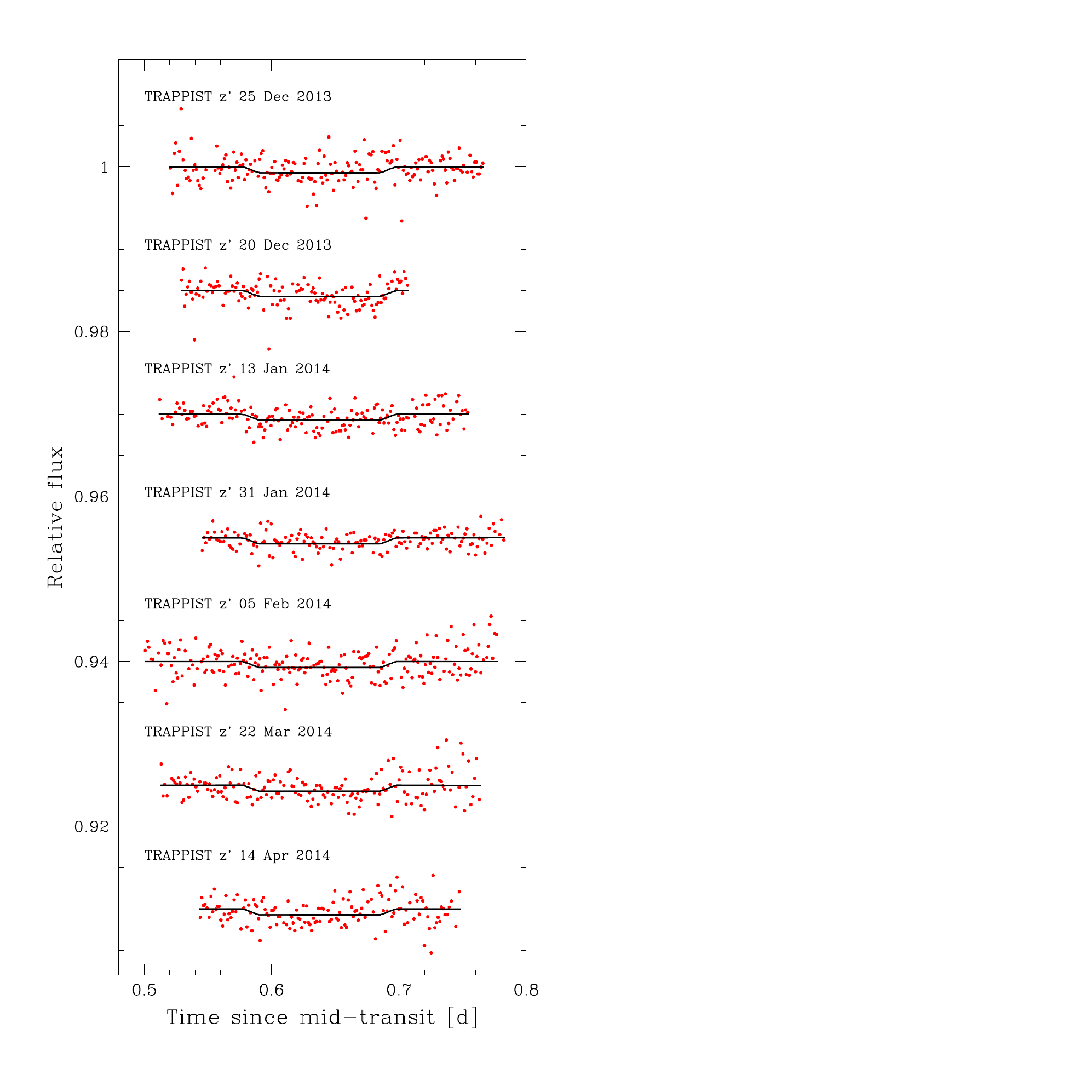}  
\vspace{0.4cm}
\includegraphics[bb=322 420 520 570, width=0.35\textwidth, height=5.1cm]{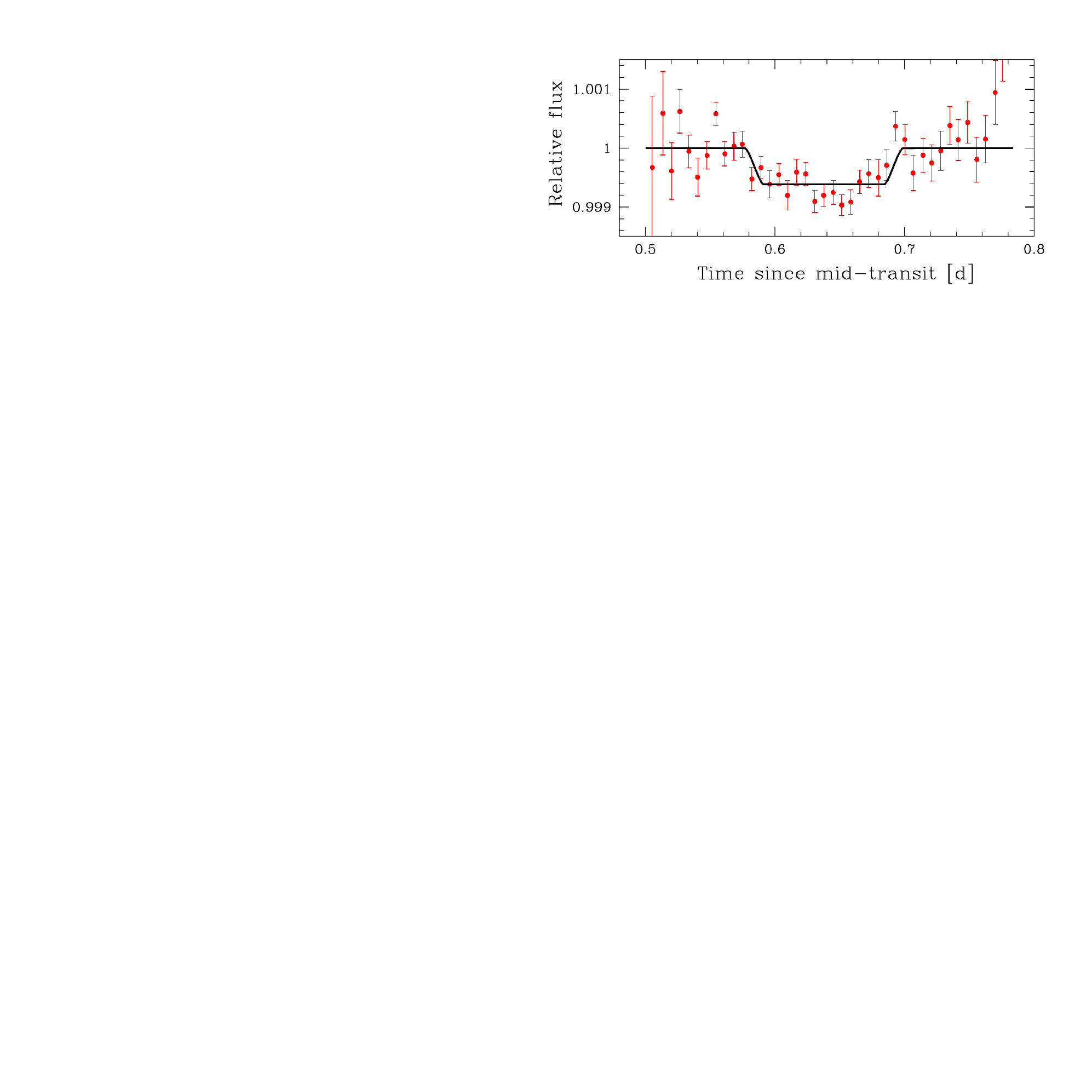}  
\vspace{-0.4cm}
\caption{\textit{Top:} Individual follow-up occultation lightcurves for \hbox{WASP-121 b}. The observations are binned per 2 min and period-folded on the best-fit transit ephemeris (see Section \ref{mcmc}). Each lightcurve has been divided by the respective best-fit photometric baseline model. For each lightcurve, the superimposed, solid black line is our best-fit occultation model. The lightcurves are shifted along the \textit{y}-axis for clarity. \textit{Bottom:} Combined follow-up occultation photometry for \hbox{WASP-121 b} (bin width=10 min).}       
\label{lcsoccs}
\end{figure}

To refine the system's parameters, high-precision eclipse (transit and occultation) lightcurves were obtained using the 60cm TRAPPIST robotic telescope (TRAnsiting Planets and PlanetesImals Small Telescope) and the EulerCam CCD camera that is mounted on the 1.2m Euler-Swiss telescope, which are both located at ESO La Silla Observatory. These follow-up lightcurves are summarized in Table \ref{obstable} and presented in Fig. \ref{lcstransits} and \ref{lcsoccs}. The transits were observed in different filters to search for a potential color dependence of the transit depth, which might have been indicative of a blend (see Section \ref{pastis}).

\vspace{-0.2cm}
\subsubsection{TRAPPIST observations} 
\label{trappist}

TRAPPIST is a 60cm robotic telescope dedicated to the detection and characterization of transiting exoplanets and to the photometric monitoring of bright comets and other small bodies. It is equipped with a thermoelectrically-cooled 2K$\times$2K CCD, which has a pixel scale of 0.65'' that translates into a 22'$\times$22' field of view. For details of TRAPPIST, see \cite{gillon} and \cite{jehin}. TRAPPIST was used to observe four transits of WASP-121\,b in a \hbox{Sloan-$z'$} filter ($\lambda_{\mathrm{eff}}=895\pm1$ nm) and one transit in a Johnson-$B$ filter ($\lambda_{\mathrm{eff}}=440\pm1$ nm). As we noticed from a preliminary analysis that \hbox{WASP-121\,b} is actually an extremely favorable target for secondary eclipse measurements (we will elaborate on this in Section \ref{discussionatm}), we also observed seven occultation windows in the Sloan-$z'$ filter. During the runs, the positions of the stars on the chip were maintained to within a few pixels thanks to a ``software guiding'' system that regularly derives an astrometric solution for the most recently acquired image and sends pointing corrections to the mount if needed. After a standard pre-reduction (bias, dark, and flatfield correction), the stellar fluxes were extracted from the images using the IRAF/DAOPHOT\footnote{IRAF is distributed by the National Optical Astronomy Observatory, which is operated by the Association of Universities for Research in Astronomy, Inc., under cooperative agreement with the National Science Foundation.} aperture photometry software (\citealt{stetson}). For each lightcurve, we tested several sets of reduction parameters and kept the one giving the most precise photometry for the stars of similar brightness as the target. After a careful selection of reference stars, the photometric lightcurves were finally obtained using differential photometry.

\vspace{-0.4cm}
\subsubsection{EulerCam observations}

EulerCam is a 4K$\times$4K E2V CCD installed at the Cassegrain focus of the 1.2m Euler-Swiss telescope. The field of view of EulerCam is 15.7'$\times$15.7', producing a pixel scale of 0.23''. To keep the stars on the same locations on the detector during the observations, EulerCam employs an ``Absolute Tracking" system that is very similar to the one of TRAPPIST, which matches the point sources in each image with a catalog, and if needed, adjusts the telescope pointing between exposures to compensate for drifts. EulerCam was used to observe two transits of \hbox{WASP-121\,b} in a Gunn-$r'$ filter ($\lambda_{\mathrm{eff}}=664\pm1$ nm) and two other transits in a \hbox{Geneva-$B$} filter\footnote{\url{http://obswww.unige.ch/gcpd/ph13.html}} ($\lambda_{\mathrm{eff}}=425\pm1$ nm). A slight defocus was applied to the telescope to optimize the observation efficiency and to minimize pixel-to-pixel effects. The reduction procedure used to extract the transit lightcurves was similar to that performed on TRAPPIST data. Further details of the EulerCam instrument and data reduction procedures can be found in \cite{lendl}.

%\vspace{-0.1cm}
\subsection{Out-of-eclipse photometric monitoring}
\label{photvar}

\begin{figure}
\centering   
\includegraphics[bb=100 30 455 550, width=0.3\textwidth, height=8cm]{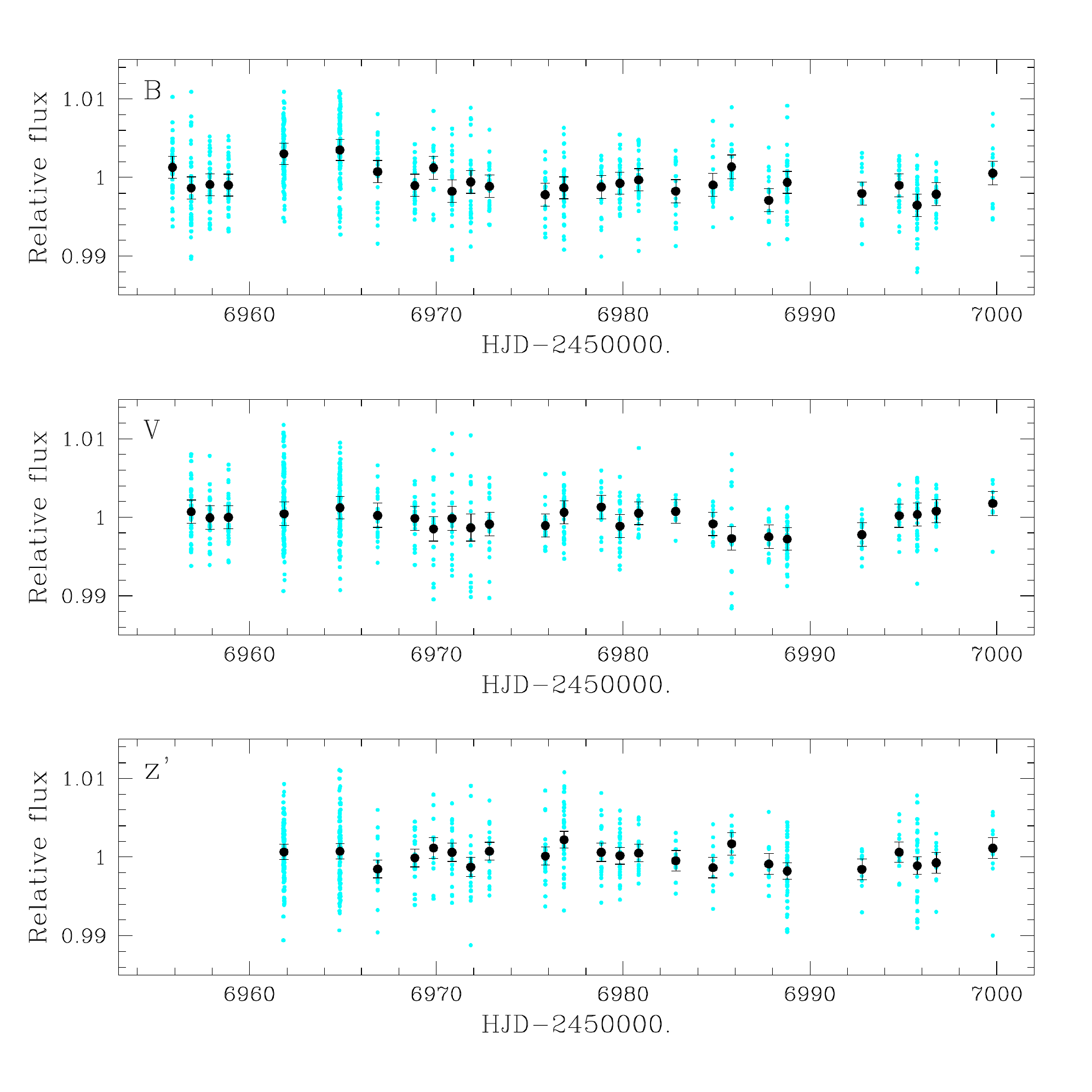} 
\caption{Out-of-eclipse photometric monitoring: globally normalized TRAPPIST differential photometry obtained for \hbox{WASP-121} in Johnson-$B$ (\textit{top}), Johnson-$V$ (\textit{middle}), and Sloan-$z'$ (\textit{bottom}) filters, unbinned (\textit{cyan}) and binned by day (\textit{black}).}       
\label{variability}
\vspace{-0.2cm}
\end{figure}

To search for potential out-of-eclipse photometric variability that would not have been detected in the WASP photometry (see Section \ref{waspphot}), we monitored WASP-121 with TRAPPIST for 27 non-consecutive nights between 2014 Oct 25 and 2014 Dec 8. This monitoring consisted in taking every night a short sequence of ten images in three filters: Johnson-$B$ ($\lambda_{\mathrm{eff}}=440\pm1$ nm), Johnson-$V$ \hbox{($\lambda_{\mathrm{eff}}=546.5\pm1$ nm)}, and Sloan-$z'$ ($\lambda_{\mathrm{eff}}=895\pm1$ nm). The data were reduced as described in Section \ref{trappist}. The globally normalized differential lightcurves obtained in each filter are shown in \hbox{Fig. \ref{variability}}. WASP-121 appears to be very quiet in photometry, the standard deviations of the binned lightcurves being 1.6 mmag ($B$), 1.3 mmag ($V$), and 1.1 mmag ($z'$).

\vspace{-0.1cm}
%%%%%%%%%%%%%%%%%%%%%%%%%%%%%%%%%%%%%%
\section{Analysis}
\label{analysis}

%------------------------------------------------------------------------
\subsection{Spectroscopic analysis}
\label{barry}

The individual CORALIE spectra were co-added to produce a single spectrum with an average S/N of around 150:1. The analysis was performed using standard pipeline reduction products and the procedures given in \cite{amanda}. The derived stellar parameters are listed in Table \textcolor{blue}{4}.\\ 
\indent
The excitation balance of the Fe~{\sc i} lines was used to determine the effective temperature $T_{\mathrm{eff}}$. The surface gravity log $g_{\star}$ was determined from the ionisation balance of Fe~{\sc i} and Fe~{\sc ii}. The Ca~{\sc i} line at 6439 {\AA} and the Na~{\sc i} D lines were also used as log $g_{\star}$ diagnostics. The iron abundance was determined from equivalent width measurements of several unblended lines and is relative to the solar value obtained by \cite{asplund}. A value for microturbulence ($\xi_{\mathrm{t}}$) was determined from Fe~{\sc i} using the method of \cite{magain}. The quoted error estimates include those given by the uncertainties in $T_{\mathrm{eff}}$ and log $g_{\star}$, as well as the scatter due to measurement and atomic data uncertainties.\\ 
\indent
The projected stellar rotation velocity $v_{\star}$ sin $i_{\star}$ was determined by fitting the profiles of several unblended Fe~{\sc i} lines. A macroturbulent velocity ($v_{\mathrm{mac}}$) of 6.0 $\pm$ 0.6 $\mathrm{km}\,\mathrm{s}^{-1}$ was assumed using the asteroseismic-based calibration of \cite{amanda2} and an instrumental resolution of 55,000. A best-fitting value of $v_{\star}$ sin $i_{\star}$ = 13.5 $\pm$ 0.7 km $\mathrm{s}^{-1}$ was obtained.\\ 
\indent
There is no significant detection of lithium in the spectra, with an abundance upper limit log A(Li)$<$1.0. The lack of any detectable lithium does not provide an age constraint as the star's $T_{\mathrm{eff}}$ places it in the lithium gap (\citealt{bohm}). There is also no significant chromospheric emission in the Ca~{\sc ii} H and K line cores.\\
\indent 
The spectral type was estimated from $T_{\mathrm{eff}}$ using the Table B.1 in \cite{gray} and the \cite{torres} calibration was used to obtain first stellar mass and radius estimates: \hbox{$M_{\star}$ = 1.37 $\pm$ 0.14 $M_{\odot}$} and \hbox{$R_{\star}$ = 1.52 $\pm$ 0.41 $R_{\odot}$}. 

\vspace{-0.3cm}
%------------------------------------------------------------------------
\subsection{Stellar jitter}
\label{jitter}

\begin{figure}
\centering                    
\includegraphics[bb=15 330 563 550, width=0.46\textwidth]{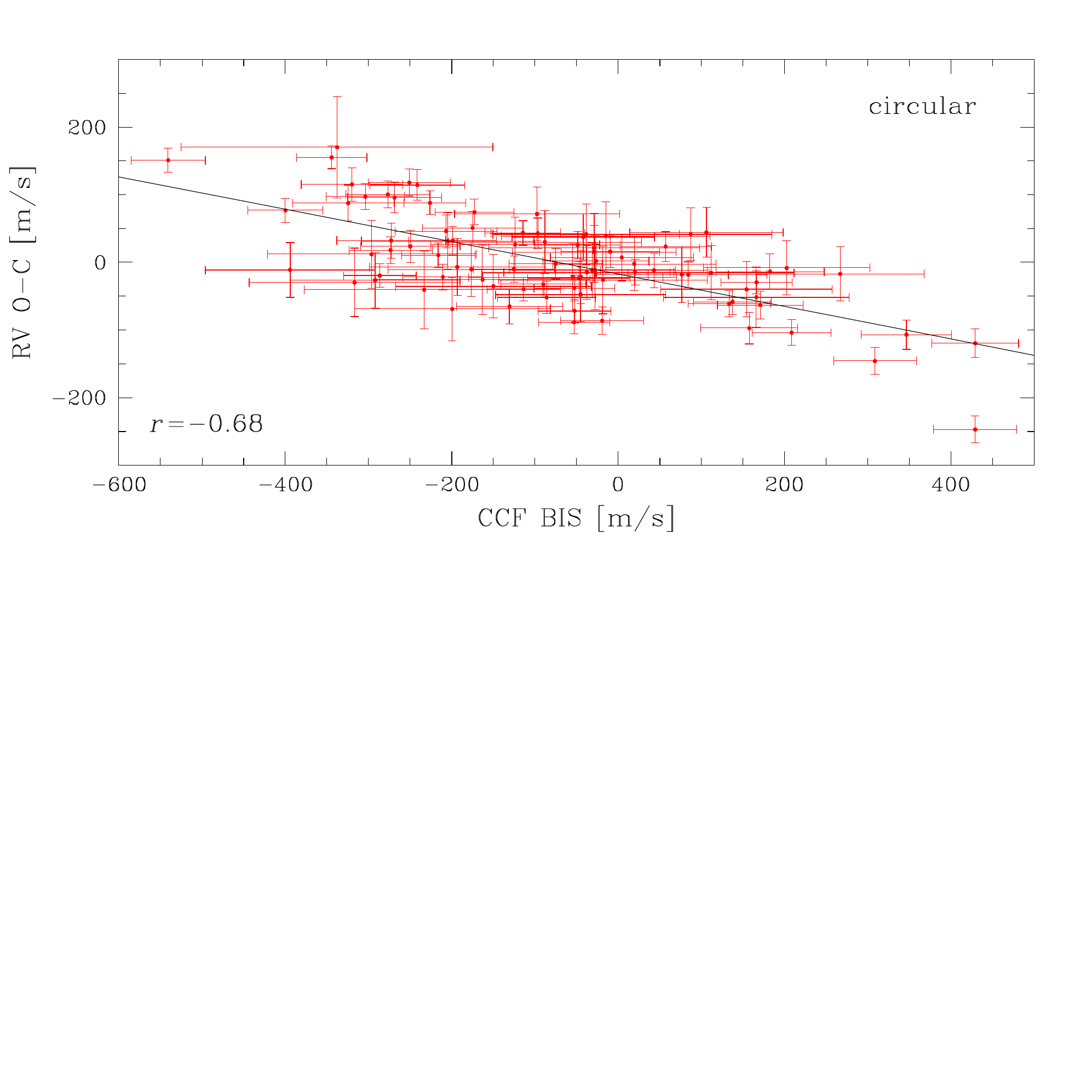}
\includegraphics[bb=15 290 563 600, width=0.46\textwidth]{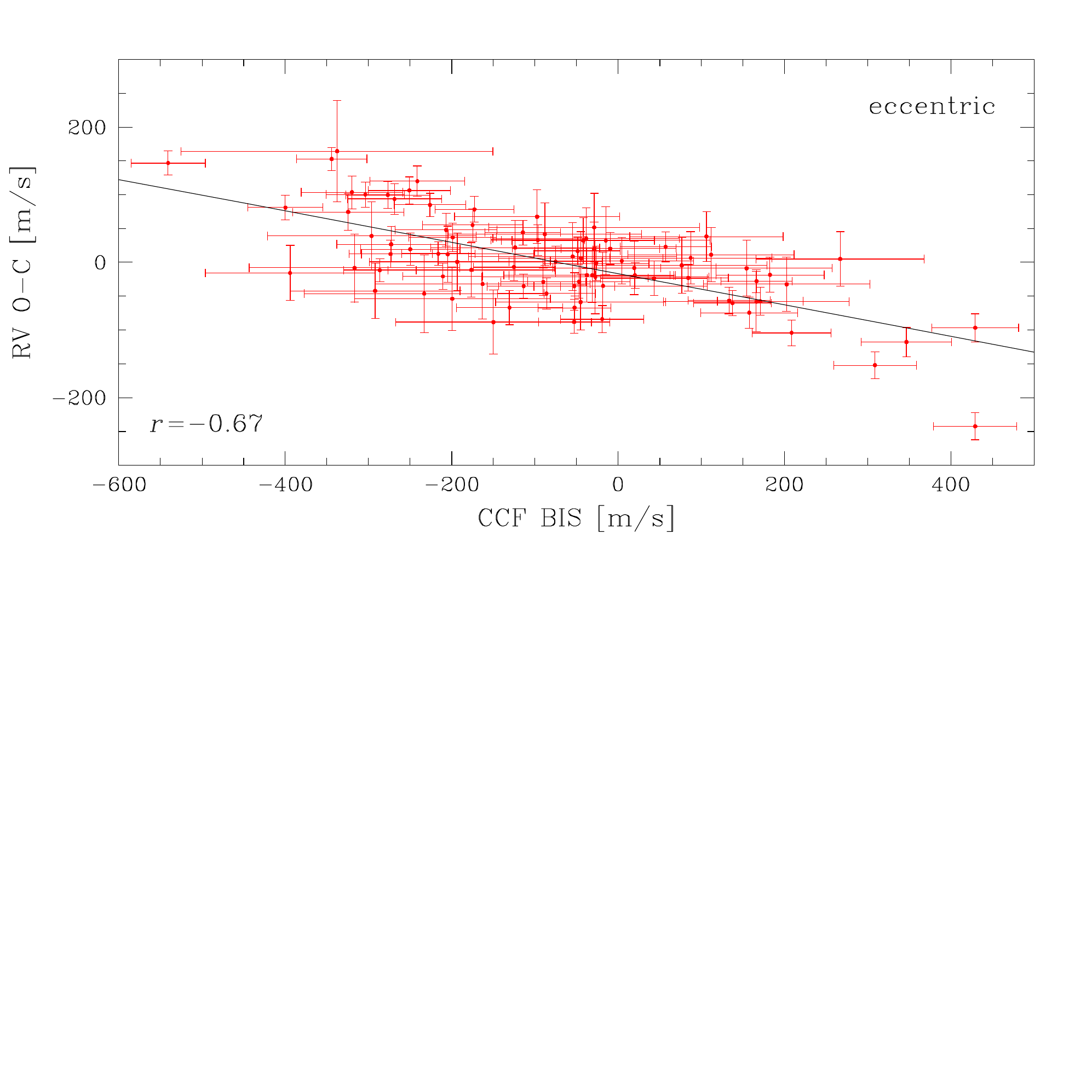}
\vspace{-0.1cm}
\caption{RV residuals from the best-fit circular (top) and eccentric (bottom) orbital models vs. CCF bisector spans. The two quantities are clearly anti-correlated. Linear fits to the data are overplotted and the correlation coefficients (\textit{r}) are given.} 
\vspace{-0.25cm}
\label{correl}
\end{figure}

As mentioned in Section \ref{rvdata}, \hbox{WASP-121} exhibits an especially high scatter in its RV residuals (Fig. \ref{rv}a), CCF bisector spans (Fig. \ref{rv}c), and FWHM (Fig. \ref{rv}d). Fig. \ref{correl} compares the RV residuals to the CCF bisector spans assuming a circular (top) and an eccentric (bottom) orbit. There is a significant anti-correlation between these two quantities, the correlation coefficients being -0.68 and -0.67 in the circular and eccentric cases, respectively. Such an anti-correlation is commonly interpreted as being a signature of stellar activity (see e.g. \citealt{queloz}, \citealt{melo}), but \cite{santerne2015} showed that it could also be produced by a blended star with a lower CCF FWHM (and thus slower rotation) than the target star. However, as detailed in \hbox{Section \ref{pastis}}, we were not able to reproduce the observed RVs, CCF bisector spans, and FWHM assuming such a scenario, thus making it likely that the high jitter measured for \hbox{WASP-121} is due to stellar activity. One might have concerns about the non-detection of this stellar activity in the form of emission in the Ca~{\sc ii} H and K line cores (see Section \ref{barry}), out-of-eclipse photometric variability (see Sections \ref{waspphot} and \ref{photvar}), or spot-crossing events during transits (see Fig. \ref{lcstransits}). A detailed study of \hbox{WASP-121's} activity being beyond the scope of this paper, we just propose here some potential explanations regarding these non-detections.\\ 
\indent
First, we note that such a situation is not atypical for an F-type star as it was also encountered for the exoplanet \hbox{F-type} host stars \hbox{HAT-P-33} (\citealt{hartman}) and WASP-111 (\citealt{w111}), which both present high activity-related RV jitter with no other apparent sign of stellar activity. More precisely, in the case of WASP-111, the scatter in the RV residuals, CCF bisector spans and FWHM dropped from one season to the next, identifying clearly stellar activity as an origin for these.

\indent 
Secondly, it can be seen from \cite{noyes84} that the chromospheric Ca~{\sc ii} emission of stars decreases with lowering $B-V$. As WASP-121's $B-V$ is only 0.43, its Ca~{\sc ii} emission could simply be too weak to be detected in our low S/N CORALIE spectra.\\
\indent
As for the non-detections of out-of-eclipse variations or spot-crossing events in the photometry, they might be explained, at least to some extent, if the star is plage-dominated, in opposition to spot-dominated (see \citealt{dumusque}). Both spots and plages are regions of strong magnetic fields that inhibit locally the convection and suppress the convective blueshift effect (see e.g. \citealt{dravins1981}). These regions are thus redshifted compared to the quiet photosphere and induce RV variations as they appear and disappear from the visible stellar disk due to the rotation of the star. The amplitude of this effect is expected to be similar for spots and plages of the same size. However, plages present a much lower flux ratio with the quiet photosphere compared to spots (see \citealt{dumusque} for details), so that they induce smaller brightness variations as they appear and disappear from the visible stellar disk.

\begin{figure}
\centering  
\includegraphics[bb=25 280 563 560, width=0.49\textwidth, height=4.5cm]{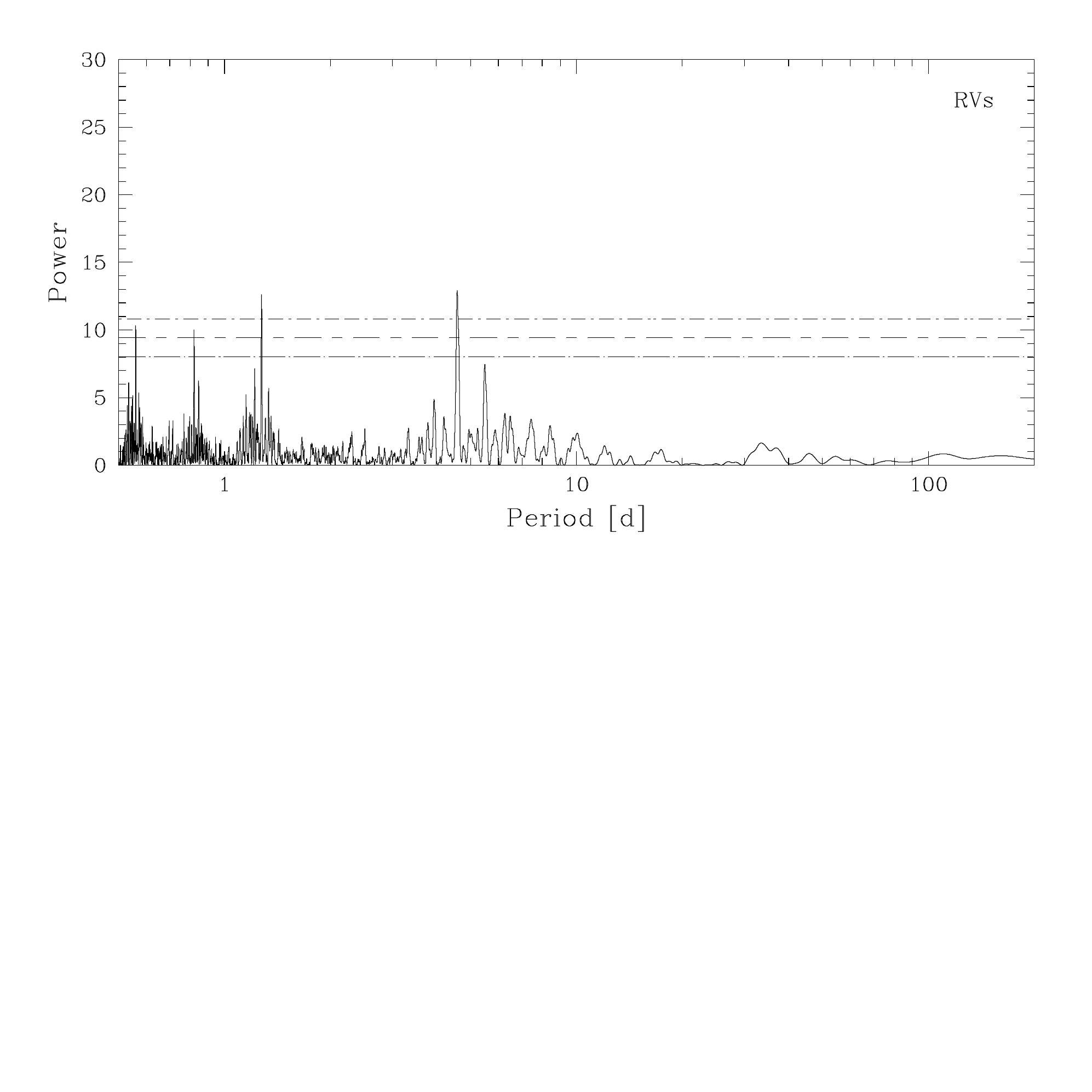}
\includegraphics[bb=25 280 563 560, width=0.49\textwidth, height=4.5cm]{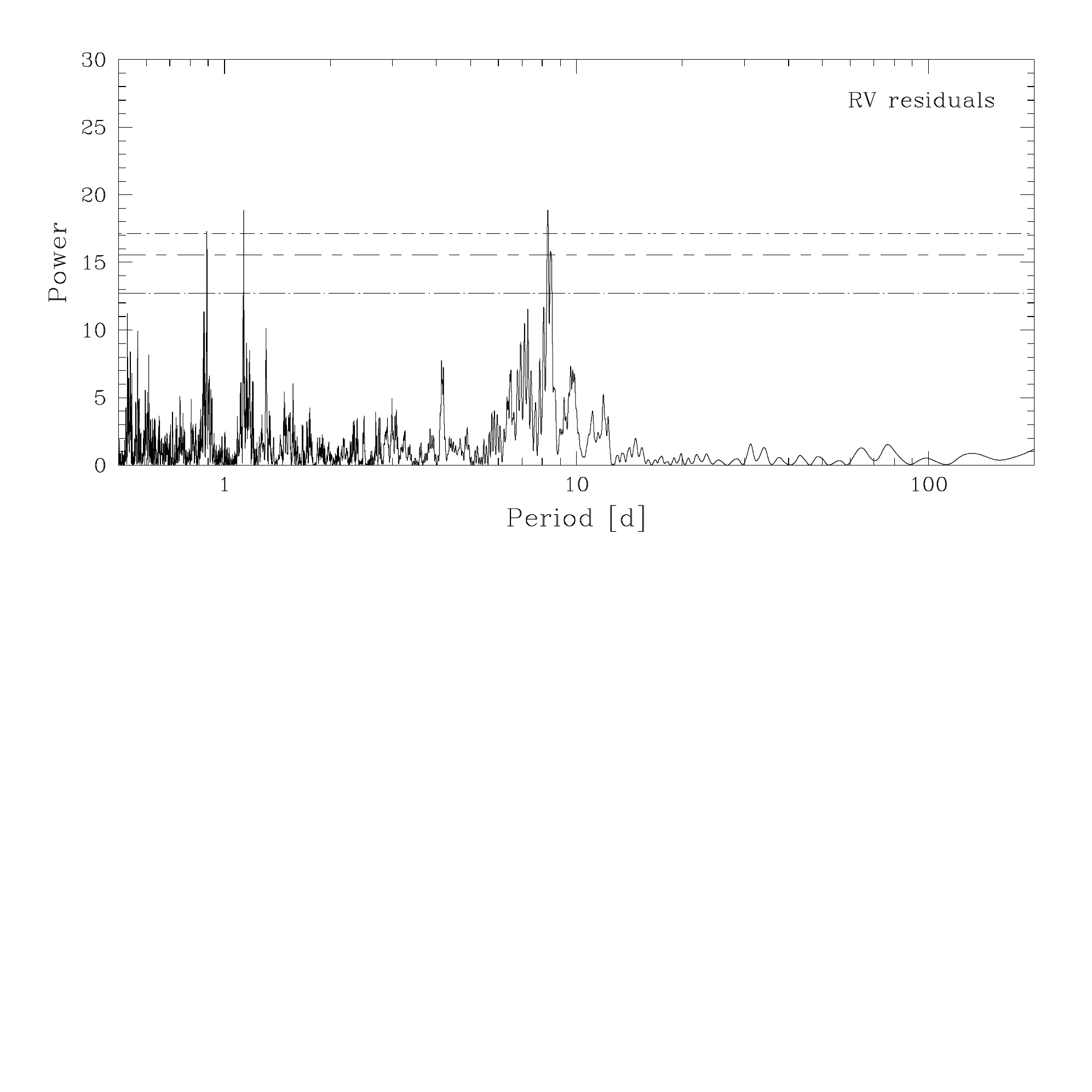}                  
\includegraphics[bb=25 330 563 550, width=0.49\textwidth, height=3.5cm]{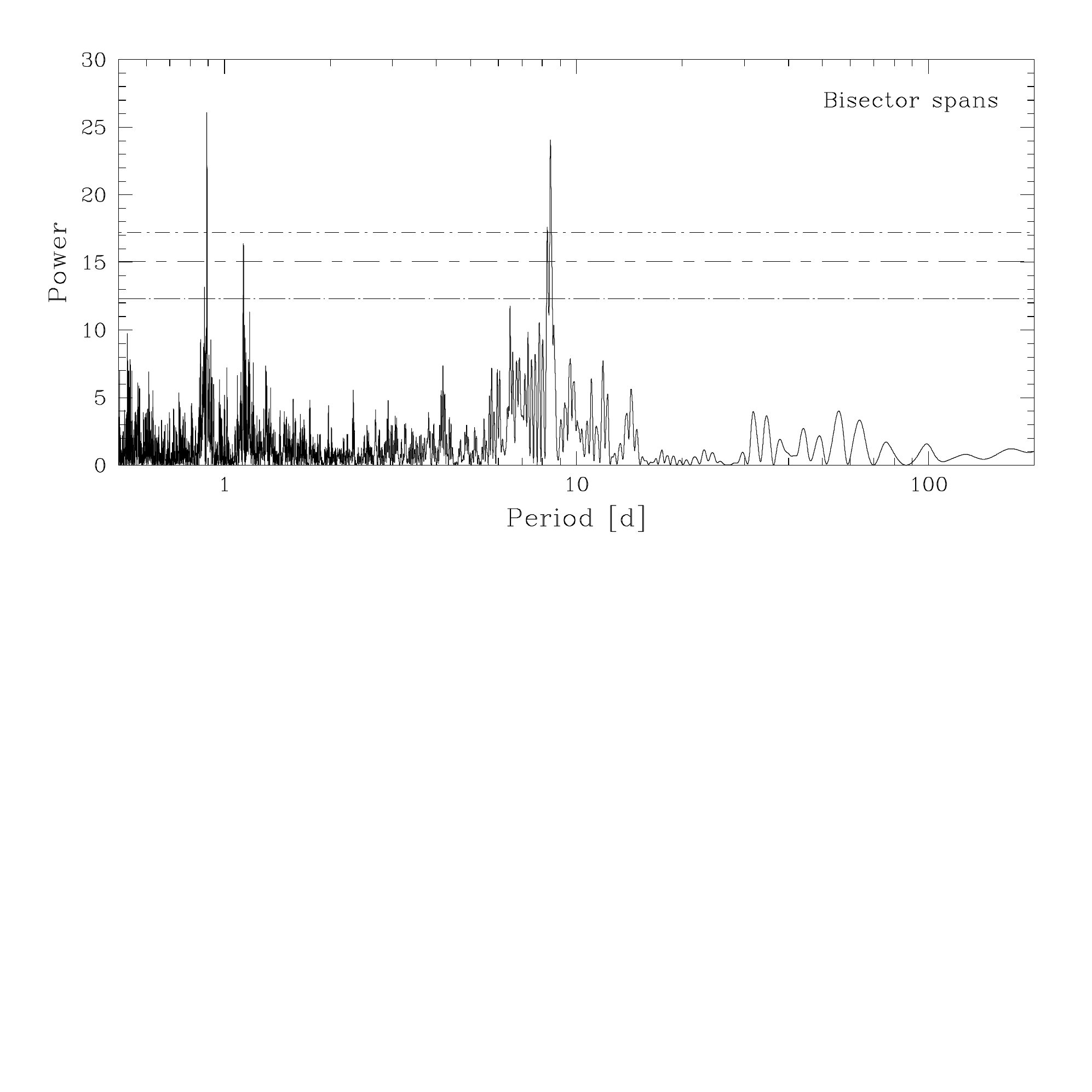}
\includegraphics[bb=25 315 563 600, width=0.49\textwidth, height=4.5cm]{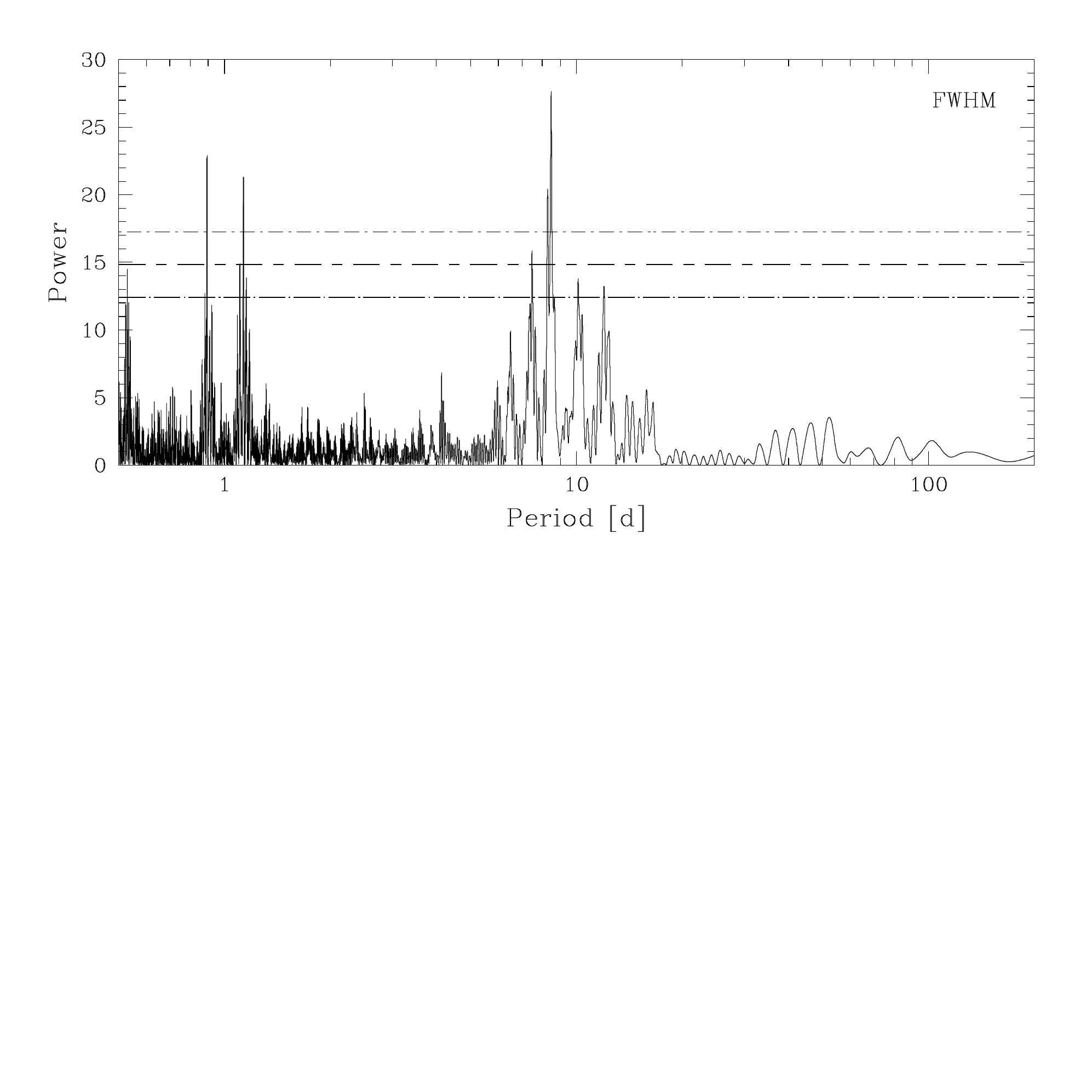}
\vspace{-0.1cm}
\caption{From top to bottom: Lomb-Scargle periodograms of the RVs, RV residuals (assuming a circular orbit), CCF bisector spans, and FWHM. Horizontal lines indicate false alarm probability levels of 0.001, 0.01, and 0.1 (from top to bottom).} 
\vspace{-0.45cm}
\label{periodograms}
\end{figure}

\indent
Fig. \ref{periodograms} shows Lomb-Scargle periodograms (\citealt{scargle82}) of the RV residuals (assuming a circular orbit, second panel from the top), CCF bisector spans (third panel from the top), and FWHM (bottom panel). In each of these periodograms, the highest peaks are found at periods of 0.89, 1.13, and 8.3 days (all with false alarm probabilities \hbox{$<$ 0.001}, except for the peak at 1.13 days in the periodogram of the bisector spans). We could assume that one of these three periods, which are each other's daily aliases, is the stellar rotation period (this supposes that the activity-induced RV signal is related to the rotation of the star). A stellar rotation period of \hbox{8.3 days} does not seem very likely, as the maximal rotation period implied by \hbox{$v_{\star}$ sin $i_{\star}$} and our final estimate of the stellar radius (\hbox{$R_{\star}$ = 1.458 $\pm$ 0.030 $R_{\odot}$}, see Section \ref{mcmc}) is \hbox{5.46 $\pm$ 0.32 days} (assuming \hbox{sin $i_{\star}$ = 1}). Stellar rotation periods of 0.89 and 1.13 days would imply, together with our measured value for $R_{\star}$, rotation velocities $v_{\star}$ of \hbox{82.9 $\pm$ 1.7 $\mathrm{km}\,\mathrm{s}^{-1}$} and 65.3 $\pm$ 1.4 $\mathrm{km}\,\mathrm{s}^{-1}$, respectively, and inclinations $i_{\star}$ of \hbox{9.4 $\pm$ 0.6 deg} and \hbox{11.9 $\pm$ 0.7 deg}, respectively, where the uncertainties are the quadratic sum of the uncertainties due to each input parameter. In both cases, the star would thus be seen nearly pole-on, i.e. with the rotation axis oriented towards the Earth. This configuration could also contribute to the non-detection of out-of-eclipse photometric variations, as the modulation in brightness due to the changing visibility of star spots or plages with rotation could then be negligible. If the star is seen nearly pole-on, WASP-121b should then be in a (nearly) polar orbit to produce transits. As mentioned previously, this is actually what we find via our observations of the RM effect (see Sections \ref{mcmc} and \ref{orbi}), which makes the stellar rotation period much likely to be 0.89 or \hbox{1.13 days}. We note that for the 1-Gyr-old cluster NGC6811 observed with \textit{Kepler} by \cite{meibom13}, the stars with \hbox{$B-V$ = 0.43} (such as WASP-121) have $\sim$1-day rotation periods (see their Fig. 1b). As the age of WASP-121 obtained by stellar evolution modeling is \hbox{1.5 $\pm$ 1.0 Gyr} (see Section \ref{val}), a $\sim$1-day rotation period would thus be consistent with their values.\\

%------------------------------------------------------------------------

\vspace{-0.6cm}
\subsection{Global modeling of the data}
\label{mcmc}

To determine the system parameters, we performed a combined analysis of the follow-up photometry and the RV data, using for this purpose the adaptive Markov Chain Monte-Carlo (MCMC) code described in \cite{gillon2} and references therein. To model the photometry, we used the eclipse model of \cite{mandel} multiplied by a different baseline model for each lightcurve. These baseline models aim to represent astrophysical, instrumental, or environmental effects, which are able to produce photometric variations and can, therefore, affect the photometric lightcurves. They can be made up of different first to fourth-order polynomials with respect to time or other variables, such as airmass, PSF full-width at half maximum, background, or stellar position on the detector. To find the optimal baseline function for each lightcurve, i.e. the model minimizing the number of parameters and the level of noise in the best-fit residuals, the Bayes factor, as estimated from the Bayesian Information Criterion (BIC, \citealt{schwarz}), was used. The best photometric baseline functions are listed in Table \ref{obstable}. For eight TRAPPIST lightcurves (see Table \ref{obstable}), a normalization offset was also part of the baseline model to represent the effect of the meridian flip; that is, the $180^{\circ}$ rotation that the German equatorial mount telescope has to undergo when the meridian is reached. This movement results in different positions of the stellar images on the detector before and after the flip, and the normalization offset allows to take into account a possible consecutive jump in the differential photometry at the time of the flip.\\ 
\indent
On their side, the RVs were modelled using a classical Keplerian model (e.g. \citealt{murray}) added to a baseline model for the stellar and instrumental variability. For the RVs obtained during a transit, the RM effect was modelled using the formulation of \cite{gimenez}. The RVs were partitioned into four datasets, each with a different baseline model: the RVs obtained before the replacement of the CORALIE optical fibre (37 RVs, dataset \#1), those obtained after (15 RVs, dataset \#2), the first RM sequence (19 RVs, dataset \#3), and the second one (18 RVs, dataset \#4). For all datasets, the minimal baseline model was a scalar $V_{\gamma}$ representing the systemic velocity of the star. For datasets \#1 and \#2, first-order polynomial functions of the CCF bisector span and FWHM were also part of the baseline models. This choice of baselines allowed to reduce the scatter in the global RV residuals from \hbox{67.1 $\mathrm{m}\,\mathrm{s}^{-1}$} to \hbox{37.7 $\mathrm{m}\,\mathrm{s}^{-1}$} (see Fig. \ref{rv}a and b) and was thus strongly favored by the BIC. These additional baseline terms were not necessary for the datasets \#3 and \#4 (RM sequences).

\begin{table}
\centering
\begin{tabular}{lccc}
  \hline
  Filter & $u_{1}$ & $u_{2}$ \\
  \hline
  Sloan-$z'$ & 0.171 $\pm$ 0.014 & 0.306 $\pm$ 0.006\\
  Gunn-$r'$ & 0.295 $\pm$ 0.015 & 0.325 $\pm$ 0.006\\
  Johnson- and Geneva-$B$ & 0.510 $\pm$ 0.027 & 0.260 $\pm$ 0.019\\
  \hline
\end{tabular}
\label{ldtable}
\caption{Expectations and standard deviations of the normal distributions used as prior distributions for the quadratic limb-darkening (LD) coefficients $u_{1}$ and $u_{2}$ in our MCMC analysis.}
\vspace{-0.3cm}
\end{table}

\indent
The basic jump parameters in our MCMC analyses, i.e. the parameters that are randomly perturbed at each step of the MCMC, were: the planet/star area ratio \hbox{$dF$ = $(R_{\mathrm{p}}/R_{\star})^{2}$}; the occultation depth in the $z'$-band $dF_{\mathrm{occ},\:z'}$; the transit impact parameter in the case of a circular orbit \hbox{$b'$ = $a$ $\mathrm{cos}$ $i_{\mathrm{p}}/R_{\star}$}, where $a$ is the orbital semi-major axis and $i_{\mathrm{p}}$ is the orbital inclination; the transit width (from 1st to 4th contact) $W$; the time of mid-transit $T_{0}$; the orbital period $P$; the stellar effective temperature $T_{\mathrm{eff}}$; the stellar metallicity [Fe/H]; the parameter \hbox{$K_{2} = K \sqrt{1-e^{2}}\:\:P^{1/3}$}, where $K$ is the RV orbital semi-amplitude and $e$ is the orbital eccentricity; the two parameters \hbox{$\sqrt{e}$ cos $\omega$} and \hbox{$\sqrt{e}$ sin $\omega$}, where $\omega$ is the argument of the periastron; and the two parameters \hbox{$\sqrt{v_{\star}\:\mathrm{sin}\:i_{\star}}$ cos $\beta$} and \hbox{$\sqrt{v_{\star}\:\mathrm{sin}\:i_{\star}}$ sin $\beta$}, where \hbox{$v_{\star}$ sin $i_{\star}$} is the projected rotational velocity of the star and $\beta$ is the sky-projected angle between the stellar spin axis and the planet's orbital axis. The reasons to use \hbox{$\sqrt{e}$ cos $\omega$} and \hbox{$\sqrt{e}$ sin $\omega$} as jump parameters instead of the more traditional \hbox{$e$ cos $\omega$} and \hbox{$e$ sin $\omega$} are detailed in \cite{triaud3}. The results obtained from the spectroscopic analysis (see Section \ref{barry}) were used to impose normal prior distributions on $T_{\mathrm{eff}}$, [Fe/H], and $v_{\star}$ sin $i_{\star}$, with expectations and standard deviations corresponding to the quoted measurements and errors, respectively. Uniform prior distributions were assumed for the other parameters. The photometric and RV baseline model parameters were not actual jump parameters; they were determined by least-square minimization from the residuals at each step of the MCMC.\\ 
\indent
The effect of stellar limb-darkening on our transit lightcurves was accounted for using a quadratic limb-darkening law, where the quadratic coefficients $u_{1}$ and $u_{2}$ were allowed to float in our MCMC analysis. However, we did not use these coefficients themselves as jump parameters but their combinations, $c_{1} = 2 \times u_{1} + u_{2}$ and $c_{2}=u_{1} - 2 \times u_{2}$, to minimize the correlation of the obtained uncertainties as introduced by \cite{holman}. To obtain a limb-darkening solution consistent with theory, we used normal prior distributions for $u_{1}$ and $u_{2}$ based on theoretical values and 1-$\sigma$ errors interpolated in the tables by \cite{claret}. These prior distributions are presented in \hbox{Table \textcolor{blue}{3}}.\\
\indent
A preliminary analysis was performed to determine the correction factors ($CF$) for our photometric errors, as described in \cite{gillon2}. For each lightcurve, $CF$ is the product of two contributions, $\beta_{w}$ and $\beta_{r}$. On one side, $\beta_{w}$ represents the under- or overestimation of the white noise of each measurement. It is computed as the ratio between the standard deviation of the residuals and the mean photometric error. On the other side, $\beta_{r}$ allows us to take into account the correlated noise present in the lightcurve (i.e., the inability of our model to perfectly fit the data). It is calculated from the standard deviations of the binned and unbinned residuals for different binning intervals ranging from 5 to 120 min with the largest value being kept as $\beta_{r}$. The standard deviation of the best-fit residuals (unbinned and binned per intervals of 2 min) and the deduced values for $\beta_{w}$, $\beta_{r}$ and $CF=\beta_{w} \times \beta_{r}$ for each lightcurve are presented in Table \ref{obstable}. This preliminary analysis also allowed to compute the jitter values that were added quadratically to the RV errors of each RV dataset to equal their mean values to the standard deviations of the best-fits residuals, and thus achieve reduced $\chi^{2}$ values of unity. These jitter values were \hbox{27.9 $\mathrm{m}\,\mathrm{s}^{-1}$}, 22.3 $\mathrm{m}\,\mathrm{s}^{-1}$, 16.9 $\mathrm{m}\,\mathrm{s}^{-1}$, and zero for the datasets \#1, \#2, \#3, and \#4, respectively.
\\
\indent
A second analysis with the updated photometric and RV errors was then performed to determine the stellar density $\rho_{\star}$, which can be derived from the Kepler's third law and the jump parameters $(R_{\mathrm{p}}/R_{\star})^{2}$, $b'$, $W$, $P$, \hbox{$\sqrt{e}$ cos $\omega$} and \hbox{$\sqrt{e}$ sin $\omega$} (see e.g. \citealt{seager} and \citealt{winn}). This analysis consisted of two Markov chains of $10^{5}$ steps, whose convergence was checked using the statistical test of \cite{GR}. The first 20\% of each chain was considered as its burn-in phase and discarded. The resulting stellar density was used as input of a stellar evolution modeling, together with the effective temperature and metallicity derived from spectroscopy, with the aim to estimate the stellar mass $M_{\star}$ and the age of the system. This stellar evolution modeling is described in details in Section \ref{val}.
\\
\indent
Two final analyses were then performed: one assuming a circular orbit ($e$ = 0) and one with a free eccentricity. Each analysis consisted of two Markov chains of $10^{5}$ steps, whose convergence was again checked with the Gelman \& Rubin test (\citealt{GR}). As previously, the first 20\% of each chain was considered as its burn-in phase and discarded. At each step of the Markov chains, $\rho_{\star}$ was computed as described above and a value for $M_{\star}$ was drawn within a normal distribution having as expectation and standard deviation the value and error obtained from the stellar evolution modeling. The stellar radius $R_{\star}$ was derived from $M_{\star}$ and $\rho_{\star}$, and the other physical parameters of the system were then deduced from the jump parameters and stellar mass and radius. It appeared that the circular orbit was strongly favored by the Bayes factor ($\sim$5000 in its favor) compared to the eccentric orbit. As there was no evidence for a significant eccentricity ($e$ $<$ 0.07 at 3$\sigma$), we thus adopted the circular orbit as our nominal solution. The corresponding derived parameters are presented in \hbox{Table \textcolor{blue}{4}}, while the best-fit models are shown in Fig. \ref{rv}b (RVs), \ref{rossiterfig} (zoom on the RM effect), \ref{lcstransits} (transit photometry), and \ref{lcsoccs} (occultation photometry).\\

\begin{table*}
\vspace{1.5cm}
\centering
\label{param}
\vspace{-0.3cm}
\begin{tabular}{llll}
  \hline
  \hline
  \multicolumn{4}{c}{\textbf{General information}} \\
  \hline
  \hline
  RA (J2000) & 07h 10m 24.06s & $V$-magnitude & 10.44  \\
  Dec (J2000) & $-39^{\circ}\,05^{\prime}\,50.55^{\prime\prime}$ & $K$-magnitude & 9.37  \\ 
  Distance $\mathrm{[pc]}^{a}$ & 270 $\pm$ 90 & & \\
  \hline
  \hline
  \multicolumn{4}{c}{\textbf{Stellar parameters from spectroscopic analysis}} \\
  \hline
  \hline
  Spectral type & F6V & Microturbulence $\xi_{\mathrm{t}}$ [km $\mathrm{s}^{-1}$] & 1.5 $\pm$ 0.1 \\
  Effective temperature $T_{\mathrm{eff}}$ [K] & 6460 $\pm$ 140 & Macroturbulence $v_{\mathrm{mac}}$ [km $\mathrm{s}^{-1}$] & 6.6 $\pm$ 0.6 \\
  Surface gravity log $g_{\star}$  [cgs] & 4.2 $\pm$ 0.2 & Proj. rot. velocity $v_{\star}$ sin $i_{\star}$ [km $\mathrm{s}^{-1}$] & 13.5 $\pm$ 0.7 \\
  Metallicity $[$Fe/H$]$ [dex] & 0.13 $\pm$ 0.09 & Lithium abundance log A(Li) [dex] & $<$ 1.0 \\
  \hline
  \hline
  \multicolumn{4}{c}{\textbf{Parameters from MCMC analysis}}\\
  \hline
  \hline
  \textbf{Jump parameters} & & & \\
  \hline
  Planet/star area ratio $(R_{\mathrm{p}}/R_{\star})^{2}$ [$\%$] & 1.551 $\pm$ 0.012 & Effective temperature $T_{\mathrm{eff}}$ $\mathrm{[K]}^{b}$ & 6459 $\pm$ 140 \\
  Occultation depth $dF_{\mathrm{occ},\:z'}$ [ppm] & 603 $\pm$ 130 & Metallicity [Fe/H] $\mathrm{[dex]}^{b}$ & 0.13 $\pm$ 0.09 \\
  $b' = a\:\mathrm{cos}\:i_{\mathrm{p}}/R_{\star}\:\: [R_{\star}]$ & $0.160_{-0.042}^{+0.040}$ & $c_{1,z'}$ = $2\:u_{1,z'} + u_{2,z'}$ & $0.637_{-0.025}^{+0.026}$ \\
  Transit width $W$ [d] & 0.1203 $\pm$ 0.0003 & $c_{2,z'}$ = $u_{1,z'} - 2\:u_{2,z'}$ & $-0.445$ $\pm$ 0.018 \\
  $T_{0}$ - 2 450 000 [$\mathrm{HJD_{TDB}}$] & $6635.70832_{-0.00010}^{+0.00011}$ & $c_{1,r'}$ = $2\:u_{1,r'} + u_{2,r'}$ & $0.904_{-0.026}^{+0.027}$ \\
  Orbital period $P$ [d] & $1.2749255_{-0.00000025}^{+0.00000020}$ & $c_{2,r'}$ = $u_{1,r'} - 2\:u_{2,r'}$ & $-0.361_{-0.018}^{+0.020}$ \\
  RV $K_{2}$ [$\mathrm{m\:s^{-1}\:d^{1/3}}$] & $196.4_{-6.9}^{+6.8}$ & $c_{1,B}$ = $2\:u_{1,B} + u_{2,B}$ & $1.269_{-0.031}^{+0.032}$ \\
  $\sqrt{e}$ cos $\omega$ & 0 (fixed) & $c_{2,B}$ = $u_{1,B} - 2\:u_{2,B}$ & $-0.012_{-0.045}^{+0.043}$ \\
  $\sqrt{e}$ sin $\omega$ & 0 (fixed) & & \\
  $\sqrt{v_{\star}\:\mathrm{sin}\:i_{\star}}$ cos $\beta$ & $-0.78 \pm 0.34$ & & \\
  $\sqrt{v_{\star}\:\mathrm{sin}\:i_{\star}}$ sin $\beta$ & $-3.59_{-0.11}^{+0.13}$ & & \\
  \hline
  \textbf{Deduced stellar parameters} & & & \\
  \hline
  Mean density $\rho_{\star}$ [$\rho_{\odot}$] & $0.437_{-0.009}^{+0.008}$ & Limb-darkening coefficient $u_{1,z'}$ & 0.166 $\pm$ 0.013 \\
  Surface gravity log $g_{\star}$ [cgs] & $4.242_{-0.012}^{+0.011}$ & Limb-darkening coefficient $u_{2,z'}$ & 0.305 $\pm$ 0.008 \\
  Mass $M_{\star}$ [$M_{\odot}$]$^{c}$ & $1.353_{-0.079}^{+0.080}$ & Limb-darkening coefficient $u_{1,r'}$ & 0.290 $\pm$ 0.014 \\
  Radius $R_{\star}$ [$R_{\odot}$] & 1.458 $\pm$ 0.030 & Limb-darkening coefficient $u_{2,r'}$ & 0.325 $\pm$ 0.007 \\
  Luminosity $L_{\star}$ [$L_{\odot}$] & 3.3 $\pm$ 0.3 & Limb-darkening coefficient $u_{1,B}$ & 0.505 $\pm$ 0.018 \\
  Proj. rot. velocity $v_{\star}$ sin $i_{\star}$ [km $\mathrm{s}^{-1}$]$^{b}$ & $13.56_{-0.68}^{+0.69}$ & Limb-darkening coefficient $u_{2,B}$ & $0.259_{-0.019}^{+0.020}$ \\
  Mean systemic velocity $V_{\gamma}$ [$\mathrm{km\:s^{-1}}$] & 38.350 $\pm$ 0.021 & & \\
  \hline
  \textbf{Deduced planet parameters} & & & \\
  \hline
  RV $K$ [$\mathrm{m\:s^{-1}}$] & $181.1_{-6.4}^{+6.3}$ & Mean density $\rho_{\mathrm{p}}$ [$\rho_{\mathrm{Jup}}$] & 0.201 $\pm$ 0.010 \\
  Planet/star radius ratio $R_{\mathrm{p}}/R_{\star}$ & $0.12454_{-0.00048}^{+0.00047}$ & Surface gravity log $g_{\mathrm{p}}$ [cgs] & 2.973 $\pm$ 0.017 \\
  $T_{\mathrm{occ}}$ - 2 450 000 [$\mathrm{HJD_{TDB}}$] & $6636.34578_{-0.00010}^{+0.00011}$ & Mass $M_{\mathrm{p}}$ [$M_{\mathrm{Jup}}$] & $1.183_{-0.062}^{+0.064}$ \\
  Scaled semi-major axis $a/R_{\star}$ & $3.754_{-0.028}^{+0.023}$ & Radius $R_{\mathrm{p}}$ [$R_{\mathrm{Jup}}$] & 1.807 $\pm$ 0.039 \\ 
  Orbital semi-major axis $a$ [AU] & $0.02544_{-0.00050}^{+0.00049}$ & Roche limit $a_{\mathrm{R}}$ $\mathrm{[AU]}^{d}$ & 0.02205 $\pm$ 0.00066 \\
  Orbital inclination $i_{\mathrm{p}}$ [deg] & 87.6 $\pm$ 0.6 & $a/a_{\mathrm{R}}$ & 1.153 $\pm$ 0.019 \\
  Orbital eccentricity $e$ & 0 (fixed) & Equilibrium temperature $T_{\mathrm{eq}}$ $\mathrm{[K]}^{e}$ & 2358 $\pm$ 52 \\
  Argument of periastron $\omega$ [deg] & - & Irradiation [erg $\mathrm{s}^{-1} \mathrm{cm}^{-2}$] & $7.1_{-1.1}^{+1.3}\:10^{9}$ \\
  Sky-projected orbital obliquity $\beta$ [deg] & $257.8_{-5.5}^{+5.3}$ \\
  \hline
  \textbf{Planet parameters corrected for asphericity} & & & \\
  \hline
  Radius $R_{\mathrm{p}}$ [$R_{\mathrm{Jup}}$] & 1.865 $\pm$ 0.044 & Mean density $\rho_{\mathrm{p}}$ [$\rho_{\mathrm{Jup}}$] & 0.183 $\pm$ 0.016 \\
  \hline
  \hline
\end{tabular}
\vspace{-0.1cm}
\caption{System parameters for WASP-121. The values given for the parameters derived from our MCMC analysis are medians and 1-$\sigma$ limits of the marginalized posterior probability distributions. $ ^{a}$From $V$ mag and estimated absolute magnitude. $ ^{b}$Using as priors the values derived from the spectroscopic analysis. $ ^{c}$Using as prior the value obtained from stellar evolution modeling. $ ^{d}$Using \hbox{$a_{\mathrm{R}}$ = 2.46 $R_{\mathrm{p}}(M_{\star}/M_{\mathrm{p}})^{1/3}$} (\citealt{chandr}). $ ^{e}$Assuming a null Bond albedo and isotropic reradiation (reradiation factor $f$=1/4, \citealt{seagerlopez}).}
\end{table*}

\vspace{-0.5cm}
%------------------------------------------------------------------------
\subsection{Stellar evolution modeling}
\label{val}

As introduced in Section \ref{mcmc}, we performed for the host star a stellar evolution modeling based on the CLES code (\citealt{scuflaire}), in order to estimate the stellar mass and the age of the system. We used as inputs the stellar density deduced from the transit lightcurves, and the effective temperature and metallicity derived from spectroscopy (see Table \textcolor{blue}{4}). We considered that [Fe/H] represents the global metallicity with respect to the Sun i.e. \hbox{[Fe/H] $=[\log (Z/X)_* - \log (Z/X)_{\odot}]$}, with $(Z/X)_{\odot} = 0.0181$ (\citealt{asplund}). The parameter of the mixing-length theory (MLT) of convection was kept fixed to the solar calibration ($\alpha_{\rm MLT} = 1.8$), and the possible convective core extra-mixing (due to overshooting, rotationally-induced mixing, etc.) and microscopic diffusion (gravitational settling) of elements were included.

\begin{figure}
\centering             
\includegraphics[trim= 240 0 260 0.5, clip=true,height=7.8cm,width=8.5cm]{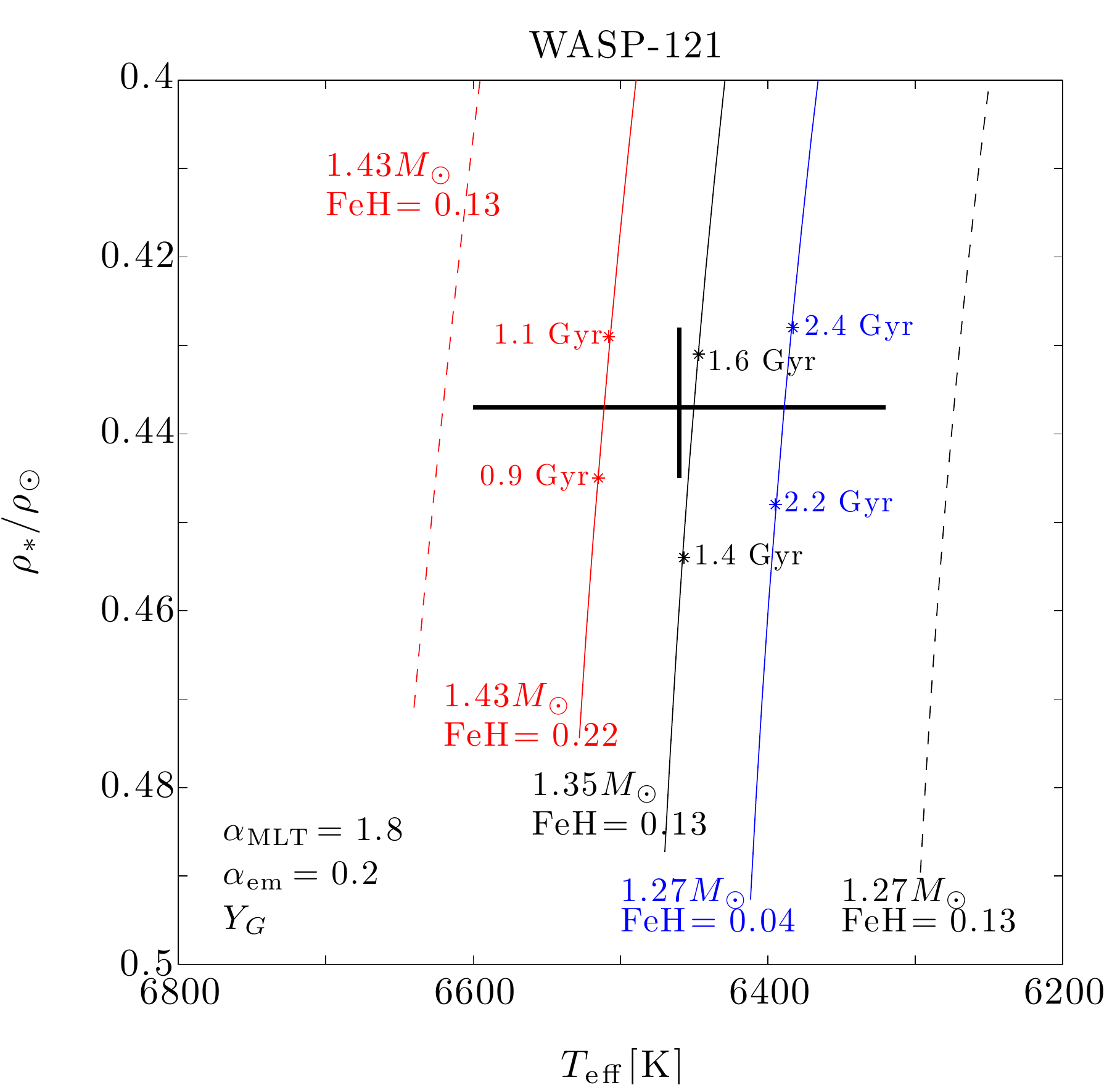}
\vspace{-0.5cm}
\caption{Evolutionary tracks in a $T_{\mathrm{eff}}-\rho_{\star}/\rho_{\odot}$ HR diagram for WASP-121, for different masses and metallicities, within (solid lines) and without (dashed lines) the 1-$\sigma$ box $T_{\mathrm{eff}}-\rho_{\star}/\rho_{\odot}$. Some stellar ages are also indicated.} 
\label{sem}
\end{figure}

\indent
We obtained a stellar mass of 1.355 $\pm$ 0.080 $M_{\odot}$ and an age of 1.5 $\pm$ 1.0 Gyr. These 1-$\sigma$ uncertainties were obtained by considering the respective 1-$\sigma$ range for the effective temperature, metallicity and stellar density, but also by varying the internal stellar physics. We computed, since the helium atmospheric abundance cannot be directly measured from spectroscopy, evolutionary tracks with two initial helium abundances: the solar value ($Y_{\odot,0}=0.27$), and a value labelled $Y_G$ that increases with $Z$ (as expected if the local medium follows the general trend observed for the chemical evolution of galaxies; \citealt{izotov}). We also investigated the effects of the possible convective core extra-mixing, by varying it between 0 and 0.3 (\citealt{noelsmontal}).\\ 
\indent
Evolutionary tracks are presented on Fig. \ref{sem} for several stellar masses and metallicities. Obtaining an accurate stellar mass from evolution modeling primarily needs accurate spectroscopic estimates for the effective temperature but also, very importantly, for the metallicity. Metallicity is a parameter that strongly determines the location of the evolutionary tracks in a HR diagram (compare in Fig. \ref{sem} the solid and dashed tracks for identical stellar masses, but with different metallicities). The stellar mass obtained from stellar evolution modeling \hbox{(1.355 $\pm$ 0.085 $M_{\odot}$)} is in excellent agreement with the first estimate derived from the \cite{torres} calibration in Section \ref{barry}. The inferred stellar age of \hbox{1.5 $\pm$ 1.0 Gyr} places \hbox{WASP-121} on the main sequence. The total lifetime on the main sequence (H-core burning) of \hbox{WASP-121} is 4.3 Gyr for a moderate extra-mixing $\alpha_{\rm em}=$ 0.2, and is reduced to 3.3 Gyr without considering any extra-mixing process in the core.

%------------------------------------------------------------------------

\vspace{-0.2cm}
\section{\texttt{PASTIS} validation}
\label{pastis}

Astrophysical false-positive scenarios such as eclipsing binaries might mimic both the transit and radial velocity signal of a planet (\citealt{torres2005}). In some particular configurations, even the Rossiter-McLaughlin effect might be mimicked (\citealt{santerne2015}). In the present case, the line bisector and the FWHM present a large variation. Moreover we were able to detect from the ground in the $z'$-band a relatively deep secondary eclipse. Both arguments prevent to secure the planetary nature of WASP-121\,b without a more careful investigation. We can imagine five scenarios which can reasonably produce the observed data: (1) a planet transiting the target star, (2) a planet transiting a chance-aligned background star, (3) a chance-aligned background eclipsing binary, (4) a planet transiting a physical companion to the target star, and (5) a star eclipsing a physical companion to the target star (i.e. a triple system). The scenarios 2 -- 5 can be split in two categories and discussed separately: the background or physical companion sources of false positive.

\vspace{-0.2cm}
\subsection{A background source}

\begin{figure}
\begin{center}
\includegraphics[bb= 25 250 640 310, width=0.49\textwidth]{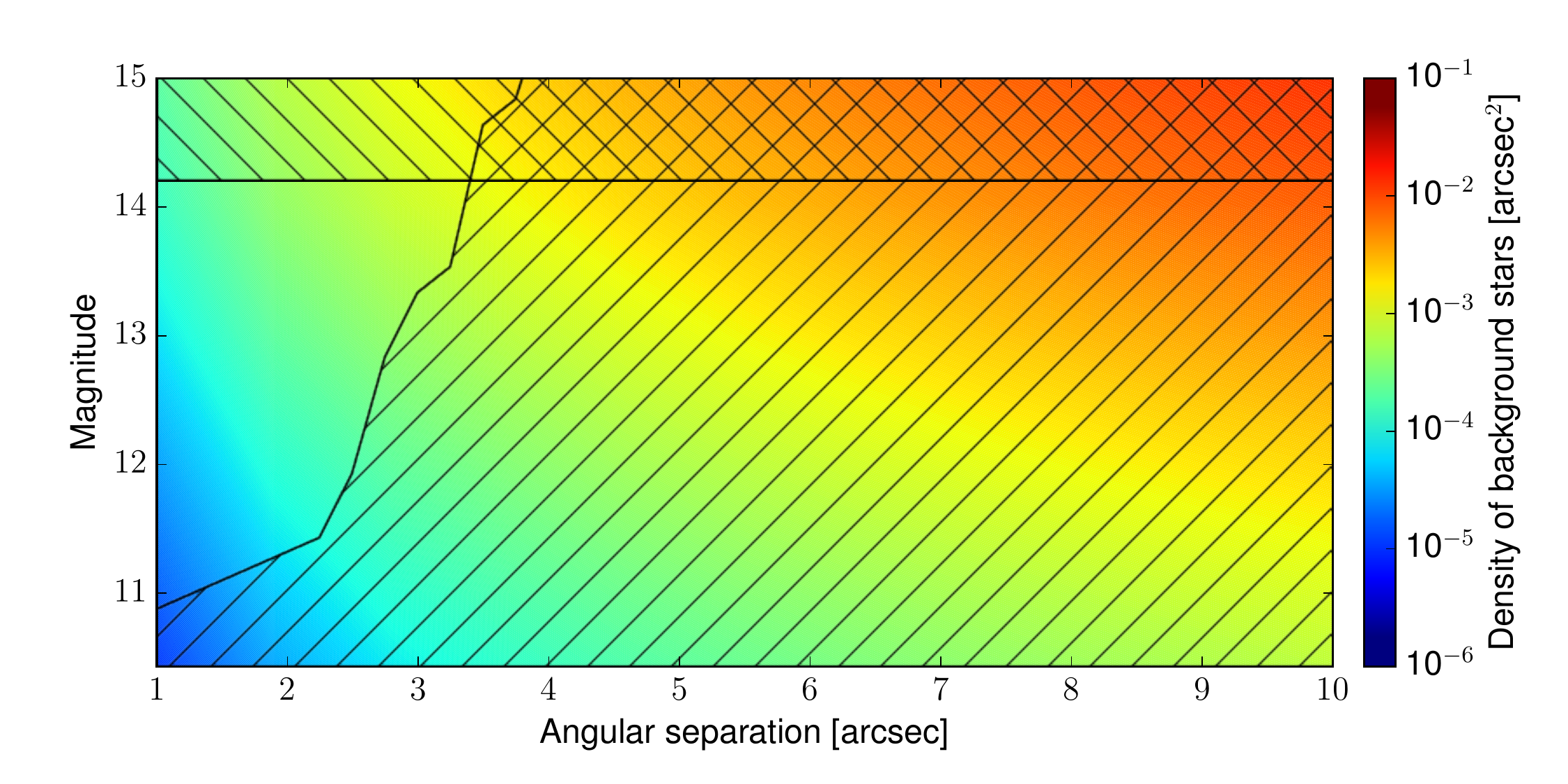}\\
\includegraphics[bb=25 5 640 560, width=0.49\textwidth]{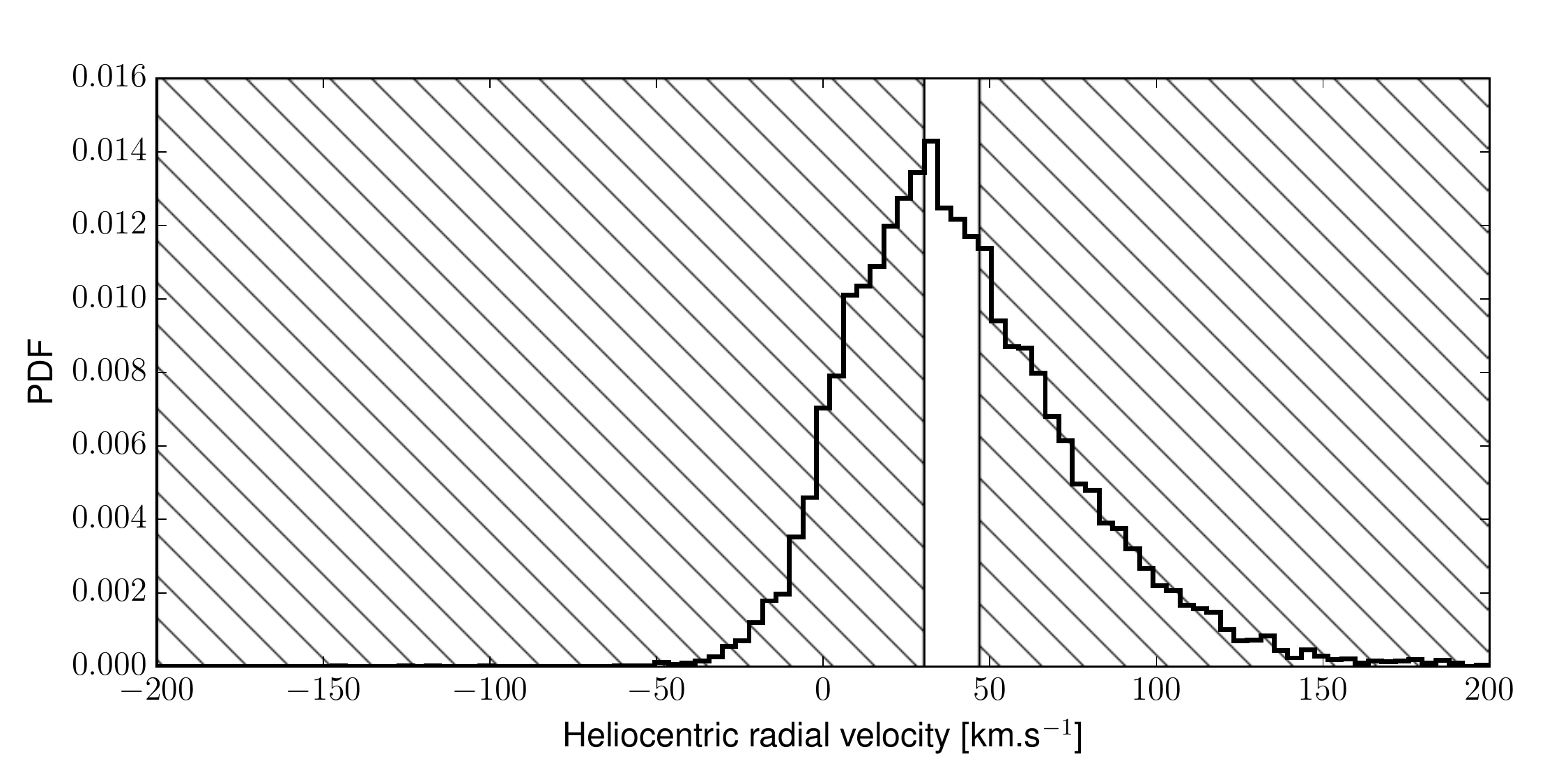}
\vspace{-0.4cm}
\caption{\textit{Upper panel:} Map of the density of background stars chance-aligned with WASP-121, integrated within an angular separation of up to 10\arcsec, as function of the $V$-band magnitude. The positive-slope hatched region displays all the stars that would have been significantly detected in the TRAPPIST data. The negative-slope hatched region displays the stars that are too faint to reproduce the observed transit depth of WASP-121. \textit{Lower panel:} Distribution of heliocentric radial velocity of stars within 10 degrees from WASP-121 as observed by the RAVE experiment. The hatched regions show the velocity of background stars that should be spectroscopically resolved with WASP-121 and thus should produce a radial velocity variation in anti-phase with the transit ephemeris, or no significant variation (\citealt{santerne2015}).}
\label{BGfig}
\vspace{-0.3cm}
\end{center}
\end{figure}

A background source of false positive can mimic the transit data of a planet within a range a magnitude $\Delta m$ defined as (\citealt{morton2011}):
\begin{equation}
\Delta m = 2.5 \log_{10}\left(\frac{\delta_{tr}}{\delta_{bg}}\right),
\end{equation}
where $\delta_{tr}$ and $\delta_{bg}$ are the depth of the transit, as measured in the lightcurve, and the depth of the background eclipse, respectively. Assuming a maximum eclipse depth of 50\% for the background star, we find that the maximum magnitude range is of 3.78. Note that a 50\% depth eclipsing binary should also produce a secondary eclipse with a similar depth, which is clearly excluded in the case of WASP-121. Therefore, this magnitude range is over-estimated, which will also over-estimate the probability of a background star as the host of the observed signal. The star is of magnitude 10.44 in the $V$-band, hence false positive can probe stars as faint as magnitude 14.22 in the same bandpass.\\
\indent
To compute the probability of having a background star chance-aligned with WASP-121, we took the APASS (\citealt{henden2015}) DR8 $V$-magnitude of all stars within 1 degree from WASP-121. The magnitude limit of this catalog is between magnitude 16 and 17, which is about 2 magnitudes above the maximum magnitude of the background star. We therefore assume that this catalog is complete in the range of magnitude we are considering here. We did not detect in the TRAPPIST data any background source that could host the transit. Thus, in this scenario, the background source should be blended within the TRAPPIST PSF. In Fig. \ref{BGfig} we display the sensitivity curve of the TRAPPIST PSF. This curve was obtained by injecting artificial stars in good-seeing TRAPPIST images at various separations and with a range of magnitude, and then attempting to detect them with IRAF/DAOPHOT. Any star brighter than magnitude 14 and separated from WASP-121 by more than 3.5\arcsec\, should have been detected. Using these constraints and assuming that the stars are randomly distributed around the target star, we find that the probability for WASP-121 to be chance-aligned with a background star brighter than magnitude 14.22 is at the level of maximum 8.10$^{-4}$. If we account for the probability that this hypothetical background star host a binary ($\sim$50\% ; \citealt{raghavan2010}) or a planet ($\sim$50\% ; \citealt{mayor2011}) and the eclipse or transit probability at 1.3 days ($\sim$25\%), we end with a \textit{a priori} probability that the transit signal is hosted by a background star lower than $\sim1.10^{-4}$.

\indent
A significant radial velocity variation has been detected in phase with the transit ephemeris and no other component has been significantly detected in the CORALIE CCFs with a flux ratio greater than 0.7\% (i.e. 5.4 magnitudes). Therefore, to mimic both the photometric and radial velocity data of WASP-121, the background source should be chance-aligned along the line of sight and in the radial velocity space. To produce a radial velocity in phase with the transit ephemeris, the background star should have a systemic velocity within the width of the line profile of \hbox{WASP-121} (\citealt{santerne2015}), i.e. the two stars should be spectroscopically unresolved. Given the systemic velocity of WASP-121 (\hbox{$\gamma \sim 38.3$~km s$^{-1}$}) and the width of its line profile observed by CORALIE (\hbox{$\sigma_{CCF} = \mathrm{FWHM}/2\sqrt{2\ln(2)} \approx 8.2$~km s$^{-1}$}), the background star should thus have a systemic radial velocity in the range [30.1 ; 46.5] km s$^{-1}$. We then compared this range of radial velocity with the distribution of stars in the vicinity of WASP-121 from the RAVE database (\citealt{kordopatis2013}). The offset between the CORALIE zero-point and the RAVE reference is expected to be less than a few hundreds of m s$^{-1}$. We took all the stars observed by this spectroscopic survey within 10 degrees of WASP-121 for which we display their radial velocity distribution in Fig. \ref{BGfig}. Only 20\% of the stars in the vicinity of WASP-121 should be spectroscopically unresolved. Assuming that the position of a star in a place of the sky is independent from its radial velocity, we find that the \textit{a priori} probability that the signal observed in WASP-121 is caused by a background source is, at most, at the level of 20 ppm. We can therefore exclude all background false-positive scenarios as the source of the detected signal.

\vspace{-0.2cm}

\subsection{A physical companion}

Apart from the planet scenario, the most likely scenario to reproduce both the photometric and spectroscopic data is a system physically bound with the target star. In such configuration, the various stellar components could be easily blended in both the plane of the sky and the radial velocity space. Then, the transit could be mimicked either by an eclipsing star or a transiting sub-stellar object.\\
\indent
To estimate the probability that WASP-121 is either a planet transiting the target, a physical companion (planet in binary) or a triple system, we used the \texttt{PASTIS} software (\citealt{diaz}, \citealt{santerne14}, \citealt{santerne2015}) to model the 11k+ photometric measurements obtained by TRAPPIST and EulerCam in different filters. The lightcurve was modelled using the \texttt{EBOP} code (\citealt{nelson1972}, \citealt{etzel1981}, \citealt{popper1981}) extracted from the \texttt{JKTEBOP} package (\citealt{southworth2008}). For the limb darkening coefficients, we used the interpolated values from \cite{claret}. To model the stars, we used the Dartmouth stellar evolution tracks of \cite{dotter2008} and the BT-SETTL stellar atmosphere models of \cite{allard2012} that we integrated in each individual bandpass. Since they are gravitationally bound, all stars were assumed to have the same metallicity and the same age. The orbits were assumed to be circular. We imposed that the physical companion is fainter by at least 1 magnitude in the $V$-band than the target star, otherwise, it would have been clearly identified in the spectral or CCF analysis. Each of the 16 lightcurves was modelled self-consistently allowing a different out-of-transit flux, contamination and an extra source of white noise (jitter). As \texttt{PASTIS} is not yet able to model the activity of stars in radial velocity data, we decided not to use these extra constraints. We also modelled the spectral energy distribution of WASP-121 composed by the magnitudes in the Johnson-$B$ and -$V$, Sloan-$g'$, -$r'$, and -$i'$, 2-MASS $J$, $H$, and $K$s, and WISE $W1$ to $W4$ bandpasses from the APASS database (\citealt{henden2015}) and the AllWISE catalog (\citealt{wright2010}).\\
\indent
We analysed the aforementioned data using a MCMC procedure described in \cite{diaz}. For the priors, we used the results from the spectroscopic analysis (\hbox{Section \ref{barry}}) for the parameters of the target star and the initial mass function from \cite{kroupa2001} for the blended stars. For the orbital ephemeris, we used Normal priors matching the ephemeris reported in Table \textcolor{blue}{4} with uncertainties boosted by 100, to avoid biasing the results with too narrow priors. For the other parameters, we chose uninformative priors. We limited the priors on the planet radius to be less than \hbox{2.2 $R_{\mathrm{Jup}}$}, which is the radius of the biggest planet found so far: \hbox{KOI-13} (\citealt{szabo2011}). The exhaustive list of parameters and their priors are reported in Table \ref{PASTISpriors}. Both the planet and triple scenarios were described by ten free parameters, while the planet in binary scenario used 11 free parameters. Among them, eight were in common (target star and orbital parameters). An additional 49 free parameters were needed to describe the data: three for each of the 16 lightcurves as already mentioned and an extra-source of white noise for the spectral energy distribution. For all scenarios, we ran 40 chains of 3.10$^{5}$ iterations, randomly started from the joint prior distribution. All chains converged toward the same maximum of likelihood. We then thinned and merged the chains to derive the posterior distributions of all scenarios. They all have a posterior distribution with more than 1000 independent samples. We report in Table \ref{PASTISresults} the median and 68.3\% confidence interval for the free parameters. All the fitted parameters for the planet scenario are compatible within 1-$\sigma$ with those derived in Section \ref{mcmc}.

\begin{figure}
\begin{center}
\includegraphics[bb=15 5 580 580, width=0.49\textwidth]{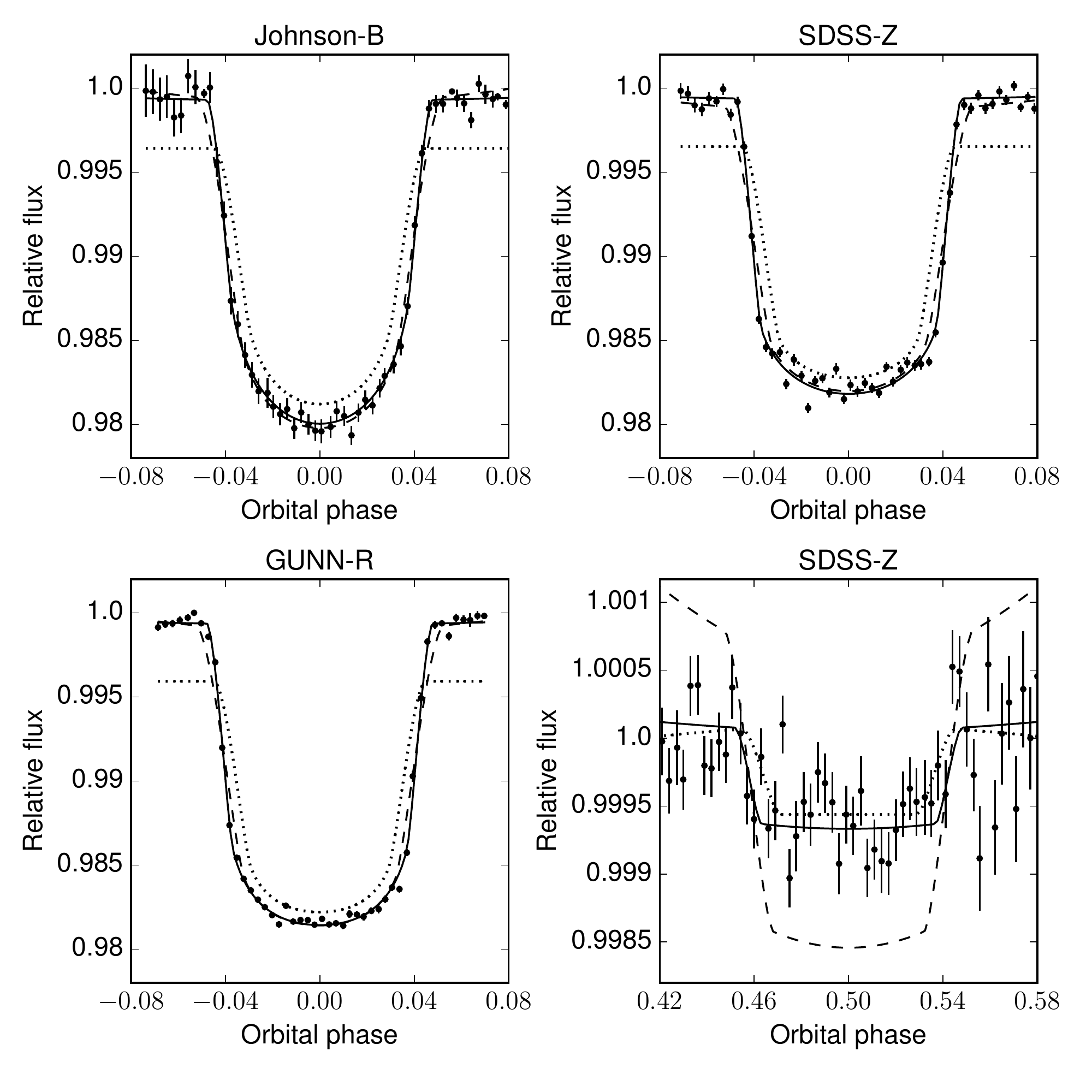}
\vspace{-0.4cm}
\caption{Phase-folded binned photometric data of WASP-121 observed in the various bandpasses together with the best planet (solid line), triple system (dash line), and planet in binary (dot line) models.}
\label{PASTISmodel}
\end{center}
\end{figure}

\begin{figure*}
\centering   
\includegraphics[bb=5 5 550 550, angle=0, width=0.48\textwidth]{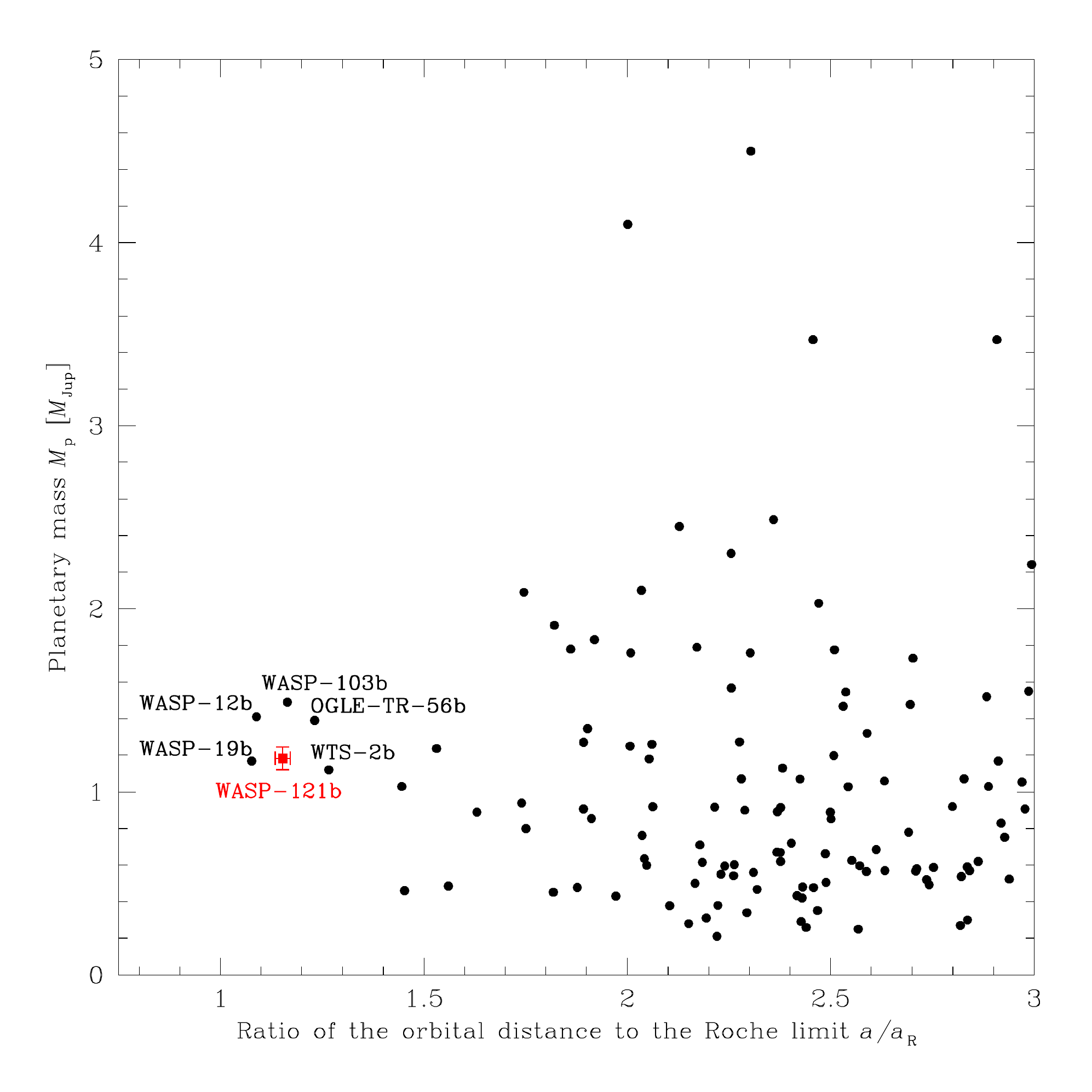}
\includegraphics[bb=5 5 550 550, angle=0, width=0.48\textwidth]{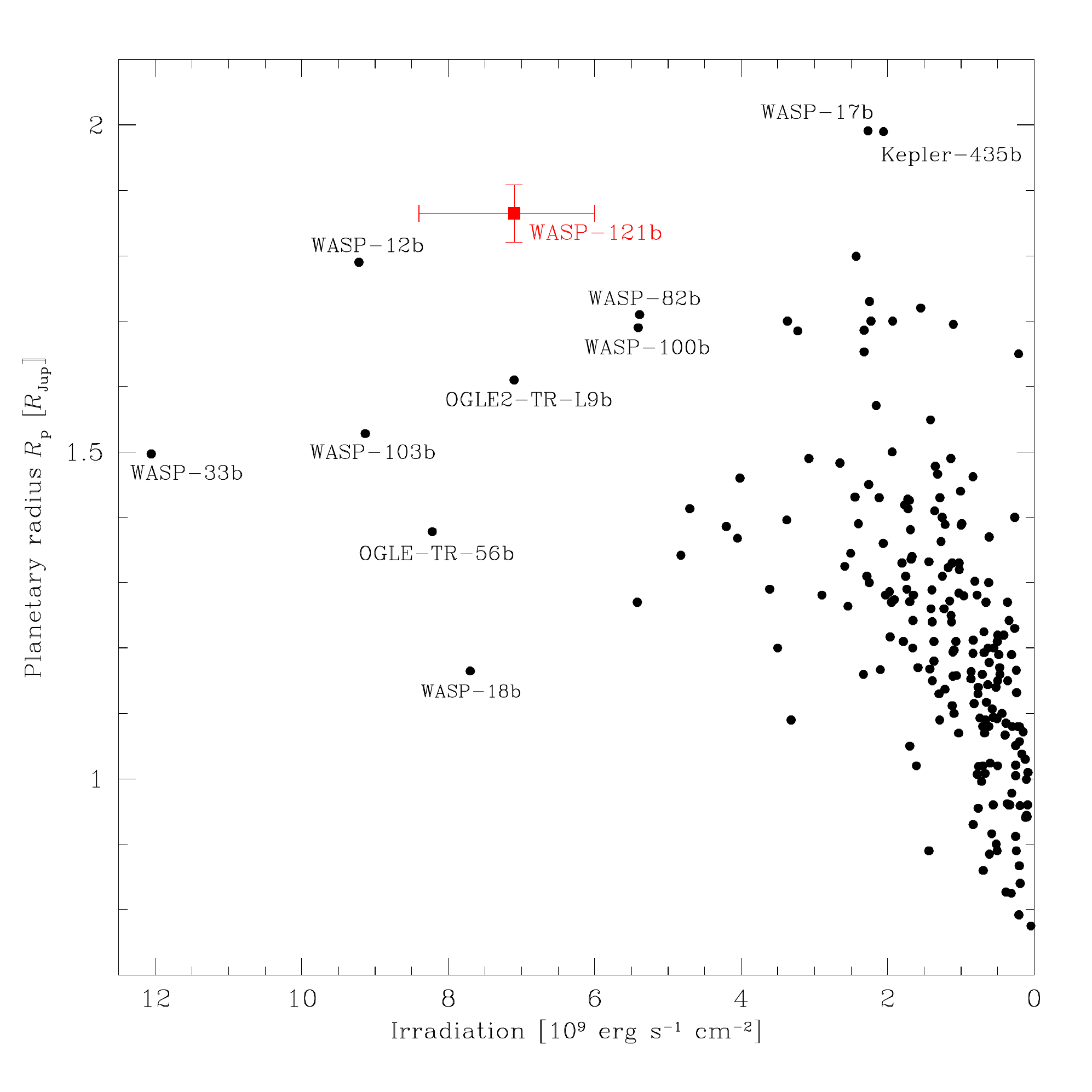}    
\vspace{-0.1cm}       
\caption{\textit{Left:} Orbital distance to the Roche limit ratio-mass diagram for the known transiting hot Jupiters with \hbox{0.2 $M_{\mathrm{Jup}}<M_{\mathrm{p}}<13$ $M_{\mathrm{Jup}}$} and $P<12\:\mathrm{d}$ (data from the NASA Exoplanet Archive). We only show planets with $a/a_{\mathrm{R}}$ $<$ 3. The planet WASP-121\,b is shown in red. \textit{Right:} Irradiation-radius diagram for the same sample of planets.}
\label{flux}
\end{figure*}

\indent
We display in Fig. \ref{PASTISmodel} the transit and occultation data of WASP-121 in the various bandpasses together with the best planet, planet in binary and triple models. While the planet scenario reproduces well all the data, the triple system scenario is not able to fit perfectly the relatively sharp ingress and egress of the transit. The main difference between the two scenarios is for the secondary eclipse. While the planet scenario is able to reproduce well the occultation data, the best triple scenario exhibits a secondary depth of more than 2000 ppm, which is clearly excluded by the data. For the planet transiting a companion star to the target, it is not possible to reproduce the observed transit depth and duration if we assume both that the companion star is fainter by at least 1 magnitude than the target star and that the maximum planetary radius allowed is 2.2 $R_{\mathrm{Jup}}$. In these conditions, to reproduce well the transit data, the planet would need to have a radius of 3.24 $R_{\mathrm{Jup}}$, which is clearly unphysical for such objects.\\
\indent
We estimated the Bayes factor, $B_{ij}$, between the three scenarios using the method of \cite{tuomi2012}. This method has some limitations that are discussed in \cite{santerne14}. However, since we tested here scenarios that have nearly the same number of free parameters and most of these parameters have the same priors in the various scenarios, we assume that these limitations have no significant impact on our results. We found that $\log_{10}B_{ij} \sim 250$ and $\log_{10}B_{ij} \sim 870$, against the triple and planet-in-binary scenarios (respectively), in favour to the planet one. This significantly validates \hbox{WASP-121~b} as a \textit{bona-fide} planet. Computing the BIC between the best models give a value of 1103 and 3905 against the triple and planet-in-binary scenarios (respectively), which confirms that \hbox{WASP-121\,b} is a transiting planet.

\vspace{-0.3cm}
%%%%%%%%%%%%%%%%%%%%%%%%%%%%%%%%%%%%%%
\section{Discussion}
\label{discussion}

WASP-121\,b is a $\sim$1.18 $M_{\mathrm{Jup}}$ planet in a 1.27 days orbit around a bright ($V$=10.4) F6V star. Its most notable property is that its orbital semi-major axis is only $\sim$1.15 times larger than its Roche limit, which suggests that the planet might be close to tidal disruption. \hbox{Fig. \ref{flux} (left)} shows the distribution of the orbital distance to the Roche limit ratio ($a/a_{\mathrm{R}}$) as a function of the planetary mass for the known transiting planets with \hbox{0.2 $M_{\mathrm{Jup}}$ $<$ $M_{\mathrm{p}}$ $<$ 13 $M_{\mathrm{Jup}}$} and \hbox{$P$ $<$ 12 d}. WASP-121\,b is one of the closest systems to tidal disruption, its direct competitors being \hbox{WASP-12\,b} (\citealt{w12}), \hbox{WASP-19\,b} (\citealt{w19}), \hbox{WASP-103\,b} (\citealt{w103}), \hbox{OGLE-TR-56\,b} (\citealt{ogle56a}, \citealt{ogle56}), and \hbox{WTS-2\,b} (\citealt{birkby}). According to \cite{matsumura}, these extreme planets are now expected to undergo tidal orbital decay through tidal dissipation inside the star only (we will elaborate on this in Section \ref{orbi} in the case of \hbox{WASP-121\,b}). The speed of this final orbital decay depends mainly on the mass of the planet and the tidal dissipation efficiency of the star. As noted by \cite{w103}, the fact that all the planets near tidal disruption found to date have similar masses (between 1.1 and \hbox{1.5 $M_{\mathrm{Jup}}$}, see the left panel of Fig. \ref{flux}) could thus suggest a narrow range of tidal dissipation efficiencies for the host stars of the known transiting hot Jupiters (mainly solar-type stars).

\subsection{Structural parameters of WASP-121\,b: correction for asphericity using Roche geometry}

\indent
Being close to its Roche limit, WASP-121\,b might be significantly deformed by the intense tidal forces it is subject to (e.g. \citealt{budaj}) and even lose some of its mass through Roche lobe overflow (e.g. \citealt{li}, \citealt{lai}). To evaluate the tidal distortion of the planet, we calculate its Roche shape by using the Roche model of \cite{budaj}, as done by \cite{southworth103} for WASP-103\,b. This model assumes that the planet is on a circular orbit and that it is rotating synchronously with its orbital period ($P_{\mathrm{rot}}=P_{\mathrm{orb}}$). This second assumption is perfectly reasonable as current theories of tidal evolution of close-in exoplanets predict synchronization times much shorter than times needed for circularization of the orbits (see e.g. \citealt{rasio96}). The model takes as main inputs the orbital semi-major axis ($a$ = 5.47 $\pm$ 0.11 $R_{\odot}$), the star/planet mass ratio ($M_{\star}/M_{\mathrm{p}} = 1198_{-128}^{+141}$), and the planetary radius ($R_{\mathrm{p}}$ = 1.807 $\pm$ 0.039 $R_{\mathrm{Jup}}$), and computes the Roche shape of the planet which would have the same cross-section during the transit as the one we inferred from our observations assuming a spherical planet (eclipse model of \citealt{mandel}, see Section \ref{mcmc}).\\ 
\indent
The results are displayed in Table \textcolor{blue}{5}. $R_{\mathrm{sub}}$, $R_{\mathrm{back}}$, $R_{\mathrm{pole}}$, and $R_{\mathrm{side}}$ are the planetary radii at the sub-stellar point, the anti-stellar point, the rotation pole, and on the side, respectively. Together, these parameters describe the Roche shape of the planet. $R_{\mathrm{cross}}$ is the cross-sectional radius, i.e. the radius of the sphere that would have the same cross-section as the Roche surface of the planet during the transit. It is the planetary radius we derived from our global analysis in Section \ref{mcmc} (i.e. the observed radius). $R_{\mathrm{mean}}$ is the radius of the sphere with the same volume as the Roche surface of the planet. The Roche-lobe filling parameter $ff$ is defined as $R_{\mathrm{sub}}/R_{\mathrm{L_{1}}}$, where $R_{\mathrm{L_{1}}}$ is the distance of the Lagrangian L1 point. The asphericity of the planet can be quantified by the ratios $R_{\mathrm{sub}}/R_{\mathrm{side}}$, $R_{\mathrm{sub}}/R_{\mathrm{pole}}$, and $R_{\mathrm{side}}/R_{\mathrm{pole}}$. Finally, the quantity $(R_{\mathrm{cross}}/R_{\mathrm{mean}})^{3}$ is the correction factor that must be applied to the planetary density $\rho_{\mathrm{p}}$ derived from our global analysis assuming a spherical planet to convert it to the density obtained using the Roche model.\\ 
\indent
With a Roche-lobe filling parameter of 0.59, \hbox{WASP-121\,b} is still well away from Roche lobe overflow, despite being significantly deformed. It would nonetheless be interesting to search for potential signatures of planetary material surrounding WASP-121 (e.g. excess transit depths in the near-UV). Such signatures have indeed been possibly detected for WASP-12 by \cite{fossati1}, \cite{haswell}, and \cite{fossati2}, although \cite{budaj} reported a value of only 0.61 for the Roche-lobe filling parameter of WASP-12\,b. If we use the $R_{\mathrm{sub}}/R_{\mathrm{pole}}$ ratio to quantify the departures from the sphere, we find that WASP-121\,b ($R_{\mathrm{sub}}/R_{\mathrm{pole}}$=1.124) is one of the most distorted planets known to date, alongside WASP-12\,b ($R_{\mathrm{sub}}/R_{\mathrm{pole}}$=1.138, \citealt{budaj}), \hbox{WASP-19\,b} ($R_{\mathrm{sub}}/R_{\mathrm{pole}}$=1.124, \citealt{budaj}), and \hbox{WASP-103\,b} ($R_{\mathrm{sub}}/R_{\mathrm{pole}}$=1.120, \citealt{southworth103}). $R_{\mathrm{mean}}$ being more representative of the physical size of the planet than $R_{\mathrm{cross}}$, we adopt it as our revised value for the planetary radius (1.865 $\pm$ 0.044 $R_{\mathrm{Jup}}$) and include it, as well as the subsequent revised value for the planetary density (0.183 $\pm$ 0.016 $\rho_{\mathrm{Jup}}$), in \hbox{Table \textcolor{blue}{4}}. We note that $R_{\mathrm{mean}}$ should be used when comparing the radius of the planet with theoretical models (e.g. \citealt{fortney}), while $R_{\mathrm{cross}}$ should rather be employed when interpreting transmission or occultation data.\\
\indent
Several works showed that hot Jupiters' radii correlate well with their incident irradiation (see e.g. \citealt{demory}, \citealt{enoch2}, or \citealt{weiss}). Fig. \ref{flux} (right) shows the position of \hbox{WASP-121\,b} in an irradiation-radius diagram for the same sample of transiting planets as previously. With a radius of 1.865 $\pm$ 0.044 $R_{\mathrm{Jup}}$ and an irradiation of \hbox{$\sim$$7.1\:10^{9}$ erg $\mathrm{s}^{-1} \mathrm{cm}^{-2}$}, WASP-121\,b joins the handful of extremely irradiated planets with super-inflated radii. Its radius is actually significantly larger than the value of \hbox{1.50 $\pm$ 0.03 $R_{\mathrm{Jup}}$} predicted by the equation derived by \cite{weiss} from a sample of 103 transiting planets with a mass between \hbox{150 $M_{\oplus}$} and \hbox{13 $M_{\mathrm{Jup}}$} and that relates \hbox{planets' sizes} to their masses and irradiations. Several physical mechanisms have been proposed to explain the inflated radii of hot Jupiters, such as tidal heating (\citealt{bodenheimer}), deposition of kinetic energy into the planets from strong winds driven by the large day/night temperature contrasts (\citealt{showman}), enhanced atmospheric opacities (\citealt{burrows}), reduced heat transport efficiency by layered convection inside the planets (\citealt{chabrier}), or Ohmic heating from currents induced through winds in the planetary atmospheres (\citealt{batygin}). As the WASP-121 system is quite young (1.5 $\pm$ 1.0 Gyr, see Section \ref{val}), tidal circularization and dissipation might have occurred recently enough to contribute to the observed inflated radius.

\begin{table}
\centering
\begin{tabular}{lc}
  \hline
  Parameter & Value \\
  \hline
  Radius at the sub-stellar point $R_{\mathrm{sub}}$ [$R_{\mathrm{Jup}}$] & 2.009 $\pm$ 0.072 \\
  Radius at the anti-stellar point $R_{\mathrm{back}}$ [$R_{\mathrm{Jup}}$] & 1.997 $\pm$ 0.069 \\
  Radius at the rotation pole $R_{\mathrm{pole}}$ [$R_{\mathrm{Jup}}$] & 1.787 $\pm$ 0.038 \\
  Radius on the side $R_{\mathrm{side}}$ [$R_{\mathrm{Jup}}$] & 1.828 $\pm$ 0.041 \\
  Cross-sectional radius $R_{\mathrm{cross}}$ [$R_{\mathrm{Jup}}$] & 1.807 $\pm$ 0.039 \\
  Mean radius $R_{\mathrm{mean}}$ [$R_{\mathrm{Jup}}$] & 1.865 $\pm$ 0.044 \\
  Roche-lobe filling parameter $ff$ & 0.591 $\pm$ 0.040 \\
  $R_{\mathrm{sub}}/R_{\mathrm{side}}$ & 1.099 $\pm$ 0.020 \\
  $R_{\mathrm{sub}}/R_{\mathrm{pole}}$ & 1.124 $\pm$ 0.026 \\
  $R_{\mathrm{side}}/R_{\mathrm{pole}}$ & 1.023 $\pm$ 0.003 \\
  Density correction factor $(R_{\mathrm{cross}}/R_{\mathrm{mean}})^{3}$ & 0.910 $\pm$ 0.013 \\
  \hline
\end{tabular}
\label{rochegeom}
\caption{Parameters describing the shape of WASP-121\,b, obtained using the Roche model of Budaj (2011). The errors are the quadratic sum of the errors due to each input parameter ($a$, $M_{\star}/M_{\mathrm{p}}$, and $R_{\mathrm{p}}$).}
\end{table}

\begin{figure*}
\centering                    
\includegraphics[bb=15 280 550 545, width=0.46\textwidth]{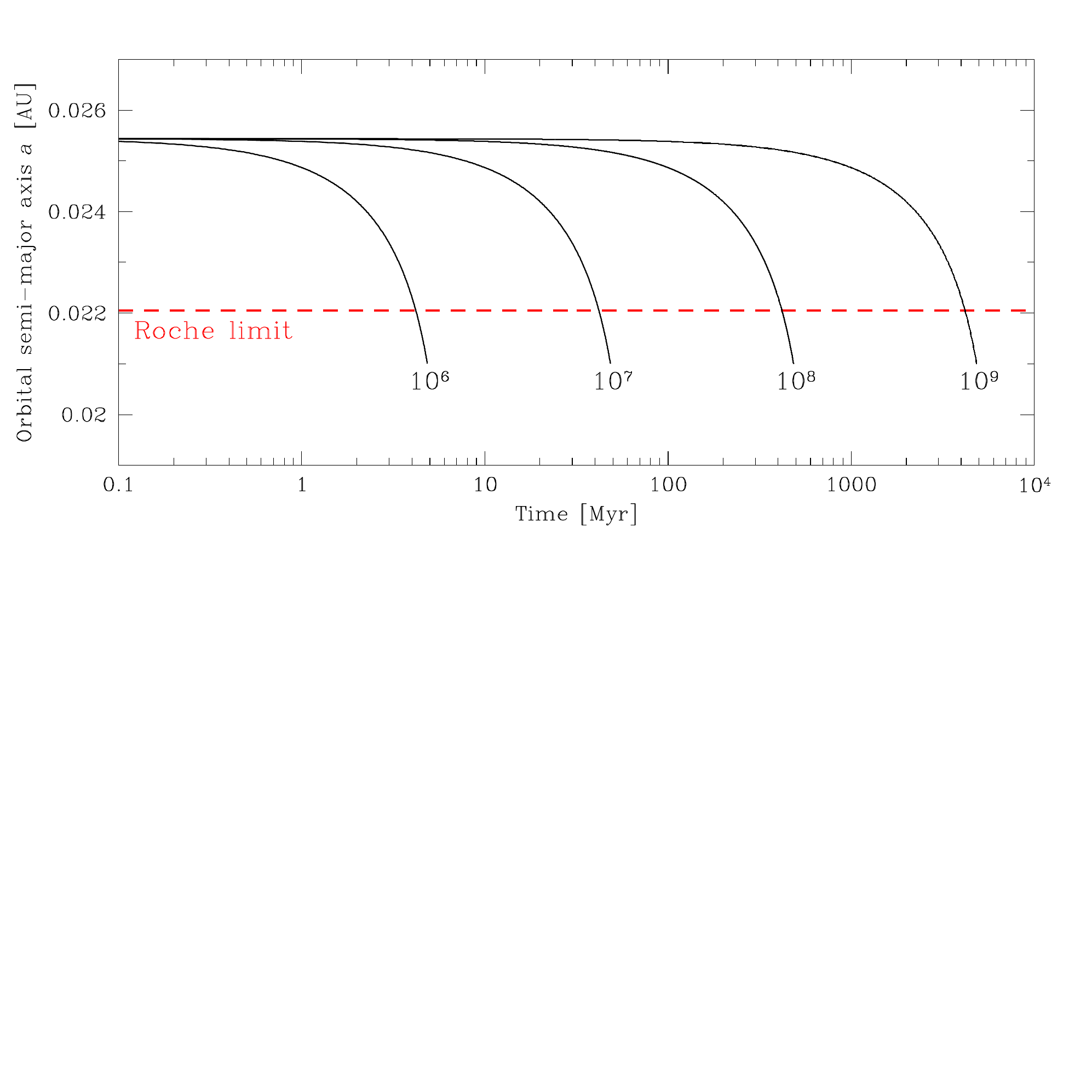}
\hspace{0.5cm}
\includegraphics[bb=15 280 550 545, width=0.46\textwidth]{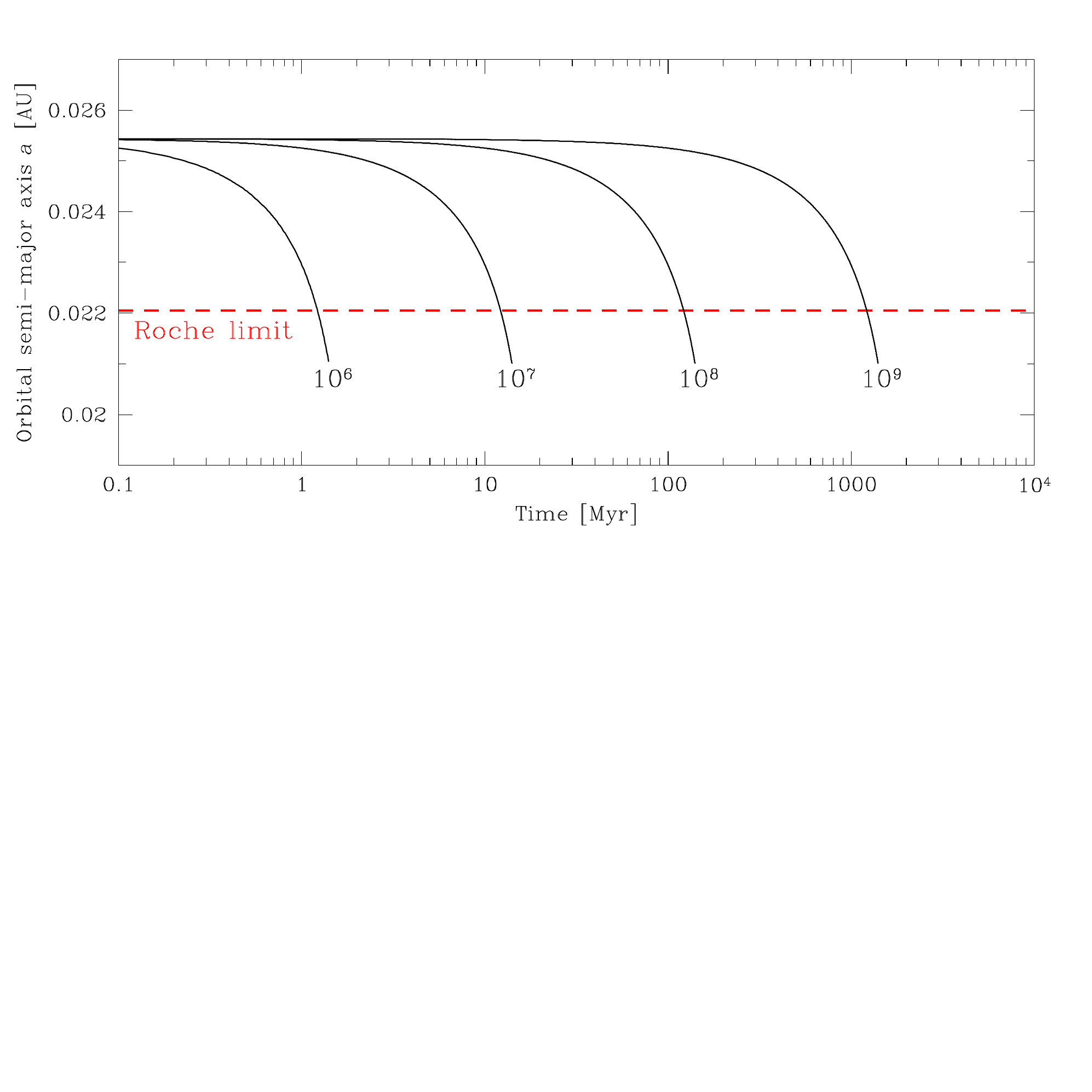}
\includegraphics[bb=15 280 550 545, width=0.46\textwidth]{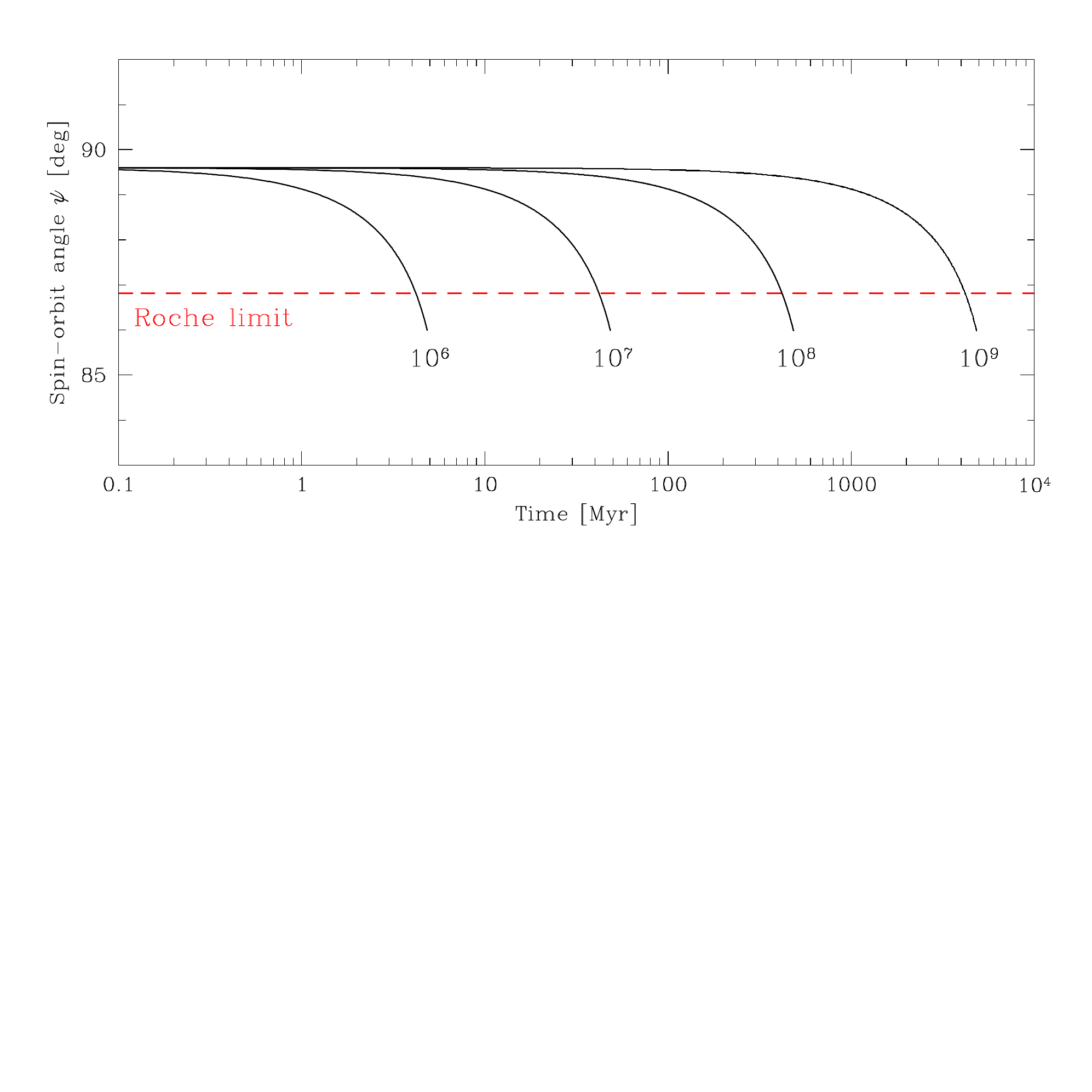}
\hspace{0.5cm}
\includegraphics[bb=15 280 550 545, width=0.46\textwidth]{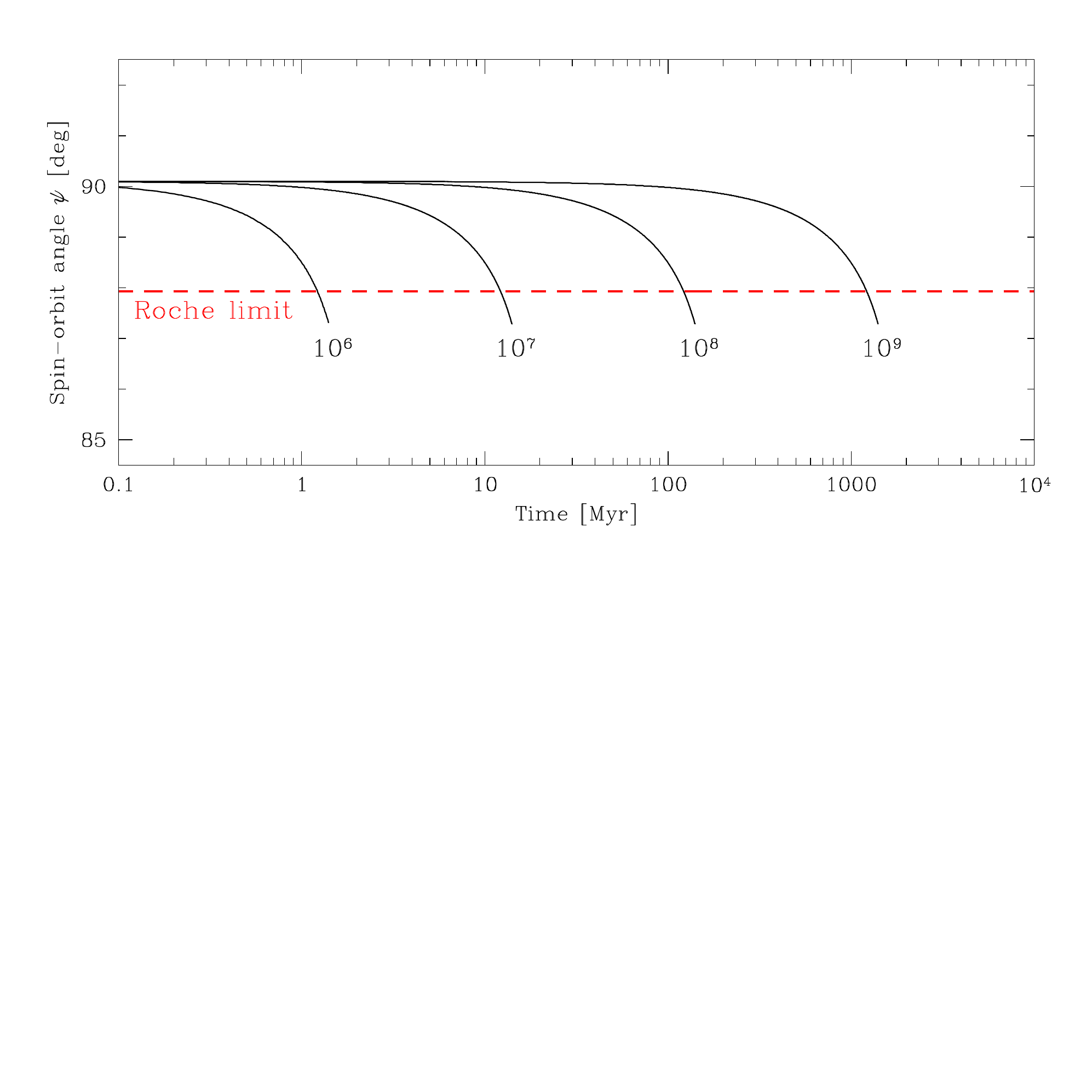}
\includegraphics[bb=15 280 550 545, width=0.46\textwidth]{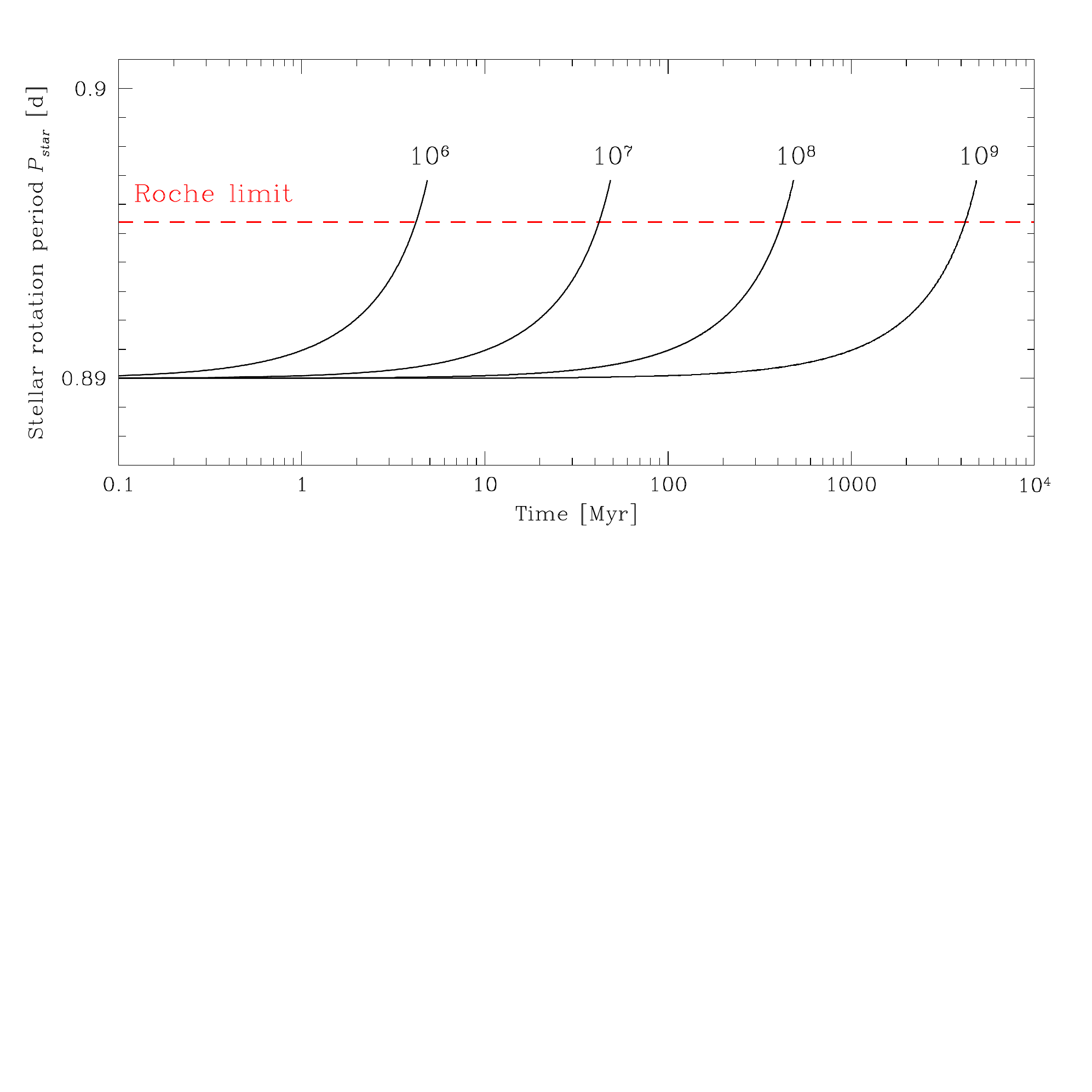}
\hspace{0.5cm}
\includegraphics[bb=15 280 550 545, width=0.46\textwidth]{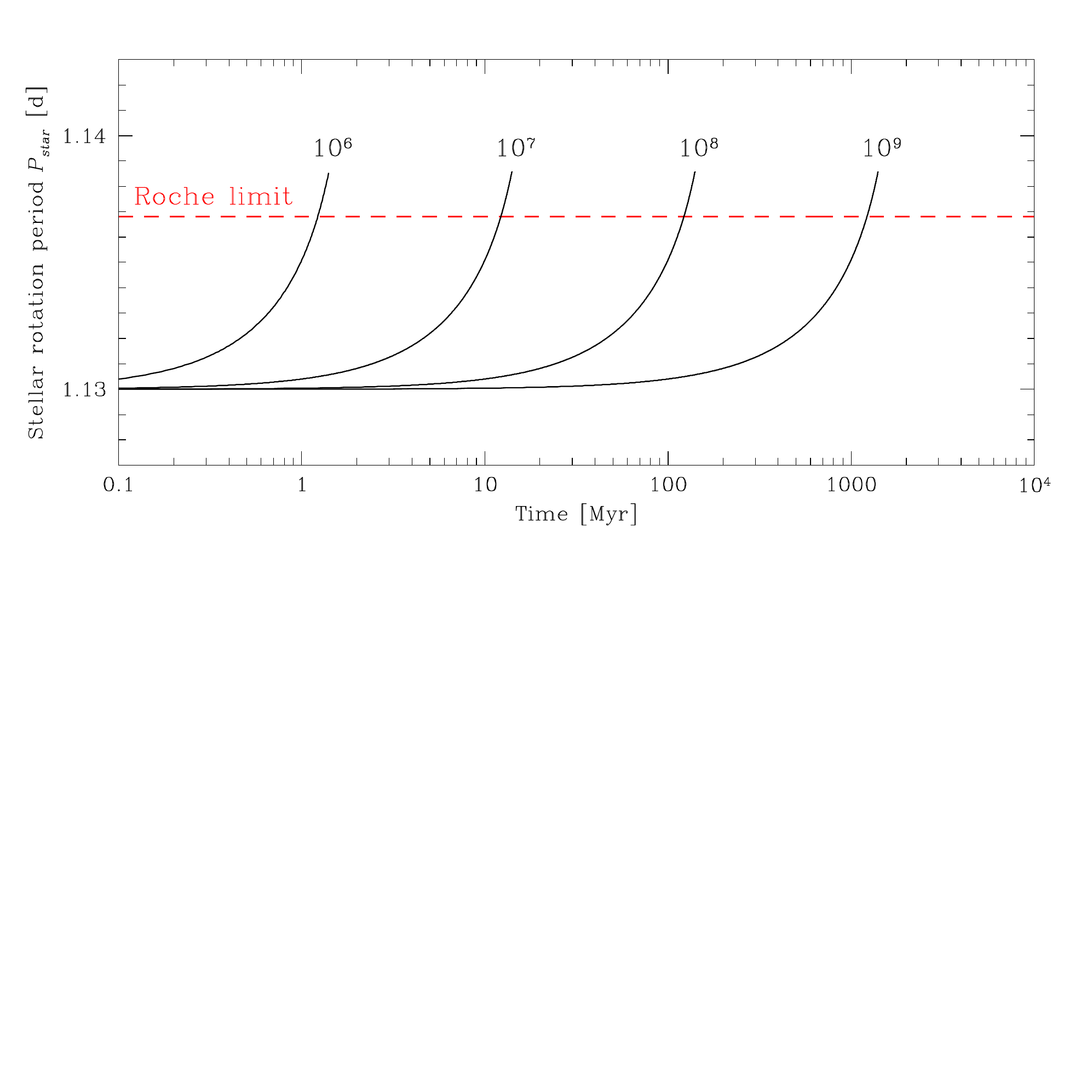}
\vspace{-0.4cm}
\caption{Future tidal evolution of the orbital semi-major axis (top), the spin-orbit angle (middle), and the stellar rotation period (bottom), assuming the current stellar rotation period $P_{\star,0}$ is 0.89 d (left panels) or 1.13 d (right panels). The labeled evolutions correspond to $Q'_{\star,0}$=$10^6$, $10^7$, $10^8$, and $10^9$. The red horizontal dashed lines represent the Roche limit.} 
\label{matsu}
\vspace{-0.2cm}
\end{figure*}

\subsection{Orbital evolution of WASP-121\,b}
\label{orbi}

\begin{figure}
\centering                    
\includegraphics[bb=15 280 550 545, width=0.46\textwidth]{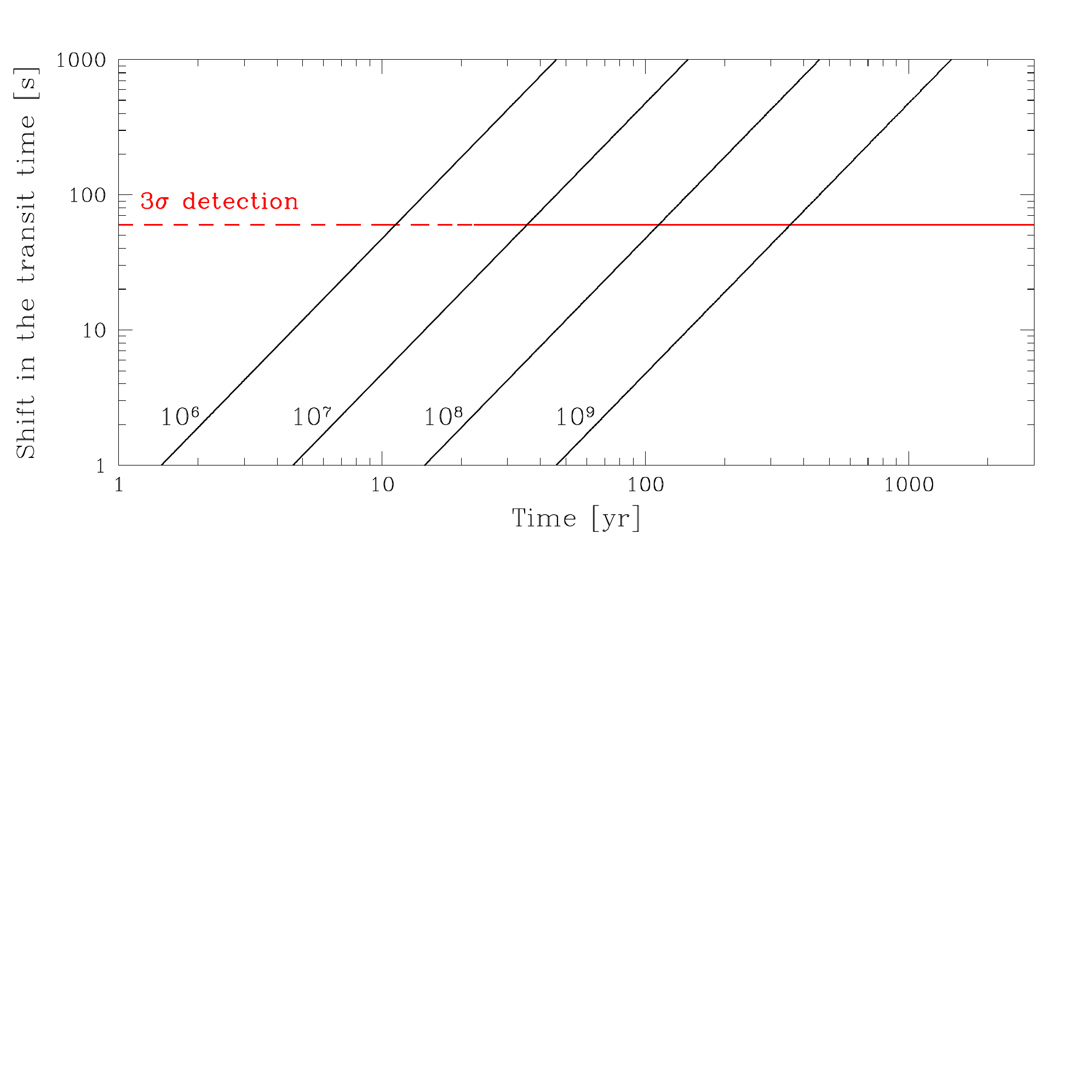}
\includegraphics[bb=15 280 550 545, width=0.46\textwidth]{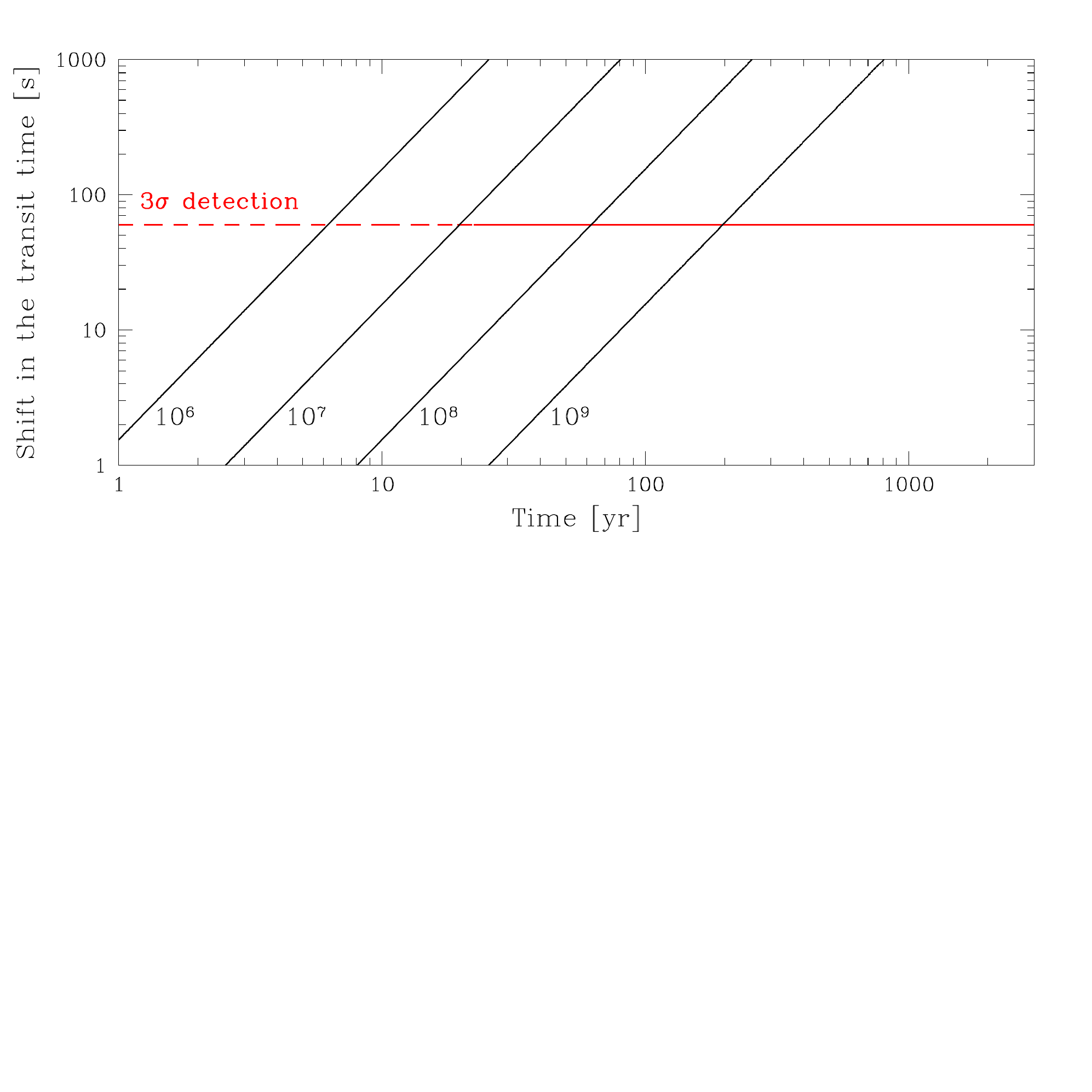}
\caption{Shift in the transit time of WASP-121\,b as a function of time, computed for different values of $Q'_{\star, 0}$, assuming the current stellar rotation period is 0.89 d (top) or 1.13 d (bottom). The labeled evolutions correspond to $Q'_{\star, 0}$=$10^6$, $10^7$, $10^8$, and $10^9$. The red horizontal line represents the 3$\sigma$ detection limit, assuming a timing accuracy of 20 s.} 
\label{ttv}
\end{figure}

\indent
In Section \ref{mcmc}, we find a sky-projected spin-orbit angle of $\beta$=$257.8_{-5.5}^{+5.3}$ deg. This high spin-orbit misalignment favors a migration of the planet involving strong dynamical events, such as planet-planet scattering (e.g. \citealt{rasio96}, \citealt{marzari}, \citealt{moorhead}, \citealt{chatterjee}) and/or Kozai-Lidov oscillations (e.g. \citealt{kozai}, \citealt{lidov}, \citealt{wu}, \citealt{fabtrem}). As shown in \cite{fabrycky09}, the true spin-orbit angle $\psi$, can be computed from $\beta$, $i_{\mathrm{p}}$ (the orbital inclination), and $i_{\mathrm{\star}}$ (the stellar inclination). If we assume that the stellar rotation period $P_{\mathrm{rot},\star}$ is either \hbox{0.89 days} or 1.13 days and we combine this information with our measured values for $R_{\star}$ and $v_{\star}$ sin $i_{\star}$, we obtain $i_{\star}$ of \hbox{9.4 $\pm$ 0.6 deg} and 11.9 $\pm$ 0.7 deg, respectively (see Section \ref{jitter}). These values, together with our measured $\beta$ and $i_{\mathrm{p}}$, yield $\psi$ of 89.6 $\pm$ 1.1 deg and 90.1 $\pm$ 1.2 deg, respectively, where the uncertainties are the quadratic sum of the uncertainties due to each input parameter. Both solutions are in excellent agreement and place the planet in a nearly polar orbit around its star, which is clearly a remarkable configuration as noted by e.g. \citealt{lai90} and \citealt{rogers90}.\\
\indent
The proximity of WASP-121\,b to its Roche limit suggests that its orbital evolution is now dominated by tidal interactions with its host star. To assess the future orbital evolution of the planet, we integrate the tidal evolution equations of \cite{matsumura} forwards in time, assuming that the planet is on a circular orbit (see \hbox{Section \ref{mcmc}}) and that its rotation is now synchronized with its orbit. Under these reasonable assumptions, the planet's orbital evolution is expected to depend only on the tidal dissipation inside the star (\citealt{matsumura}). The efficiency of this tidal dissipation is parameterized by the stellar tidal dissipation factor $Q'_{\star}$, with a higher $Q'_{\star}$ meaning a weaker tidal dissipation. We assume here that \hbox{$Q'_{\star} \propto 1/\vert2\omega_{\star}-2n\vert$}, where $\omega_{\star}$ is the stellar angular velocity and $n$ is the planet's mean motion (\hbox{$n$ = 2$\pi$/$P$}), as recommended by \cite{matsumura} for the cases where the stellar rotation is not yet synchronized with the orbit (stellar rotation period $P_{\star} \neq P$). Under this assumption, $Q'_{\star}$ changes as $Q'_{\star}=Q'_{0}\:\vert2\omega_{\star}-2n\vert/\vert2\omega_{\star,0}-2n_{0}\vert$, where 0 index indicates the current values. Fig. \ref{matsu} shows the obtained evolutions for the semi-major axis $a$ (top), the spin-orbit angle $\psi$ (middle), and the stellar rotation period $P_{\star}$ (bottom), assuming different values between $10^6$ and $10^9$ for the current stellar tidal dissipation factor $Q'_{\star,0}$ and a current stellar rotation period $P_{\star,0}$ of 0.89 d (left panels) or 1.13 d (right panels). For all cases, the model shows that the planet will continue to approach its host star until reaching its Roche limit, where it will be finally tidally disrupted. Assuming $P_{\star,0}$=0.89/1.13 d, we find remaining lifetimes of \hbox{4.20/1.22 Myr}, \hbox{42.0/12.2 Myr}, 420/122 Myr, and \hbox{4.20/1.22 Gyr} for $Q'_{\star,0}$ values of $10^6$, $10^7$, $10^8$, and $10^9$, respectively. We note that for $Q'_{\star,0}$=$10^6$, the remaining lifetime of the planet would be $<$1\% of the youngest possible age of the system (500 Myr, see Section \ref{val}), thus giving a low probability for the planet to be detected now. This result could favor $Q'_{\star,0}$ values $\gtrsim$ $10^7$ for WASP-121, which would be in line with the expectation that hot stars \hbox{($T_{\mathrm{eff}}$ $>$ 6250 K)} would have low tidal dissipation efficiencies due to their thin or quasi-nonexistent convective envelopes (see e.g. \citealt{winn10hot}, \citealt{albrecht12}). For \hbox{WASP-121}, we have \hbox{$T_{\mathrm{eff}}$ = 6460 $\pm$ 140 K} and a thin convective envelope starting at 0.88 $R_{\star}$.\\
\indent
Combining the Matsumura et al.'s expression for $\mathrm{d}a/\mathrm{dt}$ and the Kepler's third law, we can calculate the current rate of orbital period change : \hbox{(d$P$/dt)$_{0}$ = -0.0035 ($10^6$/$Q'_{\star,0}$) $\mathrm{s}\;\mathrm{yr}^{-1}$} for $P_{\star,0}$=0.89 d and \hbox{(d$P$/dt)$_{0}$ = -0.0118 ($10^6$/$Q'_{\star,0}$) $\mathrm{s}\;\mathrm{yr}^{-1}$} for $P_{\star,0}$=1.13 d. We can then estimate how long it would take to observe significant transit timing variations (TTVs) due to the orbital decay of the planet using, e.g., the equation (7) of \cite{birkby}. We note that this estimation assumes a constant rate of orbital period change d$P$/dt=(d$P$/dt)$_{0}$. \hbox{Fig. \ref{ttv}} shows the evolution of the shift in the transit time as a function of time for the different $Q'_{\star,0}$, and for $P_{\star,0}$=0.89 d (top) or $P_{\star,0}$=1.13 d (bottom). Assuming that a timing accuracy of $\sim$20 s is achievable with current instrumentation (see e.g. \citealt{gillon09}) and that \hbox{$P_{\star,0}$=0.89/1.13 d}, transit timing variations could be detected at 3$\sigma$ after \hbox{$\sim$11/6 yrs} for $Q'_{\star,0}$=$10^6$, $\sim$35/19 yrs for $Q'_{\star,0}$=$10^7$, $\sim$110/60 yrs for $Q'_{\star,0}$=$10^8$, and $\sim$348/189 yrs for $Q'_{\star,0}$=$10^9$, respectively. For low values of $Q'_{\star,0}$, the decay of WASP-121\,b's orbit could thus be detectable on the decade timescale. Alternatively, the non-detection of TTVs in the WASP-121\,b system after a certain amount of time would allow to put a lower limit on $Q'_{\star,0}$ and help constrain tidal evolution theories.

\vspace{-0.55cm}
\subsection{Validity of the spherical approximation for the star WASP-121}

\indent
The likely stellar rotation periods of 0.89 days and \hbox{1.13 days} would imply, together with our measured value for $R_{\star}$, rotation velocities $v_{\star}$ of 82.9 $\pm$ 1.7 $\mathrm{km}\,\mathrm{s}^{-1}$ and \hbox{65.3 $\pm$ 1.4 $\mathrm{km}\,\mathrm{s}^{-1}$}, respectively (see Section \ref{jitter}). Such high $v_{\star}$ might imply a non-negligible deformation of the star due to its fast rotation. Using the Roche model from \cite{maederval}, this deformation can be expressed as \hbox{$R_{\star, \mathrm{eq}}/R_{\star, \mathrm{pol}}$ = 1 $+$ (1/2) $(\omega_{\star}/\omega_{\mathrm{\star, K}})^{2}$}, where $\omega_{\star}$ is the rotation velocity; $\omega_{\star, \mathrm{K}}$ the critical break-up velocity; $R_{\mathrm{\star, eq}}$ the equatorial radius; and $R_{\mathrm{\star, pol}}$ the polar radius. We estimate the critical velocity from the Keplerian velocity $\sqrt{GM_{\star}/R_{\star, \mathrm{eq}}^{3}}$, for which the centrifugal force equals gravitational forces at the equator of the star. For WASP-121 \hbox{(Table \textcolor{blue}{4})}, $\omega_{\star}$ = 0.17 $\omega_{\star, \mathrm{K}}$, which gives $R_{\star, \mathrm{eq}}/R_{\mathrm{\star, pol}}$ = 1.015. The deformation of the star induced by its possible rapid rotation would thus not strongly bias our stellar radius measurement (which has an error bar $\sim$2$\%$), even if it was obtained assuming a spherical shape for the star (eclipse model of \citealt{mandel}, see Section \ref{mcmc}).

\subsection{First constraints on the atmospheric properties of WASP-121\,b}
\label{discussionatm}

\vspace{0.3cm}

\begin{figure}
\centering                    
\includegraphics[bb=5 5 280 230, width=0.49\textwidth]{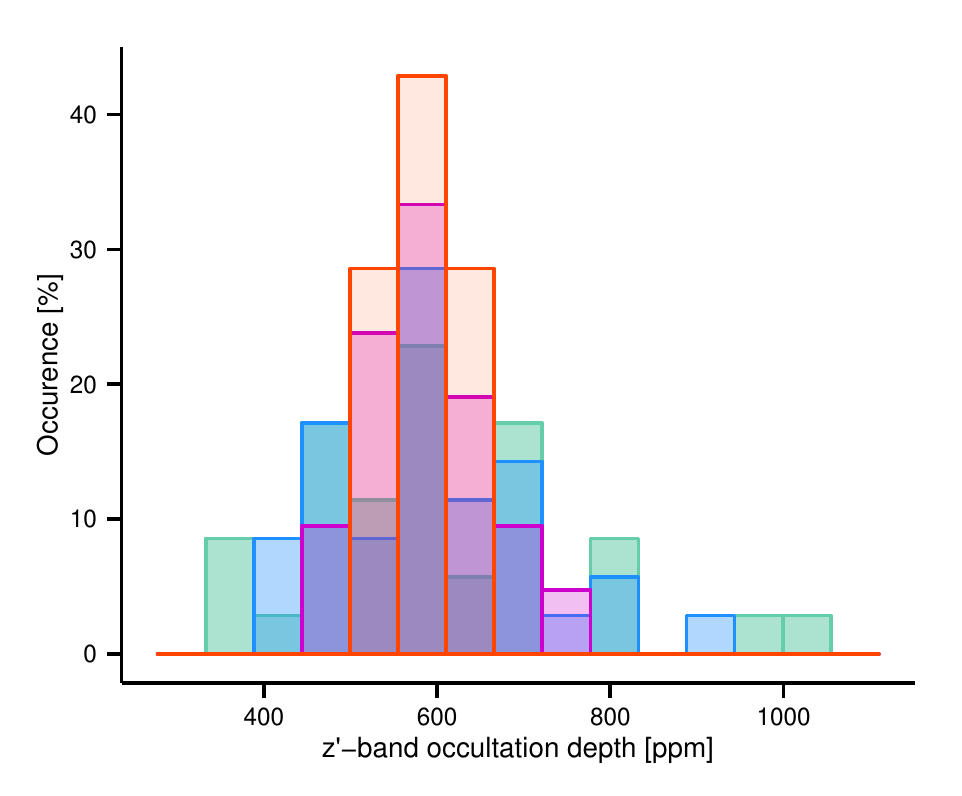}
\vspace{-0.45cm}
\caption{Histograms of the $z'$-band occultation depths derived from the MCMC analyses of subsets of three (green), four (blue), five (purple), and six (red) lightcurves chosen among the seven TRAPPIST occultation lightcurves.} 
\label{hist}
\end{figure}

\indent
The large radius of WASP-121\,b, its extreme irradiation, and the brightness of its host star make it an excellent target for atmospheric studies via secondary eclipse observations, with theoretical expectations for the planet-to-star IR flux ratio $>$0.05\% down to $\sim$0.9 $\mu$m.\\
\indent
By combining seven occultation lightcurves obtained with TRAPPIST, we detect the emission of the planet in the \hbox{$z'$-band} at better than $\sim$4$\sigma$, the measured occultation depth being \hbox{603 $\pm$ 130 ppm} (see Fig. \ref{lcsoccs}). To make sure that the detected signal does not result from a systematic effect present in one or several lightcurves, we followed the method presented in \cite{lendl19}. We thus divided our set of seven occultation lightcurves into subsets containing all possible combinations of three to six lightcurves and performed an MCMC analysis on each of them, while keeping all the jump parameters except the occultation depth fixed to the values derived from our global analysis \hbox{(Table \textcolor{blue}{4})}. Fig. \ref{hist} presents histograms of the derived occultation depths. We can see how the solutions converge towards our adopted value as we use an increasing number of occultation lightcurves.\\
\indent
Our measured occultation depth can be translated into a brightness temperature $T_{\mathrm{br}}$ of $3553_{-178}^{+160}$ K. In this calculation, we considered the planet as a blackbody of temperature $T_{\mathrm{br}}$ and used a Kurucz model spectrum for the star (\citealt{kurucz}). We also assumed that all the planet's $z'$-band flux arises from thermal emission, which is reasonable as \cite{seagerlopez} showed that reflected light contributes 10 to 20 times less than thermal emission at these wavelengths for very hot Jupiters, such as WASP-121\,b. We then defined our measured occultation depth as the product of the planet-to-star area ratio and the ratio of the TRAPPIST $z'$ bandpass-integrated planet-to-star fluxes (e.g. \citealt{charbonneau}), and adopted as brightness temperature the blackbody temperature that yielded the best match to our measured occultation depth. The uncertainty on $T_{\mathrm{br}}$ only accounts for the uncertainty on the measured occultation depth, which is the main source of error.\\
\indent
We can compare this brightness temperature to the equilibrium temperature of the planet, which is given by \hbox{$T_{\mathrm{eq}}$ = $T_{\mathrm{eff}} \sqrt{R_{\star}/a}\:\left[f\,(1-A_{\mathrm{B}})\right]^{1/4}$}, where $f$ and $A_{\mathrm{B}}$ are the reradiation factor and the Bond albedo of the planet, respectively (\citealt{seagerlopez}). The factor $f$ ranges from 1/4 to 2/3, where $f$=1/4 indicates efficient heat redistribution and isotropic reradiation over the whole planet and $f$=2/3 corresponds to instantaneous reradiation of incident radiation with no heat redistribution. The brightness temperature is significantly higher ($\sim$7$\sigma$) than the maximum equilibrium temperature of \hbox{$3013_{-75}^{+77}$ K} obtained assuming $A_{\mathrm{B}}$=0 and $f$=2/3, which could suggest that the planet has a low Bond albedo coupled to a poor heat redistribution efficiency. This would agree with the trend noted by \cite{cowan} that hottest planets are less efficient at redistributing the incident stellar energy than colder planets, the explanation proposed for this trend being that the radiative timescale is shorter than the advective timescale for hotter planets, causing these planets to reradiate the incident stellar energy rather than advecting it through winds.\\
\indent
However, we note that the comparison of brightness temperature with equilibrium temperature has only limited physical meaning here as our $z'$-band observations probe thermal emission from deep layers ($P$$\sim$1 bar) of the planetary atmosphere, that can be hotter than the maximum equilibrium temperature (e.g. \citealt{madhu}). Furthermore, the emission spectrum of a hot Jupiter can deviate significantly from that of a blackbody. For instance, the flux observed in the $z'$-band can be noticeably increased (resp. decreased) by emission (resp. absorption) due to gaseous TiO, which is expected to be abundant if the atmosphere has a carbon-to-oxygen (C/O) ratio $<$1 (O-rich atmosphere)\footnote{Whether TiO is seen in emission or absorption in that case depends on the thermal structure of the planetary atmosphere (presence or lack of temperature inversion, see \citealt{madhu} for details).}.\\
\indent
Our $z'$-band occultation measurement thus provides a first observational constraint on the emission spectrum of \hbox{WASP-121\,b}. Combined with future observations at longer wavelengths (from the ground in the near-infrared $J$-, $H$-, and $K$- bands, or from space between 1.1 and 1.7 $\mu$m with HST/WFC3 or at 3.6 and 4.5 $\mu$m with \textit{Spitzer}/IRAC), it will allow to gain insights on the planet's dayside chemical composition and thermal structure.

%%%%%%%%%%%%%%%%%%%%%%%%%%%%%%%%%%%%%%
\section*{Acknowledgements}

We would like to thank B.-O. Demory and A. Correia for their useful advices. WASP-South is hosted by the South African Astronomical Observatory and we are grateful for their ongoing support and assistance. Funding for WASP comes from consortium universities and from UK's Science and Technology Facilities Council. TRAPPIST is a project funded by the Belgian Fund for Scientific Research (Fonds National de la Recherche Scientifique, F.R.S.-FNRS) under grant FRFC 2.5.594.09.F, with the participation of the Swiss National Science Fundation (SNF). The Swiss {\it Euler} Telescope is operated by the University of Geneva, and is funded by the Swiss National Science Foundation. L. Delrez acknowledges support of the F.R.I.A. fund of the FNRS. M. Gillon, E. Jehin and V. Van Grootel are FNRS Research Associates. M. Lendl acknowledges support of the European Research Council through
the European Union’s Seventh Framework Programme (FP7/2007-2013)/ERC grant agreement number 336480. A. H. M. J. Triaud received funding from a fellowship provided by the Swiss National Science Foundation under grant number P300P2-147773. A. Santerne is supported by the European Union under a Marie Curie Intra-European Fellowship for Career Development with reference FP7-PEOPLE-2013-IEF, number 627202. We thank the anonymous referee for her/his valuable suggestions.

\bibliographystyle{mnras}
\bibliography{WASP_121}

\begin{thebibliography}{}
\makeatletter
\relax
\def\mn@urlcharsother{\let\do\@makeother \do\$\do\&\do\#\do\^\do\_\do\%\do\~}
\def\mn@doi{\begingroup\mn@urlcharsother \@ifnextchar [ {\mn@doi@}
  {\mn@doi@[]}}
\def\mn@doi@[#1]#2{\def\@tempa{#1}\ifx\@tempa\@empty \href
  {http://dx.doi.org/#2} {doi:#2}\else \href {http://dx.doi.org/#2} {#1}\fi
  \endgroup}
\def\mn@eprint#1#2{\mn@eprint@#1:#2::\@nil}
\def\mn@eprint@arXiv#1{\href {http://arxiv.org/abs/#1} {{\tt arXiv:#1}}}
\def\mn@eprint@dblp#1{\href {http://dblp.uni-trier.de/rec/bibtex/#1.xml}
  {dblp:#1}}
\def\mn@eprint@#1:#2:#3:#4\@nil{\def\@tempa {#1}\def\@tempb {#2}\def\@tempc
  {#3}\ifx \@tempc \@empty \let \@tempc \@tempb \let \@tempb \@tempa \fi \ifx
  \@tempb \@empty \def\@tempb {arXiv}\fi \@ifundefined
  {mn@eprint@\@tempb}{\@tempb:\@tempc}{\expandafter \expandafter \csname
  mn@eprint@\@tempb\endcsname \expandafter{\@tempc}}}

\bibitem[\protect\citeauthoryear{{Adams} et~al.,}{{Adams}
  et~al.}{2011}]{ogle56}
{Adams} E.~R.,  et~al., 2011, \mn@doi [The Astrophysical Journal]
  {10.1088/0004-637X/741/2/102}, \href
  {http://adsabs.harvard.edu/abs/2011ApJ...741..102A} {741, 102}

\bibitem[\protect\citeauthoryear{{Albrecht} et~al.,}{{Albrecht}
  et~al.}{2012}]{albrecht12}
{Albrecht} S.,  et~al., 2012, \mn@doi [The Astrophysical Journal]
  {10.1088/0004-637X/757/1/18}, \href
  {http://adsabs.harvard.edu/abs/2012ApJ...757...18A} {757, 18}

\bibitem[\protect\citeauthoryear{{Allard}, {Homeier}  \& {Freytag}}{{Allard}
  et~al.}{2012}]{allard2012}
{Allard} F.,  {Homeier} D.,   {Freytag} B.,  2012, in {Richards} M.~T.,
  {Hubeny} I.,  eds,  IAU Symposium Vol. 282, IAU Symposium. pp 235--242,
  \mn@doi{10.1017/S1743921311027438}

\bibitem[\protect\citeauthoryear{{Anderson} et~al.,}{{Anderson}
  et~al.}{2013}]{anderson19}
{Anderson} D.~R.,  et~al., 2013, \mn@doi [Monthly Notices of the Royal
  Astronomical Society] {10.1093/mnras/stt140}, \href
  {http://adsabs.harvard.edu/abs/2013MNRAS.430.3422A} {430, 3422}

\bibitem[\protect\citeauthoryear{{Anderson} et~al.,}{{Anderson}
  et~al.}{2014}]{w111}
{Anderson} D.~R.,  et~al., 2014, preprint, \href
  {http://adsabs.harvard.edu/abs/2014arXiv1410.3449A} {} (\mn@eprint {arXiv}
  {1410.3449})

\bibitem[\protect\citeauthoryear{{Asplund}, {Grevesse}, {Sauval}  \&
  {Scott}}{{Asplund} et~al.}{2009}]{asplund}
{Asplund} M.,  {Grevesse} N.,  {Sauval} A.~J.,   {Scott} P.,  2009, \mn@doi
  [Annual Review of Astronomy and Astrophysics]
  {10.1146/annurev.astro.46.060407.145222}, 47, 481

\bibitem[\protect\citeauthoryear{{Bakos}, {Noyes}, {Kov{\'a}cs}, {Stanek},
  {Sasselov}  \& {Domsa}}{{Bakos} et~al.}{2004}]{bakos}
{Bakos} G.,  {Noyes} R.~W.,  {Kov{\'a}cs} G.,  {Stanek} K.~Z.,  {Sasselov}
  D.~D.,   {Domsa} I.,  2004, \mn@doi [The Publications of the Astronomical
  Society of the Pacific] {10.1086/382735}, \href
  {http://adsabs.harvard.edu/abs/2004PASP..116..266B} {116, 266}

\bibitem[\protect\citeauthoryear{{Batygin} \& {Stevenson}}{{Batygin} \&
  {Stevenson}}{2010}]{batygin}
{Batygin} K.,  {Stevenson} D.~J.,  2010, \mn@doi [The Astrophysical Journal
  Letters] {10.1088/2041-8205/714/2/L238}, \href
  {http://adsabs.harvard.edu/abs/2010ApJ...714L.238B} {714, L238}

\bibitem[\protect\citeauthoryear{{Birkby} et~al.,}{{Birkby}
  et~al.}{2014}]{birkby}
{Birkby} J.~L.,  et~al., 2014, \mn@doi [Monthly Notices of the Royal
  Astronomical Society] {10.1093/mnras/stu343}, \href
  {http://adsabs.harvard.edu/abs/2014MNRAS.440.1470B} {440, 1470}

\bibitem[\protect\citeauthoryear{{Bodenheimer}, {Lin}  \&
  {Mardling}}{{Bodenheimer} et~al.}{2001}]{bodenheimer}
{Bodenheimer} P.,  {Lin} D.~N.~C.,   {Mardling} R.~A.,  2001, \mn@doi [The
  Astrophysical Journal] {10.1086/318667}, \href
  {http://adsabs.harvard.edu/abs/2001ApJ...548..466B} {548, 466}

\bibitem[\protect\citeauthoryear{{B{\"o}hm-Vitense}}{{B{\"o}hm-Vitense}}{2004}]{bohm}
{B{\"o}hm-Vitense} E.,  2004, \mn@doi [The Astronomical Journal]
  {10.1086/425053}, \href {http://adsabs.harvard.edu/abs/2004AJ....128.2435B}
  {128, 2435}

\bibitem[\protect\citeauthoryear{{Budaj}}{{Budaj}}{2011}]{budaj}
{Budaj} J.,  2011, \mn@doi [The Astronomical Journal]
  {10.1088/0004-6256/141/2/59}, \href
  {http://adsabs.harvard.edu/abs/2011AJ....141...59B} {141, 59}

\bibitem[\protect\citeauthoryear{{Burrows}, {Hubeny}, {Budaj}  \&
  {Hubbard}}{{Burrows} et~al.}{2007}]{burrows}
{Burrows} A.,  {Hubeny} I.,  {Budaj} J.,   {Hubbard} W.~B.,  2007, \mn@doi [The
  Astrophysical Journal] {10.1086/514326}, \href
  {http://adsabs.harvard.edu/abs/2007ApJ...661..502B} {661, 502}

\bibitem[\protect\citeauthoryear{{Castelli} \& {Kurucz}}{{Castelli} \&
  {Kurucz}}{2004}]{kurucz}
{Castelli} F.,  {Kurucz} R.~L.,  2004, ArXiv Astrophysics e-prints:
  astro-ph/0405087, \href {http://adsabs.harvard.edu/abs/2004astro.ph..5087C}
  {}

\bibitem[\protect\citeauthoryear{{Chabrier} \& {Baraffe}}{{Chabrier} \&
  {Baraffe}}{2007}]{chabrier}
{Chabrier} G.,  {Baraffe} I.,  2007, \mn@doi [The Astrophysical Journal]
  {10.1086/518473}, \href {http://adsabs.harvard.edu/abs/2007ApJ...661L..81C}
  {661, L81}

\bibitem[\protect\citeauthoryear{{Chandrasekhar}}{{Chandrasekhar}}{1987}]{chandr}
{Chandrasekhar} S.,  1987, {Ellipsoidal figures of equilibrium, New York :
  Dover}

\bibitem[\protect\citeauthoryear{{Charbonneau} et~al.,}{{Charbonneau}
  et~al.}{2005}]{charbonneau}
{Charbonneau} D.,  et~al., 2005, \mn@doi [The Astrophysical Journal]
  {10.1086/429991}, \href {http://adsabs.harvard.edu/abs/2005ApJ...626..523C}
  {626, 523}

\bibitem[\protect\citeauthoryear{{Chatterjee}, {Ford}, {Matsumura}  \&
  {Rasio}}{{Chatterjee} et~al.}{2008}]{chatterjee}
{Chatterjee} S.,  {Ford} E.~B.,  {Matsumura} S.,   {Rasio} F.~A.,  2008,
  \mn@doi [The Astrophysical Journal] {10.1086/590227}, \href
  {http://adsabs.harvard.edu/abs/2008ApJ...686..580C} {686, 580}

\bibitem[\protect\citeauthoryear{{Claret} \& {Bloemen}}{{Claret} \&
  {Bloemen}}{2011}]{claret}
{Claret} A.,  {Bloemen} S.,  2011, \mn@doi [Astronomy and Astrophysics]
  {10.1051/0004-6361/201116451}, 529, A75

\bibitem[\protect\citeauthoryear{{Collier Cameron} et~al.,}{{Collier Cameron}
  et~al.}{2006}]{collier}
{Collier Cameron} A.,  et~al., 2006, \mn@doi [Monthly Notices of the Royal
  Astronomical Society] {10.1111/j.1365-2966.2006.11074.x}, \href
  {http://adsabs.harvard.edu/abs/2006MNRAS.373..799C} {373, 799}

\bibitem[\protect\citeauthoryear{{Collier Cameron} et~al.,}{{Collier Cameron}
  et~al.}{2007}]{collier2}
{Collier Cameron} A.,  et~al., 2007, \mn@doi [Monthly Notices of the Royal
  Astronomical Society] {10.1111/j.1365-2966.2007.12195.x}, \href
  {http://adsabs.harvard.edu/abs/2007MNRAS.380.1230C} {380, 1230}

\bibitem[\protect\citeauthoryear{{Cowan} \& {Agol}}{{Cowan} \&
  {Agol}}{2011}]{cowan}
{Cowan} N.~B.,  {Agol} E.,  2011, \mn@doi [The Astrophysical Journal]
  {10.1088/0004-637X/729/1/54}, \href
  {http://adsabs.harvard.edu/abs/2011ApJ...729...54C} {729, 54}

\bibitem[\protect\citeauthoryear{{Cumming}}{{Cumming}}{2010}]{cumming}
{Cumming} A.,  2010, {Exoplanets, edited by S. Seager, University of Arizona
  Press}.
pp 191--214

\bibitem[\protect\citeauthoryear{{Demory} \& {Seager}}{{Demory} \&
  {Seager}}{2011}]{demory}
{Demory} B.-O.,  {Seager} S.,  2011, \mn@doi [The Astrophysical Journal
  Supplement] {10.1088/0067-0049/197/1/12}, \href
  {http://adsabs.harvard.edu/abs/2011ApJS..197...12D} {197, 12}

\bibitem[\protect\citeauthoryear{{D{\'{\i}}az}, {Almenara}, {Santerne},
  {Moutou}, {Lethuillier}  \& {Deleuil}}{{D{\'{\i}}az} et~al.}{2014}]{diaz}
{D{\'{\i}}az} R.~F.,  {Almenara} J.~M.,  {Santerne} A.,  {Moutou} C.,
  {Lethuillier} A.,   {Deleuil} M.,  2014, \mn@doi [Monthly Notices of the
  Royal Astronomical Society] {10.1093/mnras/stu601}, \href
  {http://adsabs.harvard.edu/abs/2014MNRAS.441..983D} {441, 983}

\bibitem[\protect\citeauthoryear{{Dotter}, {Chaboyer}, {Jevremovi{\'c}},
  {Kostov}, {Baron}  \& {Ferguson}}{{Dotter} et~al.}{2008}]{dotter2008}
{Dotter} A.,  {Chaboyer} B.,  {Jevremovi{\'c}} D.,  {Kostov} V.,  {Baron} E.,
  {Ferguson} J.~W.,  2008, \mn@doi [The Astrophysical Journal Supplement
  Series] {10.1086/589654}, \href
  {http://adsabs.harvard.edu/abs/2008ApJS..178...89D} {178, 89}

\bibitem[\protect\citeauthoryear{{Doyle} et~al.,}{{Doyle}
  et~al.}{2013}]{amanda}
{Doyle} A.~P.,  et~al., 2013, \mn@doi [Monthly Notices of the Royal
  Astronomical Society] {10.1093/mnras/sts267}, \href
  {http://adsabs.harvard.edu/abs/2013MNRAS.428.3164D} {428, 3164}

\bibitem[\protect\citeauthoryear{{Doyle}, {Davies}, {Smalley}, {Chaplin}  \&
  {Elsworth}}{{Doyle} et~al.}{2014}]{amanda2}
{Doyle} A.~P.,  {Davies} G.~R.,  {Smalley} B.,  {Chaplin} W.~J.,   {Elsworth}
  Y.,  2014, \mn@doi [Monthly Notices of the Royal Astronomical Society]
  {10.1093/mnras/stu1692}, \href
  {http://adsabs.harvard.edu/abs/2014MNRAS.444.3592D} {444, 3592}

\bibitem[\protect\citeauthoryear{{Dravins}, {Lindegren}  \&
  {Nordlund}}{{Dravins} et~al.}{1981}]{dravins1981}
{Dravins} D.,  {Lindegren} L.,   {Nordlund} A.,  1981, Astronomy and
  Astrophysics, \href {http://adsabs.harvard.edu/abs/1981A%26A....96..345D}
  {96, 345}

\bibitem[\protect\citeauthoryear{{Dumusque}, {Boisse}  \& {Santos}}{{Dumusque}
  et~al.}{2014}]{dumusque}
{Dumusque} X.,  {Boisse} I.,   {Santos} N.~C.,  2014, \mn@doi [The
  Astrophysical Journal] {10.1088/0004-637X/796/2/132}, \href
  {http://adsabs.harvard.edu/abs/2014ApJ...796..132D} {796, 132}

\bibitem[\protect\citeauthoryear{{Enoch}, {Collier Cameron}  \&
  {Horne}}{{Enoch} et~al.}{2012}]{enoch2}
{Enoch} B.,  {Collier Cameron} A.,   {Horne} K.,  2012, \mn@doi [Astronomy and
  Astrophysics] {10.1051/0004-6361/201117317}, 540, A99

\bibitem[\protect\citeauthoryear{{Etzel}}{{Etzel}}{1981}]{etzel1981}
{Etzel} P.~B.,  1981, in {Carling} E.~B.,  {Kopal} Z.,  eds, Photometric and
  Spectroscopic Binary Systems. p.~111

\bibitem[\protect\citeauthoryear{{Fabrycky} \& {Tremaine}}{{Fabrycky} \&
  {Tremaine}}{2007}]{fabtrem}
{Fabrycky} D.,  {Tremaine} S.,  2007, \mn@doi [The Astrophysical Journal]
  {10.1086/521702}, \href {http://adsabs.harvard.edu/abs/2007ApJ...669.1298F}
  {669, 1298}

\bibitem[\protect\citeauthoryear{{Fabrycky} \& {Winn}}{{Fabrycky} \&
  {Winn}}{2009}]{fabrycky09}
{Fabrycky} D.~C.,  {Winn} J.~N.,  2009, \mn@doi [The Astrophysical Journal]
  {10.1088/0004-637X/696/2/1230}, \href
  {http://adsabs.harvard.edu/abs/2009ApJ...696.1230F} {696, 1230}

\bibitem[\protect\citeauthoryear{{Ford} \& {Rasio}}{{Ford} \&
  {Rasio}}{2006}]{fordrasio}
{Ford} E.~B.,  {Rasio} F.~A.,  2006, \mn@doi [The Astrophysical Journal
  Letters] {10.1086/500734}, \href
  {http://adsabs.harvard.edu/abs/2006ApJ...638L..45F} {638, L45}

\bibitem[\protect\citeauthoryear{{Fortney}, {Marley}  \& {Barnes}}{{Fortney}
  et~al.}{2007}]{fortney}
{Fortney} J.~J.,  {Marley} M.~S.,   {Barnes} J.~W.,  2007, \mn@doi [The
  Astrophysical Journal] {10.1086/512120}, \href
  {http://adsabs.harvard.edu/abs/2007ApJ...659.1661F} {659, 1661}

\bibitem[\protect\citeauthoryear{{Fossati} et~al.,}{{Fossati}
  et~al.}{2010}]{fossati1}
{Fossati} L.,  et~al., 2010, \mn@doi [The Astrophysical Journal Letters]
  {10.1088/2041-8205/714/2/L222}, \href
  {http://adsabs.harvard.edu/abs/2010ApJ...714L.222F} {714, L222}

\bibitem[\protect\citeauthoryear{{Fossati}, {Ayres}, {Haswell}, {Bohlender},
  {Kochukhov}  \& {Fl{\"o}er}}{{Fossati} et~al.}{2013}]{fossati2}
{Fossati} L.,  {Ayres} T.~R.,  {Haswell} C.~A.,  {Bohlender} D.,  {Kochukhov}
  O.,   {Fl{\"o}er} L.,  2013, \mn@doi [The Astrophysical Journal Letters]
  {10.1088/2041-8205/766/2/L20}, \href
  {http://adsabs.harvard.edu/abs/2013ApJ...766L..20F} {766, L20}

\bibitem[\protect\citeauthoryear{{Gelman} \& {Rubin}}{{Gelman} \&
  {Rubin}}{1992}]{GR}
{Gelman} A.,  {Rubin} D.,  1992, Stat. Sci., 7, 457

\bibitem[\protect\citeauthoryear{{Gillon} et~al.,}{{Gillon}
  et~al.}{2009}]{gillon09}
{Gillon} M.,  et~al., 2009, \mn@doi [Astronomy and Astrophysics]
  {10.1051/0004-6361:200810929}, 496, 259

\bibitem[\protect\citeauthoryear{{Gillon} et~al.,}{{Gillon}
  et~al.}{2010}]{gilloncorot}
{Gillon} M.,  et~al., 2010, \mn@doi [Astronomy and Astrophysics]
  {10.1051/0004-6361/200913507}, \href
  {http://adsabs.harvard.edu/abs/2010A%26A...511A...3G} {511, A3}

\bibitem[\protect\citeauthoryear{{Gillon}, {Jehin}, {Magain}, {Chantry},
  {Hutsem{\'e}kers}, {Manfroid}, {Queloz}  \& {Udry}}{{Gillon}
  et~al.}{2011}]{gillon}
{Gillon} M.,  {Jehin} E.,  {Magain} P.,  {Chantry} V.,  {Hutsem{\'e}kers} D.,
  {Manfroid} J.,  {Queloz} D.,   {Udry} S.,  2011, in European Physical Journal
  Web of Conferences. p.~6002 (\mn@eprint {arXiv} {1101.5807}),
  \mn@doi{10.1051/epjconf/20101106002}

\bibitem[\protect\citeauthoryear{{Gillon} et~al.,}{{Gillon}
  et~al.}{2012}]{gillon2}
{Gillon} M.,  et~al., 2012, \mn@doi [Astronomy and Astrophysics]
  {10.1051/0004-6361/201218817}, 542, A4

\bibitem[\protect\citeauthoryear{{Gillon} et~al.,}{{Gillon}
  et~al.}{2014}]{w103}
{Gillon} M.,  et~al., 2014, \mn@doi [Astronomy and Astrophysics]
  {10.1051/0004-6361/201323014}, 562, L3

\bibitem[\protect\citeauthoryear{{Gim{\'e}nez}}{{Gim{\'e}nez}}{2006}]{gimenez}
{Gim{\'e}nez} A.,  2006, \mn@doi [The Astrophysical Journal] {10.1086/507021},
  \href {http://adsabs.harvard.edu/abs/2006ApJ...650..408G} {650, 408}

\bibitem[\protect\citeauthoryear{{Goldreich} \& {Tremaine}}{{Goldreich} \&
  {Tremaine}}{1980}]{goldreichtrem}
{Goldreich} P.,  {Tremaine} S.,  1980, \mn@doi [The Astrophysical Journal]
  {10.1086/158356}, \href {http://adsabs.harvard.edu/abs/1980ApJ...241..425G}
  {241, 425}

\bibitem[\protect\citeauthoryear{{Gray}}{{Gray}}{2008}]{gray}
{Gray} D.~F.,  2008, {The Observation and Analysis of Stellar Photospheres,
  Cambridge, UK: Cambridge University Press, 2008}

\bibitem[\protect\citeauthoryear{{Guillochon}, {Ramirez-Ruiz}  \&
  {Lin}}{{Guillochon} et~al.}{2011}]{guillochon}
{Guillochon} J.,  {Ramirez-Ruiz} E.,   {Lin} D.,  2011, \mn@doi [The
  Astrophysical Journal] {10.1088/0004-637X/732/2/74}, \href
  {http://adsabs.harvard.edu/abs/2011ApJ...732...74G} {732, 74}

\bibitem[\protect\citeauthoryear{{Hartman} et~al.,}{{Hartman}
  et~al.}{2011}]{hartman}
{Hartman} J.~D.,  et~al., 2011, \mn@doi [The Astrophysical Journal]
  {10.1088/0004-637X/742/1/59}, \href
  {http://adsabs.harvard.edu/abs/2011ApJ...742...59H} {742, 59}

\bibitem[\protect\citeauthoryear{{Haswell} et~al.,}{{Haswell}
  et~al.}{2012}]{haswell}
{Haswell} C.~A.,  et~al., 2012, \mn@doi [The Astrophysical Journal]
  {10.1088/0004-637X/760/1/79}, \href
  {http://adsabs.harvard.edu/abs/2012ApJ...760...79H} {760, 79}

\bibitem[\protect\citeauthoryear{{Hebb} et~al.,}{{Hebb} et~al.}{2009}]{w12}
{Hebb} L.,  et~al., 2009, \mn@doi [The Astrophysical Journal]
  {10.1088/0004-637X/693/2/1920}, \href
  {http://adsabs.harvard.edu/abs/2009ApJ...693.1920H} {693, 1920}

\bibitem[\protect\citeauthoryear{{Hebb} et~al.,}{{Hebb} et~al.}{2010}]{w19}
{Hebb} L.,  et~al., 2010, \mn@doi [The Astrophysical Journal]
  {10.1088/0004-637X/708/1/224}, \href
  {http://adsabs.harvard.edu/abs/2010ApJ...708..224H} {708, 224}

\bibitem[\protect\citeauthoryear{{Hellier} et~al.,}{{Hellier}
  et~al.}{2011}]{hellierws}
{Hellier} C.,  et~al., 2011, in European Physical Journal Web of Conferences.
  p.~1004 (\mn@eprint {arXiv} {1012.2286}),
  \mn@doi{10.1051/epjconf/20101101004}

\bibitem[\protect\citeauthoryear{{Hellier} et~al.,}{{Hellier}
  et~al.}{2012}]{hellierpile}
{Hellier} C.,  et~al., 2012, \mn@doi [Monthly Notices of the Royal Astronomical
  Society] {10.1111/j.1365-2966.2012.21780.x}, \href
  {http://adsabs.harvard.edu/abs/2012MNRAS.426..739H} {426, 739}

\bibitem[\protect\citeauthoryear{{Henden}, {Levine}, {Terrell}  \&
  {Welch}}{{Henden} et~al.}{2015}]{henden2015}
{Henden} A.~A.,  {Levine} S.,  {Terrell} D.,   {Welch} D.~L.,  2015, in
  American Astronomical Society Meeting Abstracts. p. 336.16

\bibitem[\protect\citeauthoryear{{Holman} et~al.,}{{Holman}
  et~al.}{2006}]{holman}
{Holman} M.~J.,  et~al., 2006, \mn@doi [The Astrophysical Journal]
  {10.1086/508155}, \href {http://adsabs.harvard.edu/abs/2006ApJ...652.1715H}
  {652, 1715}

\bibitem[\protect\citeauthoryear{{Howard} et~al.,}{{Howard}
  et~al.}{2012}]{howard}
{Howard} A.~W.,  et~al., 2012, \mn@doi [The Astrophysical Journal Supplement]
  {10.1088/0067-0049/201/2/15}, \href
  {http://adsabs.harvard.edu/abs/2012ApJS..201...15H} {201, 15}

\bibitem[\protect\citeauthoryear{{Izotov} \& {Thuan}}{{Izotov} \&
  {Thuan}}{2010}]{izotov}
{Izotov} Y.~I.,  {Thuan} T.~X.,  2010, \mn@doi [The Astrophysical Journal
  Letters] {10.1088/2041-8205/710/1/L67}, \href
  {http://adsabs.harvard.edu/abs/2010ApJ...710L..67I} {710, L67}

\bibitem[\protect\citeauthoryear{{Jackson}, {Greenberg}  \& {Barnes}}{{Jackson}
  et~al.}{2008}]{jackson2}
{Jackson} B.,  {Greenberg} R.,   {Barnes} R.,  2008, \mn@doi [The Astrophysical
  Journal] {10.1086/529187}, \href
  {http://adsabs.harvard.edu/abs/2008ApJ...678.1396J} {678, 1396}

\bibitem[\protect\citeauthoryear{{Jehin} et~al.,}{{Jehin} et~al.}{2011}]{jehin}
{Jehin} E.,  et~al., 2011, The Messenger, \href
  {http://adsabs.harvard.edu/abs/2011Msngr.145....2J} {145, 2}

\bibitem[\protect\citeauthoryear{{Konacki}, {Torres}, {Jha}  \&
  {Sasselov}}{{Konacki} et~al.}{2003}]{ogle56a}
{Konacki} M.,  {Torres} G.,  {Jha} S.,   {Sasselov} D.~D.,  2003, \mn@doi
  [Nature] {10.1038/nature01379}, \href
  {http://adsabs.harvard.edu/abs/2003Natur.421..507K} {421, 507}

\bibitem[\protect\citeauthoryear{{Kordopatis} et~al.,}{{Kordopatis}
  et~al.}{2013}]{kordopatis2013}
{Kordopatis} G.,  et~al., 2013, \mn@doi [The Astronomical Journal]
  {10.1088/0004-6256/146/5/134}, \href
  {http://adsabs.harvard.edu/abs/2013AJ....146..134K} {146, 134}

\bibitem[\protect\citeauthoryear{{Kozai}}{{Kozai}}{1962}]{kozai}
{Kozai} Y.,  1962, \mn@doi [The Astronomical Journal] {10.1086/108790}, \href
  {http://adsabs.harvard.edu/abs/1962AJ.....67..591K} {67, 591}

\bibitem[\protect\citeauthoryear{{Kroupa}}{{Kroupa}}{2001}]{kroupa2001}
{Kroupa} P.,  2001, \mn@doi [Monthly Notices of the Royal Astronomical Society]
  {10.1046/j.1365-8711.2001.04022.x}, \href
  {http://adsabs.harvard.edu/abs/2001MNRAS.322..231K} {322, 231}

\bibitem[\protect\citeauthoryear{{Lai}}{{Lai}}{2012}]{lai90}
{Lai} D.,  2012, \mn@doi [Monthly Notices of the Royal Astronomical Society]
  {10.1111/j.1365-2966.2012.20893.x}, \href
  {http://adsabs.harvard.edu/abs/2012MNRAS.423..486L} {423, 486}

\bibitem[\protect\citeauthoryear{{Lai}, {Helling}  \& {van den Heuvel}}{{Lai}
  et~al.}{2010}]{lai}
{Lai} D.,  {Helling} C.,   {van den Heuvel} E.~P.~J.,  2010, \mn@doi [The
  Astrophysical Journal] {10.1088/0004-637X/721/2/923}, \href
  {http://adsabs.harvard.edu/abs/2010ApJ...721..923L} {721, 923}

\bibitem[\protect\citeauthoryear{{Lendl} et~al.,}{{Lendl} et~al.}{2012}]{lendl}
{Lendl} M.,  et~al., 2012, \mn@doi [Astronomy and Astrophysics]
  {10.1051/0004-6361/201219585}, \href
  {http://adsabs.harvard.edu/abs/2012A26A...544A..72L} {544, A72}

\bibitem[\protect\citeauthoryear{{Lendl}, {Gillon}, {Queloz}, {Alonso},
  {Fumel}, {Jehin}  \& {Naef}}{{Lendl} et~al.}{2013}]{lendl19}
{Lendl} M.,  {Gillon} M.,  {Queloz} D.,  {Alonso} R.,  {Fumel} A.,  {Jehin} E.,
    {Naef} D.,  2013, \mn@doi [Astronomy and Astrophysics]
  {10.1051/0004-6361/201220924}, \href
  {http://adsabs.harvard.edu/abs/2013A%26A...552A...2L} {552, A2}

\bibitem[\protect\citeauthoryear{{Li}, {Miller}, {Lin}  \& {Fortney}}{{Li}
  et~al.}{2010}]{li}
{Li} S.-L.,  {Miller} N.,  {Lin} D.~N.~C.,   {Fortney} J.~J.,  2010, \mn@doi
  [Nature] {10.1038/nature08715}, \href
  {http://adsabs.harvard.edu/abs/2010Natur.463.1054L} {463, 1054}

\bibitem[\protect\citeauthoryear{{Lidov}}{{Lidov}}{1962}]{lidov}
{Lidov} M.~L.,  1962, \mn@doi [Planetary and Space Science]
  {10.1016/0032-0633(62)90129-0}, 9, 719

\bibitem[\protect\citeauthoryear{{Lin} \& {Papaloizou}}{{Lin} \&
  {Papaloizou}}{1986}]{linpapa}
{Lin} D.~N.~C.,  {Papaloizou} J.,  1986, \mn@doi [The Astrophysical Journal]
  {10.1086/164653}, \href {http://adsabs.harvard.edu/abs/1986ApJ...309..846L}
  {309, 846}

\bibitem[\protect\citeauthoryear{{L{\'o}pez-Morales} \&
  {Seager}}{{L{\'o}pez-Morales} \& {Seager}}{2007}]{seagerlopez}
{L{\'o}pez-Morales} M.,  {Seager} S.,  2007, \mn@doi [The Astrophysical Journal
  Letters] {10.1086/522118}, \href
  {http://adsabs.harvard.edu/abs/2007ApJ...667L.191L} {667, L191}

\bibitem[\protect\citeauthoryear{{Lubow} \& {Ida}}{{Lubow} \&
  {Ida}}{2010}]{lubow}
{Lubow} S.~H.,  {Ida} S.,  2010, {Planet Migration, Exoplanets, edited by
  S.~Seager. University of Arizona Press}.
pp 347--371

\bibitem[\protect\citeauthoryear{{Madhusudhan}}{{Madhusudhan}}{2012}]{madhu}
{Madhusudhan} N.,  2012, \mn@doi [The Astrophysical Journal]
  {10.1088/0004-637X/758/1/36}, \href
  {http://adsabs.harvard.edu/abs/2012ApJ...758...36M} {758, 36}

\bibitem[\protect\citeauthoryear{{Maeder}}{{Maeder}}{2009}]{maederval}
{Maeder} A.,  2009, {Physics, Formation and Evolution of Rotating Stars.
  ~Springer Berlin Heidelberg}, \mn@doi{10.1007/978-3-540-76949-1.
}

\bibitem[\protect\citeauthoryear{{Magain}}{{Magain}}{1984}]{magain}
{Magain} P.,  1984, Astronomy and Astrophysics, 134, 189

\bibitem[\protect\citeauthoryear{{Mandel} \& {Agol}}{{Mandel} \&
  {Agol}}{2002}]{mandel}
{Mandel} K.,  {Agol} E.,  2002, \mn@doi [The Astrophysical Journal]
  {10.1086/345520}, \href {http://adsabs.harvard.edu/abs/2002ApJ...580L.171M}
  {580, L171}

\bibitem[\protect\citeauthoryear{{Marcy}, {Butler}, {Fischer}  \&
  {Vogt}}{{Marcy} et~al.}{2004}]{marcy}
{Marcy} G.~W.,  {Butler} R.~P.,  {Fischer} D.~A.,   {Vogt} S.~S.,  2004, in
  {Beaulieu} J.,  {Lecavelier Des Etangs} A.,   {Terquem} C.,  eds,
  Astronomical Society of the Pacific Conference Series Vol. 321, Extrasolar
  Planets: Today and Tomorrow. p.~3

\bibitem[\protect\citeauthoryear{{Matsumura}, {Peale}  \& {Rasio}}{{Matsumura}
  et~al.}{2010}]{matsumura}
{Matsumura} S.,  {Peale} S.~J.,   {Rasio} F.~A.,  2010, \mn@doi [The
  Astrophysical Journal] {10.1088/0004-637X/725/2/1995}, \href
  {http://adsabs.harvard.edu/abs/2010ApJ...725.1995M} {725, 1995}

\bibitem[\protect\citeauthoryear{{Maxted} et~al.,}{{Maxted}
  et~al.}{2011}]{maxted}
{Maxted} P.~F.~L.,  et~al., 2011, \mn@doi [The Publications of the Astronomical
  Society of the Pacific] {10.1086/660007}, \href
  {http://adsabs.harvard.edu/abs/2011PASP..123..547M} {123, 547}

\bibitem[\protect\citeauthoryear{{Mayor} et~al.,}{{Mayor}
  et~al.}{2011}]{mayor2011}
{Mayor} M.,  et~al., 2011, preprint, \href
  {http://adsabs.harvard.edu/abs/2011arXiv1109.2497M} {} (\mn@eprint {arXiv}
  {1109.2497})

\bibitem[\protect\citeauthoryear{{McLaughlin}}{{McLaughlin}}{1924}]{mclaughlin}
{McLaughlin} D.~B.,  1924, \mn@doi [The Astrophysical Journal]
  {10.1086/142826}, \href {http://adsabs.harvard.edu/abs/1924ApJ....60...22M}
  {60, 22}

\bibitem[\protect\citeauthoryear{{Meibom} et~al.,}{{Meibom}
  et~al.}{2013}]{meibom13}
{Meibom} S.,  et~al., 2013, \mn@doi [Nature] {10.1038/nature12279}, \href
  {http://adsabs.harvard.edu/abs/2013Natur.499...55M} {499, 55}

\bibitem[\protect\citeauthoryear{{Melo} et~al.,}{{Melo} et~al.}{2007}]{melo}
{Melo} C.,  et~al., 2007, \mn@doi [Astronomy and Astrophysics]
  {10.1051/0004-6361:20066845}, 467, 721

\bibitem[\protect\citeauthoryear{{Moorhead} \& {Adams}}{{Moorhead} \&
  {Adams}}{2005}]{moorhead}
{Moorhead} A.~V.,  {Adams} F.~C.,  2005, \mn@doi [Icarus]
  {10.1016/j.icarus.2005.05.005}, \href
  {http://adsabs.harvard.edu/abs/2005Icar..178..517M} {178, 517}

\bibitem[\protect\citeauthoryear{{Morton} \& {Johnson}}{{Morton} \&
  {Johnson}}{2011}]{morton2011}
{Morton} T.~D.,  {Johnson} J.~A.,  2011, \mn@doi [The Astrophysical Journal]
  {10.1088/0004-637X/738/2/170}, \href
  {http://adsabs.harvard.edu/abs/2011ApJ...738..170M} {738, 170}

\bibitem[\protect\citeauthoryear{{Murray} \& {Correia}}{{Murray} \&
  {Correia}}{2010}]{murray}
{Murray} C.~D.,  {Correia} A.~C.~M.,  2010, {Keplerian Orbits and Dynamics of
  Exoplanets, Exoplanets, edited by S. Seager. University of Arizona Press}.
pp 15--23

\bibitem[\protect\citeauthoryear{{Nelson} \& {Davis}}{{Nelson} \&
  {Davis}}{1972}]{nelson1972}
{Nelson} B.,  {Davis} W.~D.,  1972, \mn@doi [Astrophysical Journal]
  {10.1086/151524}, \href {http://adsabs.harvard.edu/abs/1972ApJ...174..617N}
  {174, 617}

\bibitem[\protect\citeauthoryear{{Noels} \& {Montalb{\'a}n}}{{Noels} \&
  {Montalb{\'a}n}}{2013}]{noelsmontal}
{Noels} A.,  {Montalb{\'a}n} J.,  2013, in {Shibahashi} H.,  {Lynas-Gray}
  A.~E.,  eds,  Astronomical Society of the Pacific Conference Series Vol. 479,
  Astronomical Society of the Pacific Conference Series. p.~435

\bibitem[\protect\citeauthoryear{{Noyes}, {Hartmann}, {Baliunas}, {Duncan}  \&
  {Vaughan}}{{Noyes} et~al.}{1984}]{noyes84}
{Noyes} R.~W.,  {Hartmann} L.~W.,  {Baliunas} S.~L.,  {Duncan} D.~K.,
  {Vaughan} A.~H.,  1984, \mn@doi [Astrophysical Journal] {10.1086/161945},
  \href {http://adsabs.harvard.edu/abs/1984ApJ...279..763N} {279, 763}

\bibitem[\protect\citeauthoryear{{Ogilvie}}{{Ogilvie}}{2009}]{ogilvie}
{Ogilvie} G.~I.,  2009, \mn@doi [Monthly Notices of the Royal Astronomical
  Society] {10.1111/j.1365-2966.2009.14814.x}, \href
  {http://adsabs.harvard.edu/abs/2009MNRAS.396..794O} {396, 794}

\bibitem[\protect\citeauthoryear{{Penev}, {Jackson}, {Spada}  \&
  {Thom}}{{Penev} et~al.}{2012}]{penev}
{Penev} K.,  {Jackson} B.,  {Spada} F.,   {Thom} N.,  2012, \mn@doi [The
  Astrophysical Journal] {10.1088/0004-637X/751/2/96}, \href
  {http://adsabs.harvard.edu/abs/2012ApJ...751...96P} {751, 96}

\bibitem[\protect\citeauthoryear{{Pepe}, {Mayor}, {Galland}, {Naef}, {Queloz},
  {Santos}, {Udry}  \& {Burnet}}{{Pepe} et~al.}{2002}]{pepe2}
{Pepe} F.,  {Mayor} M.,  {Galland} F.,  {Naef} D.,  {Queloz} D.,  {Santos}
  N.~C.,  {Udry} S.,   {Burnet} M.,  2002, \mn@doi [Astronomy and Astrophysics]
  {10.1051/0004-6361:20020433}, 388, 632

\bibitem[\protect\citeauthoryear{{Pollacco} et~al.,}{{Pollacco}
  et~al.}{2006}]{pollacco}
{Pollacco} D.~L.,  et~al., 2006, \mn@doi [The Publications of the Astronomical
  Society of the Pacific] {10.1086/508556}, \href
  {http://adsabs.harvard.edu/abs/2006PASP..118.1407P} {118, 1407}

\bibitem[\protect\citeauthoryear{{Popper} \& {Etzel}}{{Popper} \&
  {Etzel}}{1981}]{popper1981}
{Popper} D.~M.,  {Etzel} P.~B.,  1981, \mn@doi [Astronomical Journal]
  {10.1086/112862}, \href {http://adsabs.harvard.edu/abs/1981AJ.....86..102P}
  {86, 102}

\bibitem[\protect\citeauthoryear{{Queloz} et~al.,}{{Queloz}
  et~al.}{2001}]{queloz}
{Queloz} D.,  et~al., 2001, \mn@doi [Astronomy and Astrophysics]
  {10.1051/0004-6361:20011308}, \href
  {http://adsabs.harvard.edu/abs/2001A26A...379..279Q} {379, 279}

\bibitem[\protect\citeauthoryear{{Raghavan} et~al.,}{{Raghavan}
  et~al.}{2010}]{raghavan2010}
{Raghavan} D.,  et~al., 2010, \mn@doi [The Astrophysical Journal Supplement]
  {10.1088/0067-0049/190/1/1}, \href
  {http://adsabs.harvard.edu/abs/2010ApJS..190....1R} {190, 1}

\bibitem[\protect\citeauthoryear{{Rasio} \& {Ford}}{{Rasio} \&
  {Ford}}{1996}]{rasio96}
{Rasio} F.~A.,  {Ford} E.~B.,  1996, \mn@doi [Science]
  {10.1126/science.274.5289.954}, \href
  {http://adsabs.harvard.edu/abs/1996Sci...274..954R} {274, 954}

\bibitem[\protect\citeauthoryear{{Rogers} \& {Lin}}{{Rogers} \&
  {Lin}}{2013}]{rogers90}
{Rogers} T.~M.,  {Lin} D.~N.~C.,  2013, \mn@doi [The Astrophysical Journal
  Letters] {10.1088/2041-8205/769/1/L10}, \href
  {http://adsabs.harvard.edu/abs/2013ApJ...769L..10R} {769, L10}

\bibitem[\protect\citeauthoryear{{Rossiter}}{{Rossiter}}{1924}]{rossiter}
{Rossiter} R.~A.,  1924, \mn@doi [The Astrophysical Journal] {10.1086/142825},
  \href {http://adsabs.harvard.edu/abs/1924ApJ....60...15R} {60, 15}

\bibitem[\protect\citeauthoryear{{Santerne} et~al.,}{{Santerne}
  et~al.}{2014}]{santerne14}
{Santerne} A.,  et~al., 2014, \mn@doi [Astronomy and Astrophysics]
  {10.1051/0004-6361/201424158}, \href {http://adsabs.harvard.edu/abs/2014A}
  {571, A37}

\bibitem[\protect\citeauthoryear{{Santerne} et~al.,}{{Santerne}
  et~al.}{2015}]{santerne2015}
{Santerne} A.,  et~al., 2015, ArXiv e-prints 1505.02663, \href
  {http://adsabs.harvard.edu/abs/2015arXiv150502663S} {}

\bibitem[\protect\citeauthoryear{{Santos} et~al.,}{{Santos}
  et~al.}{2002}]{santos}
{Santos} N.~C.,  et~al., 2002, \mn@doi [Astronomy and Astrophysics]
  {10.1051/0004-6361:20020876}, 392, 215

\bibitem[\protect\citeauthoryear{{Scargle}}{{Scargle}}{1982}]{scargle82}
{Scargle} J.~D.,  1982, \mn@doi [Astrophysical Journal] {10.1086/160554}, \href
  {http://adsabs.harvard.edu/abs/1982ApJ...263..835S} {263, 835}

\bibitem[\protect\citeauthoryear{{Schwarz}}{{Schwarz}}{1978}]{schwarz}
{Schwarz} G.~E.,  1978, Ann. Statist., 6, 461

\bibitem[\protect\citeauthoryear{{Scuflaire}, {Th{\'e}ado}, {Montalb{\'a}n},
  {Miglio}, {Bourge}, {Godart}, {Thoul}  \& {Noels}}{{Scuflaire}
  et~al.}{2008}]{scuflaire}
{Scuflaire} R.,  {Th{\'e}ado} S.,  {Montalb{\'a}n} J.,  {Miglio} A.,  {Bourge}
  P.-O.,  {Godart} M.,  {Thoul} A.,   {Noels} A.,  2008, \mn@doi [Astrophysics
  and Space Science] {10.1007/s10509-007-9650-1}, 316, 83

\bibitem[\protect\citeauthoryear{{Seager} \& {Deming}}{{Seager} \&
  {Deming}}{2010}]{seagerdeming}
{Seager} S.,  {Deming} D.,  2010, \mn@doi [Annual Review of Astronomy and
  Astrophysics] {10.1146/annurev-astro-081309-130837}, \href
  {http://adsabs.harvard.edu/abs/2010ARA%26A..48..631S} {48, 631}

\bibitem[\protect\citeauthoryear{{Seager} \& {Mall{\'e}n-Ornelas}}{{Seager} \&
  {Mall{\'e}n-Ornelas}}{2003}]{seager}
{Seager} S.,  {Mall{\'e}n-Ornelas} G.,  2003, \mn@doi [The Astrophysical
  Journal] {10.1086/346105}, \href
  {http://adsabs.harvard.edu/abs/2003ApJ...585.1038S} {585, 1038}

\bibitem[\protect\citeauthoryear{{Showman} \& {Guillot}}{{Showman} \&
  {Guillot}}{2002}]{showman}
{Showman} A.~P.,  {Guillot} T.,  2002, \mn@doi [Astronomy and Astrophysics]
  {10.1051/0004-6361:20020101}, 385, 166

\bibitem[\protect\citeauthoryear{{Southworth}}{{Southworth}}{2008}]{southworth2008}
{Southworth} J.,  2008, \mn@doi [Monthly Notices of the Royal Astronomical
  Society] {10.1111/j.1365-2966.2008.13145.x}, \href
  {http://adsabs.harvard.edu/abs/2008MNRAS.386.1644S} {386, 1644}

\bibitem[\protect\citeauthoryear{{Southworth} et~al.,}{{Southworth}
  et~al.}{2015}]{southworth103}
{Southworth} J.,  et~al., 2015, \mn@doi [Monthly Notices of the Royal
  Astronomical Society] {10.1093/mnras/stu2394}, \href
  {http://adsabs.harvard.edu/abs/2015MNRAS.447..711S} {447, 711}

\bibitem[\protect\citeauthoryear{{Stetson}}{{Stetson}}{1987}]{stetson}
{Stetson} P.~B.,  1987, \mn@doi [The Publications of the Astronomical Society
  of the Pacific] {10.1086/131977}, \href
  {http://adsabs.harvard.edu/abs/1987PASP...99..191S} {99, 191}

\bibitem[\protect\citeauthoryear{{Szab{\'o}} et~al.,}{{Szab{\'o}}
  et~al.}{2011}]{szabo2011}
{Szab{\'o}} G.~M.,  et~al., 2011, \mn@doi [The Astrophysical Journal Letters]
  {10.1088/2041-8205/736/1/L4}, \href
  {http://adsabs.harvard.edu/abs/2011ApJ...736L...4S} {736, L4}

\bibitem[\protect\citeauthoryear{{Tanaka}, {Takeuchi}  \& {Ward}}{{Tanaka}
  et~al.}{2002}]{tanaka}
{Tanaka} H.,  {Takeuchi} T.,   {Ward} W.~R.,  2002, \mn@doi [The Astrophysical
  Journal] {10.1086/324713}, \href
  {http://adsabs.harvard.edu/abs/2002ApJ...565.1257T} {565, 1257}

\bibitem[\protect\citeauthoryear{{Torres}, {Konacki}, {Sasselov}  \&
  {Jha}}{{Torres} et~al.}{2005}]{torres2005}
{Torres} G.,  {Konacki} M.,  {Sasselov} D.~D.,   {Jha} S.,  2005, \mn@doi [The
  Astrophysical Journal] {10.1086/426496}, \href
  {http://adsabs.harvard.edu/abs/2005ApJ...619..558T} {619, 558}

\bibitem[\protect\citeauthoryear{{Torres}, {Andersen}  \&
  {Gim{\'e}nez}}{{Torres} et~al.}{2010}]{torres}
{Torres} G.,  {Andersen} J.,   {Gim{\'e}nez} A.,  2010, \mn@doi [The Astronomy
  and Astrophysics Review] {10.1007/s00159-009-0025-1}, 18, 67

\bibitem[\protect\citeauthoryear{{Triaud} et~al.,}{{Triaud}
  et~al.}{2010}]{triaudrm}
{Triaud} A.~H.~M.~J.,  et~al., 2010, \mn@doi [Astronomy and Astrophysics]
  {10.1051/0004-6361/201014525}, 524, A25

\bibitem[\protect\citeauthoryear{{Triaud} et~al.,}{{Triaud}
  et~al.}{2011}]{triaud3}
{Triaud} A.~H.~M.~J.,  et~al., 2011, \mn@doi [Astronomy and Astrophysics]
  {10.1051/0004-6361/201016367}, 531, A24

\bibitem[\protect\citeauthoryear{{Tuomi} \& {Jones}}{{Tuomi} \&
  {Jones}}{2012}]{tuomi2012}
{Tuomi} M.,  {Jones} H.~R.~A.,  2012, \mn@doi [Astronomy and Astrophysics]
  {10.1051/0004-6361/201118114}, \href
  {http://adsabs.harvard.edu/abs/2012A%26A...544A.116T} {544, A116}

\bibitem[\protect\citeauthoryear{{Weidenschilling} \&
  {Marzari}}{{Weidenschilling} \& {Marzari}}{1996}]{marzari}
{Weidenschilling} S.~J.,  {Marzari} F.,  1996, \mn@doi [Nature]
  {10.1038/384619a0}, \href {http://adsabs.harvard.edu/abs/1996Natur.384..619W}
  {384, 619}

\bibitem[\protect\citeauthoryear{{Weiss} et~al.,}{{Weiss} et~al.}{2013}]{weiss}
{Weiss} L.~M.,  et~al., 2013, \mn@doi [The Astrophysical Journal]
  {10.1088/0004-637X/768/1/14}, \href
  {http://adsabs.harvard.edu/abs/2013ApJ...768...14W} {768, 14}

\bibitem[\protect\citeauthoryear{{Winn}}{{Winn}}{2010}]{winn}
{Winn} J.~N.,  2010, {Transits and Occultations, Exoplanets, edited by S.
  Seager. University of Arizona Press}.
pp 55--77

\bibitem[\protect\citeauthoryear{{Winn}, {Fabrycky}, {Albrecht}  \&
  {Johnson}}{{Winn} et~al.}{2010}]{winn10hot}
{Winn} J.~N.,  {Fabrycky} D.,  {Albrecht} S.,   {Johnson} J.~A.,  2010, \mn@doi
  [The Astrophysical Journal Letters] {10.1088/2041-8205/718/2/L145}, \href
  {http://adsabs.harvard.edu/abs/2010ApJ...718L.145W} {718, L145}

\bibitem[\protect\citeauthoryear{{Wright} et~al.,}{{Wright}
  et~al.}{2010}]{wright2010}
{Wright} E.~L.,  et~al., 2010, \mn@doi [The Astronomical Journal]
  {10.1088/0004-6256/140/6/1868}, \href
  {http://adsabs.harvard.edu/abs/2010AJ....140.1868W} {140, 1868}

\bibitem[\protect\citeauthoryear{{Wu} \& {Murray}}{{Wu} \& {Murray}}{2003}]{wu}
{Wu} Y.,  {Murray} N.,  2003, \mn@doi [The Astrophysical Journal]
  {10.1086/374598}, \href {http://adsabs.harvard.edu/abs/2003ApJ...589..605W}
  {589, 605}

\makeatother
\end{thebibliography}

\newpage
\appendix

\section{\texttt{PASTIS} MCMC analysis of the photometric data}

\begin{table*}
\caption{Priors used in the \texttt{PASTIS} analyses: $\mathcal{U}(a;b)$ represents a Uniform prior between $a$ and $b$; $\mathcal{N}(\mu;\sigma^{2})$ represents a Normal distribution with a mean of $\mu$ and a width of $\sigma^{2}$; $\mathcal{P}(\alpha; x_{min}; x_{max})$ represents a Power Law distribution with an exponent $\alpha$ computed between $x_{min}$ and $x_{max}$ ; $\mathcal{P}_{2}(\alpha_{1}; \alpha_{2}; x_{0}; x_{min}; x_{max})$ represents a double Power Law distribution with an exponent $\alpha_{1}$ computed between $x_{min}$ and $x_{0}$ and an exponent $\alpha_{2}$ computed between $x_{0}$ and $x_{max}$; and finally $\mathcal{S}(a;b)$ represents a Sine distribution between $a$ and $b$. \textbf{Notes.} $^{(a)}$ In \texttt{PASTIS}, planets are considered as non-self-emitting objects, hence no thermal emission. Only the reflected light is considered. Here, we used the albedo of the planet as a proxy for the depth of the occultation. The upper limit in the prior corresponds to a geometric albedo of 1 and a brightness temperature of 3500K. $^{(b)}$ The TRAPPIST photometry was performed on focus and no stellar contamination is detected within the photometric aperture. We defined a prior for the contamination centered on zero with an uncertainty of 1\% (positive and negative values are allowed) to account for possible variation of the sky background flux between the observations or for stellar brightness variation due to non-occulting spots. The EulerCam photometry was performed slightly out of focus, which means the target should be blended with the light from a $\sim$7.4\arcsec\, nearby star about 6.8 magnitudes fainter ($R$-band), resulting in a contamination of 0.2\%. The prior for the EulerCam contamination was thus centered on this value.}
\begin{center}
\begin{tabular}{lccc}
\hline
\hline
Parameter & Planet & Planet in Binary & Triple \\
\hline
\multicolumn{4}{l}{\textit{Target parameters}}\\
& & & \\
Effective temperature T$_{\rm eff}$ [K] & $\mathcal{N}(6460; 140)$ & $\mathcal{N}(6460; 140)$ & $\mathcal{N}(6460; 140)$ \\
Surface gravity $\log g$ [cm.s$^{-2}$] & $\mathcal{N}(4.2; 0.2)$ & $\mathcal{N}(4.2; 0.2)$ & $\mathcal{N}(4.2; 0.2)$ \\
Iron abondance [Fe/H] [dex] & $\mathcal{N}(0.13; 0.09)$ & $\mathcal{N}(0.13; 0.09)$ & $\mathcal{N}(0.13; 0.09)$ \\
Distance $d$ [pc] & $\mathcal{P}(2.0; 10; 10000)$ & $\mathcal{P}(2.0; 10; 10000)$ & $\mathcal{P}(2.0; 10; 10000)$\\
Interstellar extinction E(B-V) [mag] & $\mathcal{U}(0; 0.1)$ & $\mathcal{U}(0; 0.1)$ & $\mathcal{U}(0; 0.1)$ \\
\hline
\multicolumn{4}{l}{\textit{Planet parameters}}\\
& & & \\
Radius R$_{p}$ [R$_{jup}$] & $\mathcal{U}(0; 2.2)$ & $\mathcal{U}(0; 2.2)$ & --\\
Albedo$^{a}$ A$_{g}$ & $\mathcal{U}(0; 2.5)$ & $\mathcal{U}(0; 2.5)$ & --\\
\hline
\multicolumn{4}{l}{\textit{Binary parameters}}\\
& & & \\
Mass of stellar host M$_{2}$ [M$_{\odot}$] & -- & $\mathcal{P}_{2}(-1.3; -2.3; 0.5; 0.1; 2)$ & $\mathcal{P}_{2}(-1.3; -2.3; 0.5; 0.1; 2)$\\
Mass of stellar companion M$_{3}$ [M$_{\odot}$] & -- & -- & $\mathcal{P}_{2}(-1.3; -2.3; 0.5; 0.1; 2)$\\
\hline
\multicolumn{4}{l}{\textit{Orbital parameters}}\\
& & & \\
Orbital period $P$ [d] & $\mathcal{N}(1.2749281; 5.10^{-4})$ & $\mathcal{N}(1.2749281; 5.10^{-4})$ & $\mathcal{N}(1.2749281; 5.10^{-4})$ \\
Transit epoch $T_{0}$ [BJD - 2450000] & $\mathcal{N}(6635.70839; 0.03)$ & $\mathcal{N}(6635.70839; 0.03)$ & $\mathcal{N}(6635.70839; 0.03)$ \\
Orbital inclination $i$ [\degr] & $\mathcal{S}(60; 90)$ & $\mathcal{S}(60; 90)$ & $\mathcal{S}(60; 90)$ \\
\hline
\multicolumn{4}{l}{\textit{TRAPPIST data parameters} ($\times 12$)}\\
& & & \\
Contamination$^{b}$ & $\mathcal{N}(0; 0.01)$ & $\mathcal{N}(0; 0.01)$ & $\mathcal{N}(0; 0.01)$ \\
Out-of-transit flux & $\mathcal{U}(0.9; 1.1)$ & $\mathcal{U}(0.9; 1.1)$ & $\mathcal{U}(0.9; 1.1)$ \\
Jitter & $\mathcal{U}(0; 0.1)$ & $\mathcal{U}(0; 0.1)$ & $\mathcal{U}(0; 0.1)$ \\
\hline
\multicolumn{4}{l}{\textit{EulerCam data parameters} ($\times 4$)}\\
& & & \\
Contamination$^{b}$ & $\mathcal{N}(0.002; 0.01)$ & $\mathcal{N}(0.002; 0.01)$ & $\mathcal{N}(0.002; 0.01)$ \\
Out-of-transit flux & $\mathcal{U}(0.9; 1.1)$ & $\mathcal{U}(0.9; 1.1)$ & $\mathcal{U}(0.9; 1.1)$ \\
Jitter & $\mathcal{U}(0; 0.1)$ & $\mathcal{U}(0; 0.1)$ & $\mathcal{U}(0; 0.1)$ \\
\hline
\multicolumn{4}{l}{\textit{Spectral Energy Distribution parameter}}\\
& & & \\
Jitter [mag] & $\mathcal{U}(0; 1)$ & $\mathcal{U}(0; 1)$ & $\mathcal{U}(0; 1)$ \\
\hline
\hline
\end{tabular}
\end{center}
\label{PASTISpriors}
\end{table*}

\begin{table*}
\caption{Posterior distributions results of the \texttt{PASTIS} analyses. \textbf{Notes.} $^{(a)}$ This corresponds to an occultation depth of 620 $\pm$ 140 ppm. $^{(b)}$ This corresponds to an occultation depth of 2470 $\pm$ 370 ppm. $^{(c)}$ The error does not account for the uncertainty of the stellar models.}
\begin{center}
\begin{tabular}{lccc}
\hline
\hline
Parameter & Planet & Planet in Binary & Triple \\
\hline
\multicolumn{4}{l}{\textit{Target parameters}}\\
& & & \\
Effective temperature T$_{\rm eff}$ [K] & 6650 $\pm$ 60 & 6140 $\pm$ 150 & 6763$^{_{+46}}_{^{-160}}$\\
Surface gravity $\log g$ [cm.s$^{-2}$] & 4.245 $\pm$ 0.006 & 3.949$^{_{+0.031}}_{^{-0.014}}$ & 3.770$^{_{+0.002}}_{^{-0.004}}$\\
Iron abondance [Fe/H] [dex] & 0.1 $\pm$ 0.1 & -0.51$^{_{+0.01}}_{^{-0.22}}$ & -0.21$^{_{+0.05}}_{^{-0.01}}$\\
Distance $d$ [pc] & 261 $\pm$ 4.6 & 334 $\pm$ 12 & 578$^{_{+7}}_{^{-10}}$\\
Interstellar extinction E(B-V) [mag] & 0.009$^{_{+0.012}}_{^{-0.007}}$ & 0.0077$^{_{+0.014}}_{^{-0.006}}$ & 0.048$^{_{+0.014}}_{^{-0.029}}$\\
\hline
\multicolumn{4}{l}{\textit{Planet parameters}}\\
& & & \\
Radius R$_{p}$ [R$_{jup}$] & 1.76 $\pm$ 0.02 & 2.1998$^{_{+0.0001}}_{^{-0.0005}}$  & --\\
Albedo A$_{g}$ & 0.55 $\pm$ 0.12$^{a}$ & 1.14 $\pm$ 0.18$^{b}$ & --\\
\hline
\multicolumn{4}{l}{\textit{Binary parameters}}\\
& & & \\
Mass of eclipse host M$_{2}$ [M$_{\odot}$] & -- & 0.947 $\pm$ 0.007$^{c}$ & 1.41$^{_{+0.01}}_{^{-0.03}}$$^{c}$\\
Mass of eclipse companion M$_{3}$ [M$_{\odot}$] & -- & -- & 0.367$^{_{+0.003}}_{^{-0.008}}$$^{c}$\\
\hline
\multicolumn{4}{l}{\textit{Orbital parameters}}\\
& & & \\
Orbital period $P$ [d] & 1.27492477 $\pm$ 5.4 10$^{-7}$ & 1.2749280 $\pm$ 1.7 10$^{-6}$ & 1.27492495 $\pm$ 7.4 10$^{-7}$\\
Transit epoch $T_{0}$ [BJD - 2450000] & 6635.7077 $\pm$ 0.0001 & 6635.7077 $\pm$ 0.0034 & 6635.7077 $\pm$ 0.0002\\
Orbital inclination $i$ [\degr] & 89.13$^{_{+0.60}}_{^{-0.94}}$ & 89.90 $\pm$ 0.09 & 89.78$^{_{+0.16}}_{^{-0.23}}$\\
\hline
\multicolumn{4}{l}{\textit{Spectral Energy Distribution parameter}}\\
& & & \\
Jitter [mag] & 0.013 $\pm$ 0.013 & 0.170 $\pm$ 0.072 & 0.014$^{_{+0.014}}_{^{-0.009}}$\\
\hline
\hline
\end{tabular}
\end{center}
\label{PASTISresults}
\end{table*}

\label{lastpage}

\end{document}